\documentclass[prd,aps,amsmath,twocolumn,superscriptaddress,amssymb,reprint,nofootinbib,floatfix]{revtex4-2}
\pdfoutput=1
\usepackage{natbib}
\usepackage{graphicx}
\graphicspath{{Long_paper/}}
\usepackage{hyperref}
\usepackage{color}
\usepackage{amsmath,mathrsfs,amssymb,slashed,latexsym}
\usepackage{bm}
\usepackage{braket}
\usepackage{placeins}
\usepackage{adjustbox}
\usepackage{enumitem}
\usepackage{comment}
\usepackage{ragged2e}
\usepackage[absolute,overlay]{textpos}

\def\beq{\begin{equation}}
\def\eeq{\end{equation}}

\definecolor{darkblue}{cmyk}{1,0.4,0,0.3}
\definecolor{violet}{cmyk}{0,1,0,0.2}
\hypersetup{colorlinks, bookmarksnumbered, citecolor=darkblue, linkcolor=darkblue, pdfstartview=FitH, urlcolor=darkblue, linktocpage}

\begin{document}

\begin{textblock*}{9cm}(10.8cm,0.8cm)
\raggedleft
\small
CERN-TH-2026-048,
MIT-CTP/5978
\end{textblock*}
\title{Shocks from Exploding Primordial Black Holes in the Early Universe}

\author{Miguel Vanvlasselaer}
% \email{miguel.vanvlasselaer@vub.be}
\affiliation{Theoretische Natuurkunde and IIHE/ELEM, Vrije Universiteit Brussel,
\& The International Solvay Institutes, Pleinlaan 2, B-1050 Brussels, Belgium}
\affiliation{Departament de F\'isica Qu\`antica i Astrof\'isica and Institut de Ci\`encies del Cosmos (ICC), 
Universitat de Barcelona, Mart\'i i Franqu\`es 1, ES-08028, Barcelona, Spain.}

\author{Sokratis Trifinopoulos}
% \email{trifinos@mit.edu}
%\affiliation{Center for Theoretical Physics -- a Leinweber Institute, Massachusetts Institute of Technology, Cambridge, MA 02139, USA}
\affiliation{Department of Physics, Massachusetts Institute of Technology, Cambridge, MA 02139, USA}
\affiliation{Theoretical Physics Department, CERN, 1211 Geneva 23, Switzerland}
\affiliation{Physik-Institut, Universit\"at Z\"urich, 8057 Z\"urich, Switzerland
}%

\author{Alexandra P.~Klipfel}
% \email{}
\affiliation{Department of Physics, Massachusetts Institute of Technology, Cambridge, MA 02139, USA}

\author{David I.~Kaiser}
% \email{dikaiser@mit.edu}
\affiliation{Department of Physics, Massachusetts Institute of Technology, Cambridge, MA 02139, USA}

\begin{abstract}
% We present a comprehensive analysis of the energy injection from the explosion of primordial black holes (PBHs) in the early-universe plasma. Using hydrodynamical simulations and analytical computations, we first study the hydrodynamical evolution of the fluid around the PBH  long before the final explosion, and we observe that the fluid profile reaches a steady profile that we characterize in detail. Next we turn to an in-depth study of the hydrodynamical evolution which follows the very fast energy injection due to the runaway Hawking radiation at the end of a PBH's lifetime. We observe the formation and the propagation of a shock wave sweeping through the plasma, which locally reheats the plasma. Using analytical and numerical methods, we investigate the precise behaviour of this shock wave. 

We investigate how Hawking radiation from low-mass primordial black holes deposits energy into the early-universe plasma and show that the resulting phenomena are hydrodynamic rather than purely diffusive. Combining analytic arguments with relativistic hydrodynamic simulations, we find that the plasma first develops a quasi-steady outflow during the slow evaporation stage, while the final runaway phase of evaporation produces an expanding fireball that launches a shock wave into the surrounding medium. We characterize the thermalization scale of the Hawking products, the conditions under which shocks form, and the evolution and propagation of shocks. Additionally, we show that these shocks can locally restore electroweak symmetry, identifying exploding PBHs as a potentially important source of out-of-equilibrium dynamics in the early universe with profound phenomenological implications.
\end{abstract}

\maketitle
%\tableofcontents

\section{Introduction}
\label{sec:Intro}

Primordial black holes (PBHs) are hypothesized to have formed in the early universe from the gravitational collapse of overdense regions, rather than through stellar evolution~\cite{Zeldovich:1967lct,Hawking:1971ei,Carr:1974nx} (for comprehensive reviews, see Refs.~\cite{Khlopov:2008qy,Carr:2020xqk,Green:2020jor,Escriva:2022duf,Carr:2023tpt}.) Their mass spectrum spans an extraordinary range---from Planck-scale micro black holes ($\sim 10^{-5}\,\mathrm{g}$) to supermassive counterparts ($\sim 10^{38} \, \mathrm{g}$)---with formation mechanisms sensitive to the power spectrum of primordial perturbations and the equation of state of the fluid filling the early universe. 

Hawking radiation~\cite{Hawking:1974rv,Hawking:1975vcx,Page:1976df,MacGibbon:1990zk}, a semi-classical process arising from vacuum fluctuations near the event horizon, causes PBHs to lose mass and to inject energy into the early-universe plasma, leading to complete evaporation on timescales $\tau \sim M^3$ (in Planck units). For PBHs that form with masses $M\lesssim 5 \times 10^{14}\,\mathrm{g}$, this process would have concluded by the present epoch~\cite{Klipfel:2025bvh}, producing a possibly detectable signature in the extragalactic gamma-ray background or cosmic microwave background (CMB) distortions~\cite{Carr:2020xqk,Green:2020jor,Escriva:2022duf,gorton_how_2024,Carr:2009jm,Carr:2020gox,DelaTorreLuque:2024qms,Balaji:2025afr}. In addition to serving as a dark matter candidate, the existence of PBHs could provide constraints on inflationary models and the thermal history of the universe, while their Hawking radiation offers a unique probe of quantum-mechanical effects in strong gravitational fields~\cite{Khlopov:2008qy,Carr:2020xqk,Green:2020jor,Escriva:2022duf,Carr:2023tpt}.

In this paper, we investigate the impact of the energy radiated by an evaporating PBH on the surrounding early-universe plasma. In particular, a sufficiently low-mass PBH embedded in a plasma of ambient temperature $T_{\rm b}$ during the radiation-dominated epoch %emits 
can emit particles with energies $E \gg T_{\rm b}$, thereby transferring energy to the region in the immediate vicinity of the PBH. Previous studies~\cite{Das:2021wei,He:2022wwy,Levy:2025lyj,Altomonte:2025hpt,Gunn:2024xaq} have typically modeled the redistribution of this energy through diffusive heat transport, effectively assuming that the injected energy spreads locally through thermal conduction. Our study shows that the dynamics depart from this picture: Instead of producing a static hot, dense region~\cite{He:2022wwy}, the large pressure gradients generated by rapid localized heating during the final phases of evaporation trigger a hydrodynamical response of the plasma, leading to the expansion of the plasma overpressure and the formation of a growing relativistic \emph{fireball}.

We therefore treat the problem within the framework of relativistic hydrodynamics, which provides an effective description of the long-wavelength, low-frequency regime of an interacting plasma. At leading order (\emph{ideal} hydrodynamics), strong pressure gradients act as the driving source of fluid motion and can naturally lead to the formation of discontinuities in the fluid variables known as \emph{shocks}. Consequently, we identify several distinct phases in the evolution of the system. While the PBH is sufficiently massive, its temperature and emission rate are low enough for a phase of slow, quasi-steady outflow. Then, as the PBH emission ramps up near the end of its life, it enters a regime of shock formation in which continuous energy injection drives the formation of an expanding fireball. As the evaporation enters its final runaway phase, all of the remaining energy is released in a final \emph{explosion}. The fireball at this stage sets the initial conditions for the generation of a shock wave that propagates outward into the surrounding medium. The shock passes through an ultrarelativistic expansion phase (the \emph{Blandford--McKee} regime~\cite{10.1063/1.861619}), subsequently transitions to a non-relativistic \emph{Sedov--Taylor} blast wave~\cite{RezBook}, and is finally dissipated by viscosity. 

This phenomenon is conceptually similar to the case of heavy-ion collisions, in which the collisions of ultrarelativistic nuclei create a quark-gluon plasma (QGP) with extreme energy density ($\rho_{\rm QGP} \sim 10\text{--}100~\mathrm{GeV/fm^3}$), which generates an initial overpressure that drives explosive hydrodynamic expansion~\cite{Casalderrey-Solana:2011dxg}. The QGP behaves as a near-perfect fluid and cools via radial and anisotropic flow, converting pressure gradients into collective motion while expanding supersonically. At critical temperature $T_c \approx 155~\mathrm{MeV}$, hadronization occurs, terminating the high-pressure phase as the system transitions to a hadron gas with diminishing pressure. In addition to heavy-ion collisions, our study shares several similarities with previous studies of the hydrodynamical properties of astrophysical catastrophic events~\cite{2013ApJ...768..113M, 10.1063/1.861619}, such as gamma-ray bursts and supernovae, and uses several hydrodynamical results computed in those settings.   

The study of energy injection from PBHs into the early-universe plasma is important for developing constraints on PBH populations as well as for several cosmological scenarios. Such applications include scenarios in which Hawking emission from PBHs could have played a role in primordial baryogenesis~\cite{Nagatani:1998rt,Rangarajan:1999zp,Dolgov:2000ht,Nagatani:2001nz,Bugaev:2001xr,Baumann:2007yr,Hook:2014mla,Aliferis:2014ofa,Banks:2015xsa,Hamada:2016jnq,Morrison:2018xla,Carr:2019hud,Garcia-Bellido:2019vlf,Aliferis:2020dxr,Hooper:2020otu,Boudon:2020qpo,Datta:2020bht,Bernal:2022pue,Bhaumik:2022pil,Gehrman:2022imk,Barman:2022pdo,Borah:2024bcr,Calabrese:2025sfh,IguazJuan:2025vmd} as well as during big bang nucleosynthesis~\cite{Kohri:1999ex,Lemoine:2000sq,Carr:2009jm,Poulin:2016anj,Carr:2020gox,Keith:2020jww,Domenech:2020ssp,Auffinger:2022khh,He:2022wwy,Boccia:2024nly,Altomonte:2025hpt,Montefalcone:2025akm,Chaudhuri:2025asm,Wu:2025ovd,Wang:2025pum,Kalita:2025fcs,Klipfel:2026aug}. In this work, we foreshadow the broader phenomenological implications of this mechanism through one illustrative application, namely the restoration of electroweak (EW) symmetry in the regions reheated by the expanding shock wave. In a companion paper~\cite{BAUinprep}, we show that this effect constitutes a novel mechanism to produce the baryon asymmetry of the universe in non-fine-tuned scenarios of PBH formation via gravitational collapse.

The paper is organized as follows. Following a brief review of notation and conventions in Sec.~\ref{sec:notation}, Sec.~\ref{sec:Hawking_rad} presents the basic features of Hawking radiation from PBHs and the resulting timescales for PBH evaporation. In Sec.~\ref{sec:energy_inj}, we study how the energy emitted by the PBH is deposited in the plasma, and in Sec.~\ref{sec:hydro_motion} we illustrate how this energy is subsequently redistributed by hydrodynamical motion. We then study two scenarios for the rate of energy injection using both analytical methods as well as dedicated numerical simulations: a simplified case of instantaneous injection in Sec.~\ref{sec:inst}, and a more realistic approximation of an evaporating PBH in Sec.~\ref{sec:PBH-like_inj}. Finally, in Sec.~\ref{sec:Pheno_impl} we discuss EW symmetry restoration due to this mechanism, before concluding in Sec.~\ref{sec:conclusion} with an outlook on further possible directions. 

\section{Notation and conventions}
 \label{sec:notation}

In Table~\ref{tab:symbols_jump}, we present a summary of the notation we use throughout this paper. We use the term \emph{fireball} (subscript ``fb'') for the phase in which Hawking radiation from a PBH is actively injecting energy and building up a locally reheated region, and we reserve the term \emph{shock front} for the discontinuous jump in the fluid quantities. Thermodynamic quantities may be evaluated in the unshocked ambient plasma (subscript ``b'' for background) or in the freshly shocked region directly behind the shock front (subscript ``sh'' for shocked). The shock is followed by a \emph{rarefaction wave} (subscript ``rear''), along which the pressure decreases. The expanding wave develops a \emph{shell} of high-pressure and high-temperature plasma, which, depending on the context, is either defined as the region between the shock front and the rarefaction wave or the region between the shock front and the downstream region that has already cooled back toward the ambient state. Even though temperature and pressure vary continuously throughout this shell, we use the subscript ``shell'' to refer to shell-averaged quantities. We use subscript ``inst'' to refer to quantities derived for the simplified case of \emph{instantaneous} emission, which is presented in comparison to a more detailed model of extended \emph{PBH-like} energy injection. We further use subscript ``phys'' to differentiate (dimensionful) \emph{physical} quantities from (dimensionless) \emph{numerical} quantities in our simulations (subscript ``num''). Throughout, $p$ denotes fluid pressure while $P$ denotes radiated power.

\begin{table*}[ht!]
\centering
\caption{Summary of different %subscripts and 
symbols appearing in this paper and their meanings.
}
\label{tab:symbols_jump}
\renewcommand{\arraystretch}{1.2}
\begin{tabular}{|l|l|l|}
\hline
\textbf{Symbol} & \textbf{Definition} & \textbf{First appears in} \\
\hline
\hline
\multicolumn{3}{|l|}{\textit{Thermodynamic quantities}} \\
\hline
$p_{\rm b}$ & Background (upstream) pressure & \eqref{eq:p_sh_cont} \\
$p_{\rm sh}$ & Shocked (downstream) pressure & \eqref{eq:p_sh_cont} \\
$w_{\rm b}$ & Background enthalpy density & \eqref{eq:sh_cont_1} \\
$w_{\rm sh}$ & Shocked enthalpy density & \eqref{eq:sh_cont_1} \\
$\rho_{\rm b}$ & Background energy density & \eqref{eq:rhobTb} \\
$T_{\rm b}$ & Background plasma temperature & Sec. \ref{sec:Intro} \\
$T_H$ & Hawking temperature & \eqref{eq:BHTemp} \\
$T_f$ & Background temperature at PBH explosion & \eqref{eq:PowerTimeTemp} \\
$T_{\rm LPM}$ & LPM temperature & \eqref{eq:ThsEqn} \\
$T_{\rm fb}$ & Fireball temperature  & Sec.~\ref{sec:PBH_phase0} \\
$T_{\rm EW}$ & EW symmetry restoration temperature & Sec.~\ref{sec:Pheno_impl} \\
%$T_{\rm spha}$ & Sphaleron decoupling temperature & Sec.~\ref{sec:phaseA} \\
$l_{\rm mfp}$ & Mean free path of particles in the plasma  & \eqref{eq:lmfp} \\
\hline
\multicolumn{3}{|l|}{\textit{Velocities and Lorentz factors}} \\
\hline
$\gamma_{\rm b}$ & Upstream Lorentz factor (PBH frame) & \eqref{eq:gamma_up} \\
$\gamma_{\rm sh}$ & Downstream Lorentz factor (PBH frame) & \eqref{eq:E_inj_ins} \\
$\Gamma$ & Shock Lorentz factor in PBH frame & \eqref{eq:p_sh_cont} \\
$\gamma_{\rm damp}$ & Wave damping rate & Sec.~\ref{sec:inst} \\
\hline
\multicolumn{3}{|l|}{\textit{Shock geometry}} \\
\hline
$R$ & Shock radius (PBH frame) & Sec. \ref{sec:phaseA} \\
$R_{\rm max}$ & Maximum shock propagation distance & \eqref{eq:Rmax} \\
$L_{\rm fb}$  & Fireball size (PBH-like injection) &  \eqref{eq:Lfb} \\
$L_{\rm shell}$ & Shell thickness (PBH-like injection) & \eqref{eq:PBH_inj_law} \\
$\delta$ & Microscopic shock-front thickness & \eqref{eq:energy_diss} \\
\hline
\multicolumn{3}{|l|}{\textit{PBH and energy-injection quantities}} \\
\hline
$P_{\rm tot}$, $P$ & Total radiated power from PBH & \eqref{eq:TotalInjectionM}, \eqref{eq:PBH_inject} \\
$\mathcal{P}$ & Power density 
% (energy per unit volume per unit time) 
& \eqref{eq:toyeq} \\
$E_{\rm inj}$ & Total injected energy & \eqref{eq:ThsEqn} \\
$E_{\rm peak}$ & Peak of primary Hawking spectrum & \eqref{Epeak} \\
$K$ & Ratio of $E_{\rm inj}$ to $M_{\rm thres}$ & \eqref{eq:T_hs_init} \\
$t_{\rm inj}$       & Duration of energy injection window  & \eqref{eq:deltat} \\
$M_{\rm thres}$ & Threshold PBH mass for instantaneous injection & \eqref{eq:Mthres} \\
$f$, $f_{\rm max}$ & Page factor & \eqref{eq:MassEvolution}, \eqref{fMAX} \\
$\tau$ & PBH remaining lifetime & \eqref{TauSmallM} \\
\hline
\multicolumn{3}{|l|}{\textit{Thermalization and transport}} \\
\hline
$\hat{q}$ & Jet-quenching parameter & \eqref{eq:qhat_def} \\
$q_\perp$  & Transverse momentum transfer per soft scattering & \eqref{eq:qhat_def} \\
$k_\perp$ & Accumulated transverse momentum & \eqref{eq:kperp_broad} \\
$\Gamma_{\rm LPM}$  & LPM splitting rate & \eqref{eq:GammaLPM} \\
$L_{\rm LPM}$  & LPM thermalization length  & \eqref{eq:LPM_length} \\
$\mathrm{Pe}$ & P\'{e}clet number  & \eqref{eq:Pe-crit}, \eqref{eq:PecletSlow}, \eqref{eq:Pc_fb} \\
\hline
\end{tabular}
\end{table*}

Throughout this article we adopt natural units, with $c = \hbar = k_B = 1$. We retain Newton's gravitational constant $G$, usually working in terms of the (reduced) Planck mass, $M_{\rm pl} \equiv 1 / \sqrt{ 8 \pi G} = 2.43 \times 10^{18} \, {\rm GeV} = 4.33 \times 10^{-6} \, {\rm g}$. The speed of sound is $c_s= 1/\sqrt{3}$.

\section{Hawking evaporation of primordial black holes}
\label{sec:Hawking_rad}

In this section, we briefly review the formalism for calculating the Hawking radiation spectra emitted by PBHs and their resulting evaporation rates. Whether and how the semi-classical Hawking-radiation formalism might need to be modified at late stages of black hole evaporation remains an open question~\cite{Dvali:2018xpy,Dvali:2020wft,Montefalcone:2025akm}. As a conservative analysis, we work with the standard formalism~\cite{Hawking:1974rv,Hawking:1975vcx,Page:1976df,Page:1976ki,Page:1977um,MacGibbon:1990zk,MacGibbon:1991tj}, updated as in Refs.~\cite{Klipfel:2025jql,Klipfel:2025bvh,Klipfel:2026aug} to include the present-day set of Standard Model (SM) degrees of freedom.

For all regimes of interest in this article---ranging from PBH evaporation to the formation and late-stage evolution of fireballs in the surrounding plasma---we may approximate the plasma dynamics of the hot medium as if all dynamics occur within a Minkowskian background spacetime. %That is 
This is applicable because we have an exponential hierarchy among the relevant length-scales:
%%%
\beq
r_s (M) \ll L (T) \ll R_H (t)\,,
\label{eq:LengthHierarchy}
\eeq
where $r_s (M)$ is the Schwarzschild radius of a PBH of mass $M$, $L(T)$ is the typical thermal length-scale for hard modes within the plasma at temperature $T$, and $R_H (t)$ is the Hubble radius. For the PBHs of interest, the Schwarzschild radius is
%%%%
\beq
r_s = 2 G M = 1.5 \times 10^{-30} \, {\rm m} \left( \frac{ M}{1 \, {\rm g}} \right) \sim {\cal O} (10^{-24} \, {\rm m})\,,
\label{eq:rs}
\eeq
for PBH mass $M \sim 10^6 \, {\rm g}$. Note that we focus only on PBHs with $M\lesssim 10^8\,{\rm g}$, which would evaporate completely before BBN. Meanwhile, the typical hard thermal length within the plasma is given by \cite{Blaizot:2001nr}
%%%%
\beq
L (T) \sim \frac{1}{T} \sim {\cal O} (10^{-18} \, {\rm m})
\label{eq:Lthermal}
\eeq
for $T \sim {\cal O} (100 \, {\rm GeV})$. And finally, if we assume that the universe remains radiation-dominated at all times of interest, the Hubble radius is given by
%%%%
\beq
\begin{split}
R_H (t) &= 1 / H = 2  t \sim {\cal O } ( 10^{-1} \, {\rm m} )\,,
\label{RHubble}
\end{split}
\eeq
for $t \sim \tau (M=10^6\,{\rm g}) \sim {\cal O} (10^{-10} \, {\rm s})$, with $\tau$ the PBH lifetime. Hence we may safely neglect {\it both} the spacetime curvature effects on the plasma in the vicinity of the PBH {\it and} the overall cosmic expansion when considering the formation and evolution of fireballs in the plasma at times when the temperature of the plasma (far from a given PBH) is in the regime $T \sim {\cal O} (10^{-2} - 10^2 \, {\rm GeV})$. That means that we may safely adopt expressions for such quantities as integrals of distribution functions over momenta $\int d^3 p \, f ( p)$ and the covariant conservation of the energy-momentum tensor $\partial_\mu T^{\mu\nu}$, which are covariant with respect to a Minkowskian background spacetime. We do not need to make use of the more sophisticated formalism of relativistic kinetic theory in curved spacetimes~\cite{Acuna-Cardenas:2021nkj,Alonso-Monsalve:2023jfq}.

We assume the PBHs of interest have no charge and no spin. PBHs form from the collapse of (scalar) curvature perturbations, so they typically begin with little or no spin~\cite{Garc_a_Bellido_2017}.  Similarly, although small-mass PBHs could have formed with large initial charge~\cite{Alonso-Monsalve:2023brx}, such short-lived objects are expected to discharge very rapidly (if charged only under SM gauge groups ~\cite{Baker:2025zxm,Santiago:2025rzb}). Meanwhile, following their formation, black holes that happened to form with charge and/or spin will preferentially emit particles to reduce those quantities~\cite{carterChargeParticleConservation1974,gibbonsVacuumPolarizationSpontaneous1975,Page:1976df}. Furthermore, PBHs within the mass range we consider here will not spin up over time: their extraordinarily small radii preclude efficient accretion~\cite{chibaSpinDistributionPrimordial2017,delucaEvolutionPrimordialBlack2020,Jaraba:2021ces,chongchitnanExtremeValueStatisticsSpin2021}. We therefore focus on emission from Schwarzschild PBHs and their evaporation over time.

Black holes couple to every form of matter, due to the universal nature of gravitation. Hence PBHs can emit every fundamental degree of freedom during Hawking emission, with emission rates determined by each particle's mass and spin. Particles of mass $m$ may be emitted efficiently from black holes with Hawking temperature $T_H > m$. Thus, a PBH of mass $M \leq 5.8 \times 10^{10} \, {\rm g}$, with associated Hawking temperature $T_H (M) \geq 180 \, {\rm GeV}$, can emit all 17 SM particle species at high rates.

The fundamental particle emission rates, referred to as \textit{primary emission spectra}, have the form of a modified blackbody curve. A Schwarzschild black hole of mass $M$ will emit fundamental particles of species $j$ at a rate~\cite{Page:1976df, MacGibbon:1990zk}: 
\begin{equation}
    \label{eq:PrimarySpectraSchwarzschild}
    \frac{d^2N_j^{(1)}}{dtdE} = g_j \frac{\Gamma_{s_j}}{2 \pi} \left[\exp\left( \frac{E}{T_H ( M )} \right) - (-1)^{2s_j} \right]^{-1}\,,
\end{equation}
where the particle has spin $s_j$, energy $E$, 
and total quantum degrees of freedom $g_j$. (See the Supplemental Materials of Ref.~\cite{Klipfel:2025jql} for appropriate parameters $\{g_j, s_j, m_j\}$ for all SM particles.) To compute the primary emission spectrum of the $u$ quark, for example, we must use $g_{u}=4$ to account for two helicity degrees of freedom for both $u$ and $\bar{u}$. The bracketed term has the familiar form of a blackbody with temperature 
\begin{equation}
    \label{eq:BHTemp}
    T_H(M) = \frac{1}{8 \pi G M}\,.
\end{equation}
Ref.~\cite{MacGibbon:1991tj} gives a relation between PBH temperature and the energy $E_{{\rm peak}, \, s}$ of the emitted particles that maximizes the spectrum for a spin-$s$ particle:
\begin{equation}
    E_{{\rm peak}, \, s}(M) = \beta_{s} \, T_H(M)\, ,
    \label{Epeak}
\end{equation}
where $\beta_0=2.66$, $\beta_{1/2}=4.53$, $\beta_1=6.04$, and $\beta_2=9.56$ \cite{MacGibbon:1990zk, MacGibbon:1991tj}. Because the spectra are so sharply peaked, it is a reasonable approximation to assume that, for a PBH of mass $M$, all particles with spin $s$ are emitted at energy $E_{{\rm peak}, \, s}(M)$.

\begin{figure}[t]
     \centering
    \includegraphics[width=0.48\textwidth]{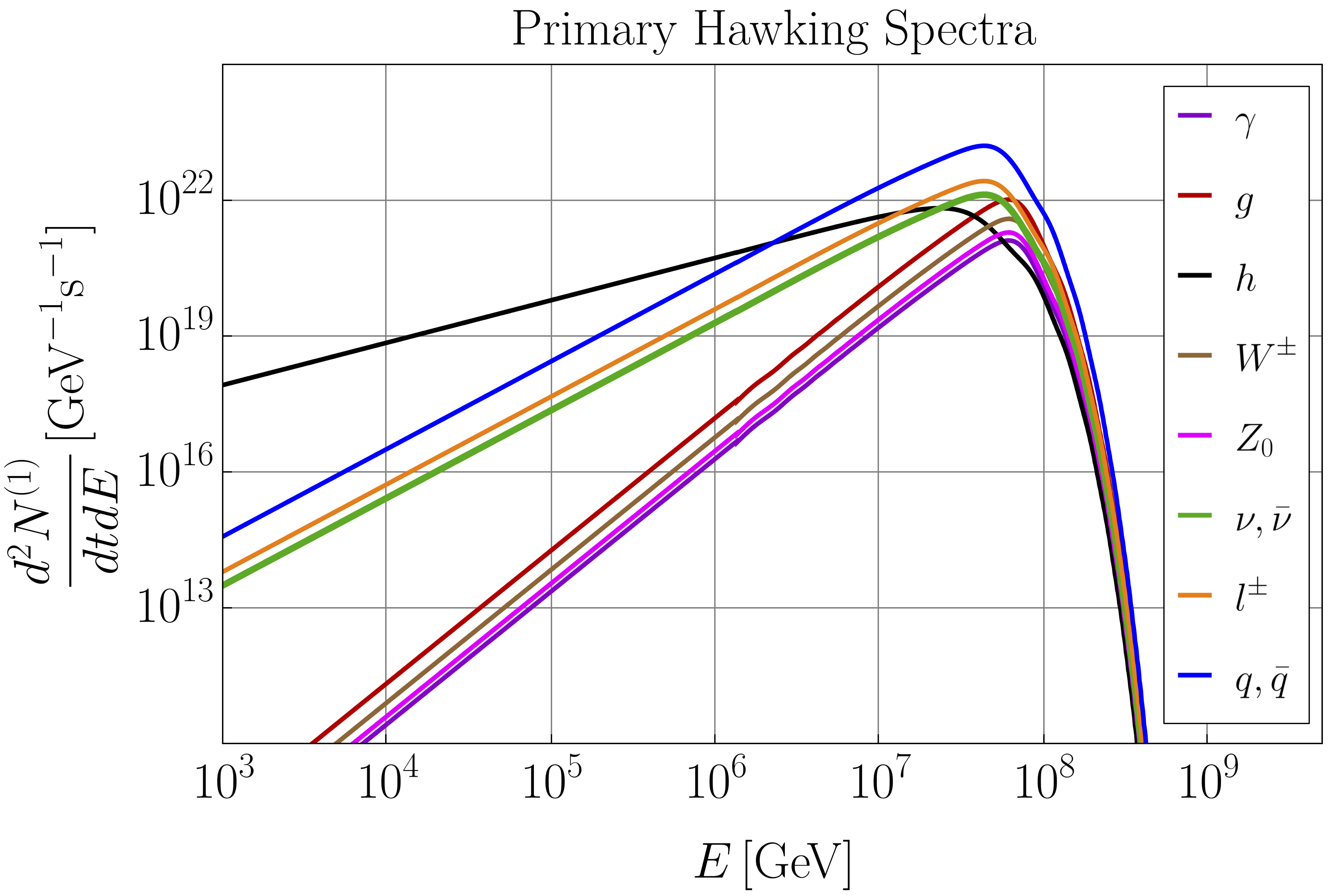} \\[0.5cm]
    \includegraphics[width=0.48\textwidth]{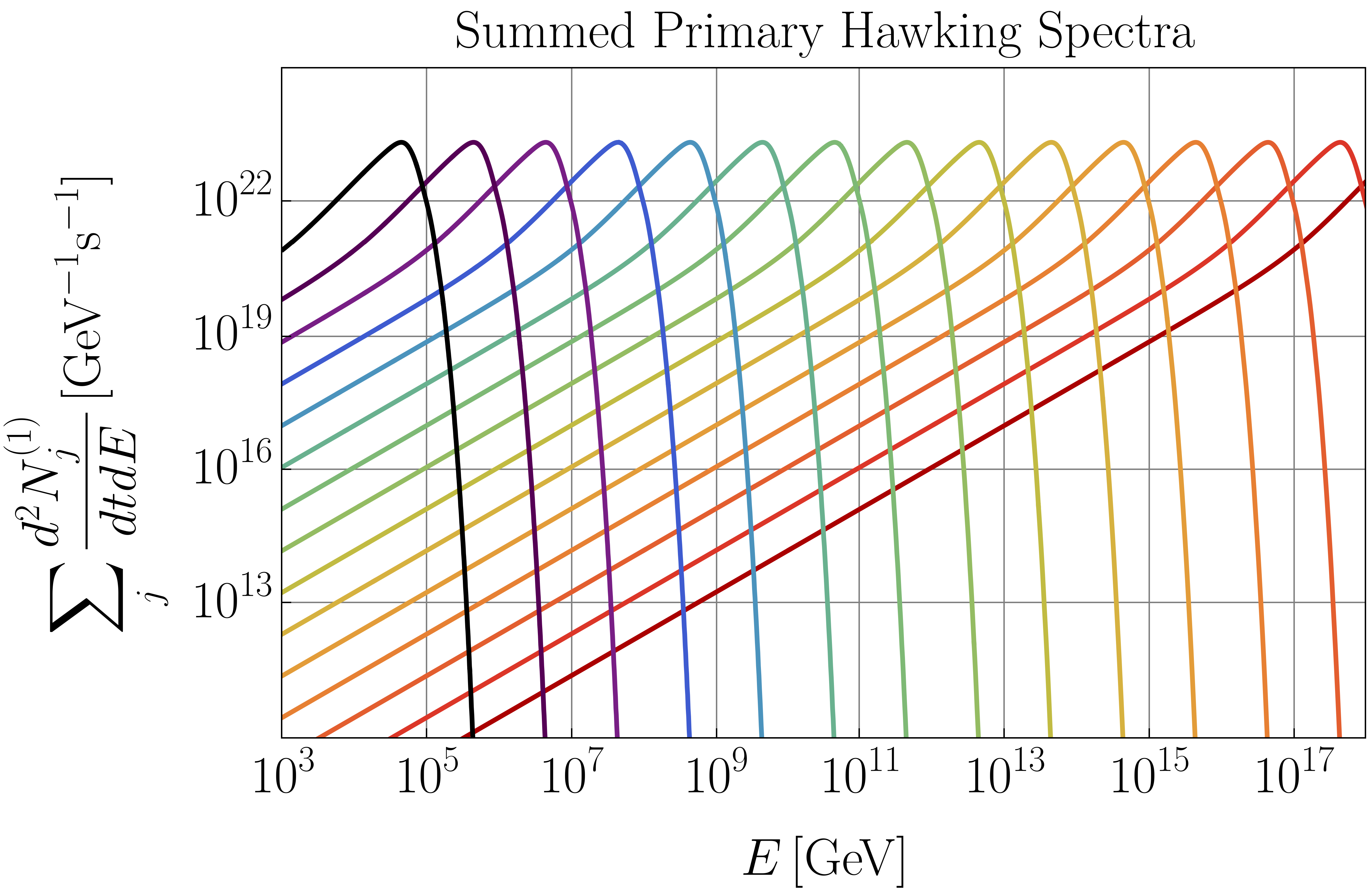}
         \label{fig:PrimSpec}
        \caption{\justifying  ({\it Top}) Primary Hawking emission spectra for a PBH with $M = 10^6 \, {\rm g}$. ({\it Bottom}) Total primary Hawking spectra summed over all SM particle species for PBH masses from $M=10^{-5} \, {\rm g}$ (dark red) to $M=10^9 \, {\rm g}$ (black). Plots prepared with \texttt{BlackHawk v2.2}~\cite{Arbey:2019mbc, Arbey:2021mbl}.}
        \label{fig:Spectra}
\end{figure}

The \textit{greybody factor} $\Gamma_{s} (M, E, m)$ is a dimensionless quantity that governs the interaction between the black hole and the emitted particles. The greybody factor encapsulates information about how the black hole's effective area scales with its mass---thus the emission spectra are not simply given by power per unit area (or alternatively per unit solid angle), as expressed for a typical blackbody. The greybody factor is defined in terms of the absorption cross section $\sigma_s$ associated with scattering an SM field of mass $m$ and spin $s$ off a black hole of mass $M$ in curved spacetime~\cite{MacGibbon:1990zk}: 
\begin{equation}
    \label{eq:GreyBody}
    \Gamma_s(M, E, m) = \frac{\sigma_s(M, E, m)}{\pi}(E^2 - m^2)\,.
\end{equation}
In practice, emission of particles of mass $m$ only becomes efficient for Hawking temperatures $T_H > m$, so one typically considers $\Gamma_s (M, E, m)$ in the ultrarelativistic limit, $E \gg m$. See Refs.~\cite{Page:1976df, MacGibbon:1990zk, Teukolsky:1973ha, Teukolsky:1974yv,Gray:2015xig} for detailed discussions of calculating greybody factors.

Figure~\ref{fig:Spectra}a shows the primary emission spectra for a black hole of $M = 10^6 \, {\rm g}$ computed numerically with \texttt{BlackHawk v2.2}~\cite{Arbey:2019mbc, Arbey:2021mbl}. The curves are summed over all relevant degrees of freedom for each species. Note the spin-dependent shape of the spectra.  Figure~\ref{fig:Spectra}b plots primary Hawking emission spectra summed over all SM particle species, $\sum_j d^2N^{(1)}_j / dtdE$, which we will use below when computing the total power emitted from a PBH. Note from Fig.~\ref{fig:Spectra}a that the main carriers of the energy emitted by the PBH are color-charged particles, namely quarks and gluons. More specifically, for PBHs with $M < 5.8 \times 10^{10} \, {\rm g}$ (and hence $T_H > 180 \, {\rm GeV}$), the fraction of the energy emitted in color-charged particles is $\sim 76\%$, with about $70\%$ in quarks and $6\%$ in gluons. This is important for the early-universe scenarios we consider here, in which PBHs emit particles into the surrounding quark-gluon plasma. 

We only consider primary Hawking emission in this analysis, because the rate at which energy is injected into the surrounding medium depends on the primary spectrum. However, it is important to note that black holes can radiate composite particles via secondary hadronization processes, and that primary spectra are modified on large distance-scales and time-scales due to particle decays~\cite{MacGibbon:1990zk, MacGibbon:1991tj}. The \textit{secondary emission spectrum} for some species $i$ is calculated via
\begin{equation}
    \label{eq:SecondaryEmisison}
    \frac{d^2N_i^{(2)}}{dt dE} = \int_0^{\infty} \sum_j \frac{d^2N_j^{(1)}}{dt dE'} \frac{dN_i^j}{dE} dE'\,,
\end{equation}
where we sum over all primary Hawking particles $j$ and weight by the differential branching ratios $dN_i^j(E', E)/dE$~\cite{Arbey:2019mbc}. The numerical calculation of secondary spectra is typically carried out with the \texttt{BlackHawk} code and various particle physics codes as afterburners that apply in different energy regimes~\cite{Bauer:2020jay, Sjostrand:2014zea,Coogan:2019qpu}. 

The total mass-loss rate due to Hawking emission for a PBH of mass $M$ is computed by integrating over the primary emission spectra for all SM degrees of freedom:
\begin{equation}
\begin{split}
    \label{eq:MassEvolution}
    \frac{dM}{dt} &= -\sum_j\int_{m_{{\rm eff},\, j}}^{\infty}dE \, E\frac{d^2N^{(1)}_j}{dtdE}\\
    &=-A \, \frac{f(M)}{M^2} \,.
    \end{split}
\end{equation}
The second line of Eq.~(\ref{eq:MassEvolution}) casts the mass loss in a condensed form in which $f(M)$ is known as the \textit{Page factor}~\cite{MacGibbon:1990zk, MacGibbon:1991tj}. We normalize the Page factor as $f (M) = 1$ for black holes that emit only photons and neutrinos (\emph{i.e.}, for $M \gtrsim 10^{18} \, {\rm g}$), and parameterize $f(M)$ as in Refs.~\cite{MacGibbon:1990zk,MacGibbon:1991tj}, updated to current best-fit values for all SM parameters as in Refs.~\cite{Klipfel:2025jql,Klipfel:2025bvh}. Upon using that parameterization, high-precision numerical integration of the primary Hawking spectra in Eq.~(\ref{eq:MassEvolution}) yields the coefficient $A = 6.036 \times 10^{72} \, {\rm GeV}^4$. Meanwhile, in the limit in which the initial PBH mass is small (i.e. it can emit all SM particles),
$M_i \lesssim 10^9 \, {\rm g}$, the Page factor reaches the constant value,
%%%%
\begin{equation}
    f_{\rm max}\equiv \lim_{M\rightarrow 0}f(M) = 15.522\,,
    \label{fMAX}
\end{equation}
if one only considers SM degrees of freedom. The Page factor grows as the PBH mass shrinks and the associated Hawking temperature $T_H (M)$ rises because smaller, hotter PBHs can efficiently emit more types of particles. Meanwhile, the numerical values of both $f_{\rm max}$ and the coefficient $A$ would change if additional heavy particles were included in some Beyond-Standard-Model dark sector(s) \cite{Baker:2021btk,Baker:2022rkn,Baker:2025ffi}. Multiple elements of the mass-loss expression in Eq.~(\ref{eq:MassEvolution}) require  careful treatment, including the lower integration bound (which relies on an implicit assumption that all emitted particles are relativistic), the explicit form of the Page factor, and the energy scales taken as inputs when calculating the Page factor. 

Solving Eq.~(\ref{eq:MassEvolution}) allows us to compute the PBH lifetime as a function of its mass. Because the Page factor $f(M)$ changes relatively slowly over the PBH lifetime, the lifetime follows an approximate scaling 
\begin{equation}
    \tau(M_i) \sim M_i^3\, ,
\end{equation}
where $M_i = M (t_i)$ is the PBH mass at the time of its formation. To exactly determine the PBH lifetime, one may solve Eq.~(\ref{eq:MassEvolution}) numerically; doing so requires high numerical precision because the gradient becomes steep as the PBH mass tends toward zero during the last stages of evaporation. The formation-time mass $M_i$ of a PBH whose lifetime is equal to the current age of the universe ($t_{0} \equiv 13.787 \pm 0.020 \, {\rm Gyr}$~\cite{Planck:2018vyg}) is defined as the \textit{cutoff mass} $M_*$. In Ref.~\cite{Klipfel:2025bvh} we find
\begin{equation}
M_* = (5.364 \pm 0.002)\times10^{14} \, {\rm g}\,.
\label{Mstar}
\end{equation}
In other words, a PBH that formed in the very early universe with initial mass $M_i = M_*$ would just be completing its Hawking evaporation today, if the PBH were only emitting SM particles.

Within the limit of small PBH masses, $M_i \lesssim 10^9\, {\rm g}$, the mass-loss equation becomes
\begin{equation}
    \label{dMdt}
    \frac{dM}{dt}=-A\frac{f_{\rm max}}{M^2}\,,
\end{equation}
which can be solved analytically for the PBH mass at later times $t > t_i$:
\begin{equation}
    \label{eq:Moft}
    M(t|M_i) = (-3Af_{\rm max}t + M_i^3)^{1/3} \,.
\end{equation}
At any time $t > t_i$, when the PBH has mass $M(t)$, the PBH's remaining lifetime is given by
\begin{equation}
    \tau(M) = \frac{M^3}{3Af_{\rm max}} = 4.14\times10^{-10} \, {\rm s} \left(\frac{M}{10^6 \, {\rm g}} \right)^3\,.
    \label{TauSmallM}
\end{equation}

\begin{figure}[t]
    \centering
\includegraphics[width=0.48\textwidth]{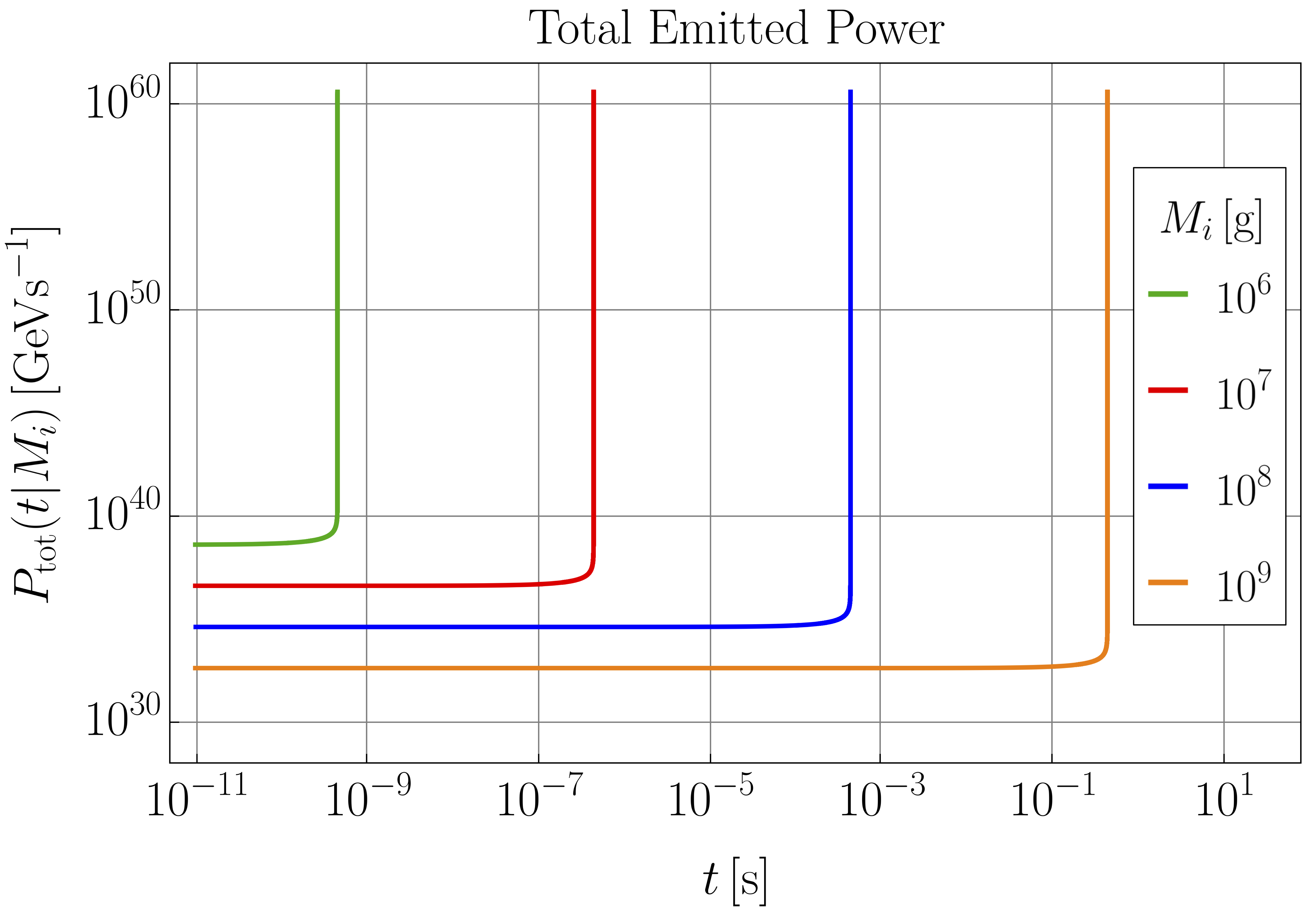}
    \caption{\justifying 
    Total emitted power from a single PBH as a function of time for several values of initial mass $M_i$. See Eq.~(\ref{eq:TotalInjectionTime}). The PBHs are initialized at $t_{\rm EW}\simeq10^{-11}\,{\rm s}$ and the emission is truncated at $M(t)=10^{-5}\,{\rm g}$ when the PBHs approach the Planck scale.}
    \label{fig:PowerTimeTemp}
\end{figure}

The power emitted by a PBH of mass $M$ via particle species $j$ is related to the primary Hawking emission spectrum as
\begin{equation}
    \label{eq:Powerj}
    P_j(M) = \int_0^{\infty}dE \frac{d^2N_j^{(1)}(M, E)}{dtdE}E\, .
\end{equation}
The total emitted power from a PBH of mass $M$---which is equivalent to the total energy injection to the medium---is thus found by summing Eq.~(\ref{eq:Powerj}) over all 17 SM particles:
\begin{equation}
    P_{\rm tot}(M) \equiv \sum_j P_j(M)\,.
    \label{eq:TotalInjectionM}
\end{equation}
We compute $P_{\rm tot}(M)$ by numerically integrating primary spectra for PBH masses in the range $10^{-5} \, {\rm g} \leq M \leq 10^9 \, {\rm g}$. (Spectra are computed with \texttt{BlackHawk v2.2}~\cite{Arbey:2019mbc, Arbey:2021mbl}, see Fig.~\ref{fig:Spectra}).
Performing a power-law fit to these values gives
\begin{equation}
    \label{eq:PowerLawFit}
    P_{\rm tot}(M) = B \left( \frac{M}{10^6\,{\rm g}} \right)^{-b}\, {\rm GeV \, s}^{-1}\,,
\end{equation}
where $B = 4.523\times10^{38}$ and $b = 2.000$. This $P\propto M^{-2}$ scaling is expected~\cite{MacGibbon:1990zk, MacGibbon:1991tj}. Combining Eqs.~(\ref{eq:Moft}) and Eq.~(\ref{eq:PowerLawFit}), we can express the total emitted power as a function of cosmological time $t \leq \tau(M_i)$ via
\begin{equation}
    \begin{split}
    P_{\rm tot}&(t|M_i)\\
    &= \sum_j P_j(M(t|M_i))\\
    & = 2.51\times10^{32}\,{\rm GeV \, s}^{-1}\left(\frac{\tau(M_i)-t}{1 \, {\rm s}}\right)^{-2/3} \,.
    \end{split}
    \label{eq:TotalInjectionTime}
\end{equation}
As above, $M_i$ is the PBH mass at the time of its formation and $\tau(M_i)$ is its lifetime, as defined in Eq.~(\ref{TauSmallM}). See Fig.~\ref{fig:PowerTimeTemp} for a plot of Eq.~(\ref{eq:TotalInjectionTime}) for several initial PBH masses.

After a population of PBHs forms at time $t_i$, the total energy density in PBHs will redshift as matter, $\rho_{\rm PBH} \sim a^{-3} (t)$, while the energy density in the hot plasma redshifts as radiation, $\rho_{\rm b} \sim a^{-4} (t)$. If the initial PBH population is sufficiently small, then the universe will remain in a radiation-dominated equation of state up to and after the time when the PBHs explode, $\tau (M_i)$. In such cases, for which $\rho_{\rm PBH} < \rho_{\rm b}$ for times $t_i \leq t \leq \tau (M_i)$, we may relate the background temperature $T_{\rm b}$ to the background energy density of the plasma:
%%%%
\begin{equation}
    \rho_{\rm b} (t) = \frac{ \pi^2}{30} g_* (T_{\rm b}) \, T_{\rm b}^4 (t)\, .
    \label{eq:rhobTb}
\end{equation}
Given that the Hubble parameter $H^2 = \rho_{\rm b} / (3 M_{\rm pl}^2)$ evolves as $H (t) = 1 / (2t)$ during radiation-domination, we may write
%%%%
\begin{equation}
    T_{\rm b} (t) = 4.83 \times 10^{-4} \, {\rm GeV} \left( \frac{ 106.75}{g_* (T_{\rm b})} \right)^{-1/4} \left( \frac{ 1 \, {\rm s} }{t} \right)^{1/2}\, ,
    \label{TbRD}
\end{equation}
where $g_* (T_{\rm b}) = 106.75$ is the maximum number of SM relativistic degrees of freedom at high temperatures. We label as $T_f$ the temperature of the background plasma at the time the PBH explodes. Then using the relationship between $T_f$ and $\tau (M_i)$ for cases in which the universe remains radiation-dominated over the PBH lifetime, we find
\begin{equation}
    \tau(M_i) = 1.22\times10^{-5}\, {\rm s} \left(\frac{T_f}{1 \, {\rm GeV}}\right)^{-2}g_*^{-1/2}(T_f)\,, 
\end{equation}
and hence, from Eq.~(\ref{eq:TotalInjectionTime}),
\begin{equation}
    \label{eq:PowerTimeTemp}
    \begin{split}
   P_{\rm tot}&(t|T_f)\\
   &=2.51\times 10^{32} \, {\rm GeV s^{-1}} \\
   &\quad\times \left[\left(\frac{3.49\times10^{-3}\,{\rm GeV}}{T_f \,g_*^{1/4}(T_f)} \right)^2  - \left(\frac{t}{1 \, {\rm s}} \right)\right]^{-2/3}  . 
   \end{split}
\end{equation}
Eq.~(\ref{eq:PowerTimeTemp}) is equivalent to Eq.~(\ref{eq:TotalInjectionTime}) for the case in which the plasma remains in a radiation-dominated equation of state between $t_i$ and $\tau (M_i)$.

Throughout this analysis, we model the energy injection and evolution of a shock from one \emph{single} PBH immersed in a hot QCD plasma in the early universe. Realistically, one would expect that an entire \emph{population} of PBHs would form at some early time $t_i$ with an extended mass distribution that is sharply-peaked at some typical mass $\bar{M}$ \cite{gorton_how_2024, Gow:2020cou}. This would result in the majority of the PBHs exploding around a time $t\simeq \tau(\bar{M})$. The cumulative effect of propagating shock fronts from many PBHs exploding over time within a Hubble volume may have interesting implications for early universe physics. We explore the role of propagating shock fronts from a \emph{population} of exploding PBHs as a novel baryogenesis mechanism in a companion paper Ref.~\cite{BAUinprep}.

\section{Energy injection in the plasma from evaporation and fluid motion}
\label{sec:energy_inj}

As described in the previous section, during their evaporation, PBHs emit hard, primarily color-charged particles with energy of the order of $E \sim \beta_s T_H$. PBHs that form with $M \sim {\cal O} (10^6 \, {\rm g})$ have a Hawking temperature much larger than the ambient temperature of the background plasma ($T_H \gg T_{\rm b}$) for much or all of their lifetime. This implies that the particles emitted by the PBH are strongly out-of-equilibrium and will reheat the surrounding plasma. In this section, we describe how the hard particles emitted by the PBH deposit their energy into the surrounding plasma, estimate the thermalization length associated with this process, and determine the resulting size and temperature of the initial reheated region around the PBH.

\subsection{Thermalization of the Hawking-radiated particles}

Thermalization of fast color-charged particles propagating in a hot QCD medium has been
studied in detail in several references. (See, for example, Refs.~\cite{Wang:1994fx,Arnold:2003zc} for an in-depth description.) In the high-energy limit ($E \gg m_{\rm eff}$ where $m_{\rm eff}$ is the effective mass of the emitted particle in the plasma), the dominant in-medium process is a nearly collinear, medium-induced bremsstrahlung splitting $E \to k + (E-k)$ generated by the cumulative effect of many soft $2 \to 2$ scatterings. Here $E$ denotes the energy of the parent hard particle and $k$ the energy of the emitted daughter particle. Moreover, we will denote by $q_\perp$ the transverse momentum (with respect to the direction of the parent particle) exchanged in a single soft elastic scattering in the medium, while the accumulated transverse momentum exchange during formation is $k_\perp^2 \sim \sum_{i} q_{\perp,i}^2$. During the splitting, the system passes through an intermediate off-shell state.  The energy difference between this intermediate state and an on-shell state is
\begin{equation}
\Delta E \simeq \frac{k_\perp^2}{2k} \, .
\label{eq:deltaE}
\end{equation}
In quantum mechanics, this energy difference produces a relative phase of order $\Delta E\, t$ between the intermediate state and an on-shell state after a time $t$. The splitting can be treated as completed, and the emitted particle can be regarded as an independent on-shell excitation, once this phase becomes of order unity.  This defines a characteristic formation time $t_{\rm form}$ via $\Delta E\, t_{\rm form} \sim 1$, which gives~\cite{Wang:1994fx}
\begin{equation}
t_{\rm form} \sim \frac{1}{\Delta E} \simeq \frac{2k}{k_\perp^2} \, .
\label{eq:formationtime_pre}
\end{equation}

Multiple soft scatterings in the medium lead to transverse momentum broadening of the emitted particle.  This is conventionally encoded in the momentum-broadening (jet-quenching) coefficient $\hat q$, defined by
\begin{equation}
\hat q \equiv \frac{d\langle k_\perp^2 \rangle}{dt} \, .
\label{eq:qhat_def}
\end{equation}
Over the formation time $t_{\rm form}$, one therefore has
\begin{equation}
k_\perp^2 \simeq \hat q\, t_{\rm form} \, .
\label{eq:kperp_broad}
\end{equation}
Combining Eqs.~\eqref{eq:formationtime_pre} and \eqref{eq:kperp_broad} and
eliminating $k_\perp^2$ gives the Landau-Pomeranchuk-Migdal (LPM) result~\cite{Wang:1994fx}
\begin{equation}
t_{\rm form} \sim \sqrt{\frac{2k}{\hat q}} \, .
\label{eq:formationtime}
\end{equation}

In a screened relativistic QCD plasma at temperature $T$, the transverse-momentum broadening arises from repeated soft elastic scatterings of the energetic particle with thermal constituents of the medium. Following Ref.~\cite{Arnold:2008vd}, these scatterings are encoded in the jet-quenching parameter $\hat q$, which may be written as an integral over the differential elastic scattering rate,
\begin{equation}
\hat q 
\equiv
\int_{q_\perp<\Lambda} d^2 q_\perp\,
\frac{d\Gamma_{\rm el}}{d^2 q_\perp}\,
q_\perp^2 \, ,
\label{eq:qhat_arnold}
\end{equation}
where $q_\perp$ is the transverse momentum transfer in a single elastic scattering and $\Lambda$ is an ultraviolet cutoff separating soft momentum transfers from rare hard scatterings.

At leading order in weak coupling, elastic scattering is dominated by screened $t$-channel gluon exchange.  The differential elastic scattering rate can be approximated by~\cite{Arnold:2008vd}
\begin{equation}
\frac{d\Gamma_{\rm el}}{d^2 q_\perp}
\simeq
\frac{C_R}{(2\pi)^2}\,
\frac{4\pi\alpha_s(T)\, T\, m_D^2}
     {q_\perp^2\left(q_\perp^2+m_D^2\right)} \, ,
\label{eq:elastic_rate}
\end{equation}
where $C_R$ is the quadratic Casimir of the energetic particle (equal to $C_A = 3$ for gluons, and
$C_F = 4/3$ for quarks) and
\begin{equation}
m_D^2 \equiv \frac{2\pi}{3}\, g_c\, \alpha_s(T)\, T^2
\label{eq:Debye_def}
\end{equation}
is the Debye screening mass, with $g_c$ the effective number of color-charged degrees of freedom in the plasma (for QCD, $g_c = 2N_c + N_f$).

Substituting Eq.~\eqref{eq:elastic_rate} into
Eq.~\eqref{eq:qhat_arnold} and performing the $q_\perp$ integral yields
\begin{equation}
\hat q 
\simeq
2\,C_R\,\alpha_s(T)\,T\,m_D^2
\ln\!\left(\frac{\Lambda}{m_D}\right) \,.
\label{eq:qhat_log}
\end{equation}
The cutoff $\Lambda$ is fixed by the transverse momentum scale accumulated during the formation time of the splitting. For an emitted particle of energy $k$, the typical total transverse momentum acquired over the formation time satisfies
\begin{equation}
\Lambda \sim \bar k_\perp
\sim \left(\hat q\, k\right)^{1/4}\, .
\label{eq:Lambda_def}
\end{equation}
Inserting Eq.~\eqref{eq:Lambda_def} into Eq.~\eqref{eq:qhat_log}, and retaining only the leading logarithm\footnote{The term proportional to $\ln\hat q$ produces only subleading $\ln\ln k$ corrections.} we obtain
\begin{equation}
\hat q
\simeq
2\,C_R\,\alpha_s(T)\,T\,m_D^2
\left[
\frac14 \ln \left(\frac{k}{T}\right) - \ln \left(\frac{m_D}{T}\right)
\right]\,.
\label{eq:qhat_LL_mD}
\end{equation}
Using Eq.~\eqref{eq:Debye_def} and again dropping the subleading logarithm (\emph{i.e.}, the second term) we may write the expression
\begin{equation}
\hat q(T)
\simeq
\frac{\pi}{3}\,
C_R\, g_c\,
\alpha_s^2(T)\, T^3\,
\ln\!\left(\frac{k}{T}\right).
\label{eq:qhat_final}
\end{equation}
The LPM splitting rate for high-energy emitted
particles of momentum $k$ in a QCD plasma of background temperature $T_{\rm b}$ is thus:
\begin{equation}
\label{eq:GammaLPM}
\Gamma_{\rm LPM}(k,T_{\rm b})\ \sim\ \frac{\alpha_s(T_{\rm b})}{t_{\rm form}(k,T_{\rm b})}
\ \sim\ \alpha_s(T_{\rm b})\,\sqrt{\frac{\hat q(T_{\rm b})}{ 2 k}}\, .
\end{equation}
In our scenario, the emitted particles are high-energy Hawking radiation from a low-mass PBH, and the plasma is the radiation-dominated fluid filling the universe during early times. Since the Hawking spectrum is sharply peaked at $E_{{\rm peak},\,s}\sim \beta_{s}T_H$, as in Eq.~(\ref{Epeak}), the typical momentum of an emitted particle is $k\sim \beta_{s} T_H$. Note that we henceforth take $\beta_s=\beta_{1/2}=4.53$ because the majority of emitted particles are quarks. 

The mean distance to the first hard split (\emph{i.e.} the microscopic thermalization length) sets the initial scale size of the fireball created by a PBH of mass $M$ in a background plasma of temperature $T_{\rm b}$, which we label $L_{\rm LPM}$:
\begin{equation}
\begin{split}
L_{\rm LPM}&(M, T_b) \\
&\equiv
\Gamma_{\rm LPM}^{-1}(E_{\rm peak}(M),T_{\rm b})\\
&\simeq
\frac{1}{\alpha_s^2(T_{\rm b})\,T_{\rm b}}\,
\sqrt{
\frac{
6\,\beta_{1/2}\,T_H
}{
\pi\,C_R\,g_c\,
T_{\rm b} \ln\left( \beta_{1/2}T_H/T_{\rm b}\right)
}
}\, .
\end{split}
\label{eq:LPM_length}
\end{equation}
In Fig.~\ref{fig:LPM}, we show the evolution of the length $L_{\rm LPM}$ as a function of the background temperature, $T_{\rm b}$ evaluated at $M=M_{\rm thres}$ as defined in Eq.~\eqref{eq:Mthres}. Importantly, we further demonstrate in Appendix \ref{app:homogen} that the reheated region is isotropic, thanks to the large number of particles emitted.

Let us now estimate the temperature at $r = L_{\rm LPM}$, which we call $T_{\rm LPM}$. This is the temperature around the PBH before the hydrodynamical response of the plasma. We assume the profile of temperature in the region $r < L_{\rm LPM}$ to be constant and equal to $T_{\rm LPM}$, and call this region the initial \emph{fireball}. It is clear that this is a very approximate assumption, and that for $r < L_{\rm LPM}$ the plasma is a complicated out-of-equilibrium fluid. However, our analysis does not attempt to describe the details of the fluid in the region $r < L_{\rm LPM}$, and for our purpose the assumption of constant temperature within that region is sufficient. This homogeneous temperature $T_{\rm LPM}$ can be obtained by equating the total energy $E_{\rm inj}$ injected 
by the exploding PBH
in a volume $\frac{4}{3}\pi L_{\rm LPM}^3(M,T_{\rm b})$ with the equivalent radiation energy density $\rho_{\rm rad}(T_{\rm LPM})$:
\begin{equation}
    \label{eq:ThsEqn}
     \frac{E_{\rm inj}}{\frac{4}{3}\pi L_{\rm LPM}^3(M,T_{\rm b})} = \frac{\pi^2}{30}g_*(T_{\rm LPM}) T_{\rm LPM}^4\, ,
\end{equation}
which yields
\begin{equation}
\label{eq:TLPMdef}
\begin{split}
    T_{\rm LPM} & (M, E_{\rm inj}, T_{\rm b}) \\
 &= \left[ \frac{ 45}{2 \pi^3} \frac{  E_{\rm inj}}{ g_* (T_{\rm LPM}) \, L_{\rm LPM}^3 (M, T_{\rm b})} \right]^{1/4} .
 \end{split}
\end{equation}

Note that the left-hand side of Eq.~(\ref{eq:ThsEqn}) depends on the background cosmological plasma temperature $T_{\rm b}$, while the right-hand side is the energy density of the fluid \emph{inside} the fireball at temperature $T_{\rm LPM}$. While our definition of the thermalization length $L_{\rm LPM}$ holds generically for a PBH with $T_H>T_{\rm b}$ and can effectively be taken to evolve as the PBH mass evolves in time because high emission rates lead to instantaneous and homogeneous thermalization within that radius (see Appendix~\ref{app:homogen}), the temperature $T_{\rm LPM}$ we derive in Eq.~\eqref{eq:TLPMdef} only holds for energy injection over sufficiently short time scales. This is because, unlike $L_{\rm LPM}$, the temperature $T_{\rm LPM}$ has an explicit dependence on $E_{\rm inj}$ which is found by time-integrating the emitted power.

In Sec.~\ref{sec:hydro_motion} we present a careful discussion of the hydrodynamics. We then determine the appropriate time scale for instantaneous energy injection, which defines the case when we \emph{can} apply the definition from Eq.~\eqref{eq:TLPMdef} to estimate the initial temperature of the fireball created by the PBH explosion.

\subsection{Heat transfer and fluid flow}
\label{sec:Peclet}

The physics of the energy distribution around the PBH is controlled by two physically distinct processes: diffusion of heat and advection. In the former mechanism, the energy is redistributed via collisions between particles and the distance over which the energy is transported is controlled by random walk, while in the latter mechanism, the energy is transported by the bulk motion of the fluid. Early studies of the behavior of fireballs around evaporating PBHs have all assumed that diffusion processes would dominate the energy transport~\cite{Das:2021wei,He:2022wwy,Levy:2025lyj,Altomonte:2025hpt, Gunn:2024xaq}, whereas, as we discuss here, it is actually {\it advection} which most often controls the heat transport. 

To build intuition, let us consider the timescales required to transport some energy over a distance $L$ in each mechanism. The timescale of advection is given by $t_{\rm ac}(L) \sim  L/v_{\rm} $, where $v$ is the typical bulk velocity of the fluid under consideration. On the other hand, the timescale of diffusion is $t_{\rm diff}(L) \sim L^2/D(T)$ where $D(T)\equiv \eta/(\rho+p) \sim 1/T$ is the diffusion constant in a relativistic plasma with $\eta$ the shear viscosity, $\rho$ the energy density and $p$ the fluid pressure. The competition between these two mechanisms can be estimated by taking the ratio of these timescales,
$t_{\rm diff}(L_{})/t_{\rm ac}(L_{\rm })$. This ratio defines the \emph{P\'eclet number}, which takes the form
\begin{equation}
\label{eq:Pe-crit}
\mathrm{Pe}_{}\;\equiv\;\frac{v_{}\,L_{}}{D(T_{})}
\,.
\end{equation}
A large P\'eclet number ensures that advection dominates transport, while a small P\'eclet number ensures that diffusion dominates. 

This simple criterion may be applied to the case of a PBH explosion. The typical length of transport that we should consider is the LPM length, given in Eq.~(\ref{eq:LPM_length}). However, the typical bulk fluid velocity is unknown at this stage. To determine it, let us distinguish two regimes of energy injection from the PBH: 1) the slow and roughly constant energy injection which occurs long before the final evaporation and 2) the final explosion, in which a large amount of energy is injected nearly instantaneously. As we will see, in each of these regimes the P\'{e}clet number is greater than one, indicating that advection dominates diffusion.

{\textbf{Slow injection}}. The case of slow injection from a PBH is discussed in detail in Sec. \ref{sec:long_before}. Assuming that the injection rate from the PBH is constant, and using both analytical arguments and simulations, we find that for distances such that $r > L_{\rm LPM}$, the velocity is
\begin{equation}
\label{eq:far_from_exp}
  v(r) = C_{\rm num}\frac{P/\rho_{\rm b}}{r^{2}} \,, \qquad C_{\rm num} = 0.005   \, ,
\end{equation}
where $C_{\rm num}$ is a numerical prefactor which has been extracted from the simulations and $P$ is the power emitted by the PBH, which can be obtained by evaluating Eq.~\eqref{eq:PowerTimeTemp} at some chosen time. A shock does not form if the velocity at $r=L_{\rm LPM}$ is smaller than the speed of sound, or in other words if 
 \begin{equation} \label{eq:no_shock_formation}
 C_{\rm num}\frac{P/\rho_{\rm b}}{L_{\rm LPM}^{2}} \lesssim 1/\sqrt{3}  \, .
 \end{equation}
 We emphasize that, in contrast to the growing fireball, which will be discussed below, the fluid velocity field in this slow injection case is \emph{static}. Finally, we can evaluate the P\'eclet number for the slow-injection case at $r= L_{\rm LPM}$, where it is expected to be maximal. We obtain 
   \begin{align}
   \begin{split}
    \mathrm{Pe}_{\rm slow} &\equiv \frac{v({r= L_{\rm LPM}})\,L_{\rm LPM}}{D(T_{\rm})} \sim \frac{C_{\rm num}}{D(T_{\rm})}  \frac{P \, L_{\rm LPM}^{-1}}{\rho_{\rm b}}
    \\
    &\sim 1.6 \times 10^7 \,C_{\rm num}\left(\frac{150\, {\rm GeV}}{T_{\rm b}} \right)^2 \\
    &\quad\times \,
\bigg(\frac{
\pi\,C_R\,g_c\,
T_{\rm b} \ln\left( \beta_{1/2}T_H/T_{\rm b}\right)
}{
6\,\beta_{1/2}\,T_H
}\bigg)^{1/2} \\
   &\quad\times  \left(\frac{150\, {\rm GeV}}{T_f} \right)^{-4/3}  \, ,
   \end{split}
   \label{eq:PecletSlow}
\end{align}
where $T_f$ is the temperature at which the PBH completes its evaporation. For the regime of interest here, involving a PBH with mass $M \lesssim {\cal O} (10^6 \, {\rm g})$ corresponding to $T_H \gtrsim {\cal O} (10^7 \, {\rm GeV})$ and $T_{\rm b} \sim T_f \sim {\cal O}(150 \, {\rm GeV})$, the term in parentheses involving $T_H$ contributes $\sim {\cal O} (10^{-2})$, and hence we find ${\rm Pe}_{\rm slow} \gg 1$. In other words, advection dominates diffusion in the slow-injection case.

{\textbf{Final explosion}}. For the very final stages of the life of a PBH, at the time of its explosion, energy injection is a runaway process. The hydrodynamical simulations described in Sec. \ref{sec:inst} indicate that when the energy injected $E_{\rm inj} \gg \rho_{b} L_{\rm LPM}^3$ is much larger than the energy of the background contained in a sphere of radius $L_{\rm LPM}$ centered on the PBH, then $v\to 1$. Concretely, reversing the inequality in Eq.~\eqref{eq:no_shock_formation}, one obtains the condition for the formation of a growing fireball:
\begin{equation}\label{eq:shock_formation}
 C_{\rm num}\frac{P/\rho_{\rm b}}{L_{\rm LPM}^{2}} \gtrsim 1/\sqrt{3}  \, . 
 \end{equation}
In this context, the P\'eclet number at the far edge of the fireball takes the value
\begin{align}
\label{eq:Pc_fb}
    \mathrm{Pe}_{\rm fb}\;&\equiv\;\frac{v_{\rm fb}\,L_{\rm LPM}}{D(T_{\rm b})} \sim L_{\rm LPM} \,T_{\rm b} \gg 1 \, , 
\end{align}
where the ``fb'' subscript refers to the heated fireball and $v_{\rm fb} \sim \mathcal O(1)$ is the plasma bulk velocity at the boundary of the fireball. We therefore have $\mathrm{Pe}_{\rm fb} \gg 1$, again indicating that advection transport dominates over diffusion.

The fact that ${\rm Pe} \gg 1$ for both the slow-injection and near-instantaneous injection cases means that we may neglect the diffusive transport in what follows and focus on the transport of energy within the plasma via flow motion.

\section{The hydrodynamical description}
\label{sec:hydro_motion}

In the previous section, we showed that, for scales larger than the fireball scale $L_{\rm LPM}$, the flow of energy is captured by advective transport. This advective transport is modeled by ideal hydrodynamics, which we discuss below.

 \subsection{Hydrodynamical evolution}

We describe the evolution of the fireball in terms of relativistic hydrodynamics, starting from the local conservation of the energy-momentum tensor,
\begin{equation}
\partial_\mu T^{\mu \nu} = 0 \, .
\end{equation}
Here $T^{\mu\nu}$ is the stress-energy tensor of the fluid, which is expressed as a function of pressure $p$ and energy density $\rho$ via 
\begin{align}
T^{\mu\nu}= ( \rho + p )u^\mu u^\nu+ \eta^{\mu\nu}p \; ,
\label{Tmnfluid}
\end{align}
where $\eta^{\mu\nu}$ is the Minkowski metric tensor and $u^\mu= \gamma (1, v, 0, 0)$ is the fluid velocity four-vector in spherical coordinates. 

Projecting this equation along and perpendicular to the local flow velocity, and assuming spherical symmetry of the solution, the system can be reduced to an effective one-dimensional form. The conservation equations take the generic form
\begin{equation}
\label{eq:hydro_eq}
\partial_t U + \partial_r f + g = 0 \, ,
\end{equation}
where $U$ denotes the conserved quantities, $f$ their fluxes, 
and $g$ represents geometric source terms arising from spherical symmetry.\footnote{Note that these variables are functions of $(r,t)$ and they describe both the shock and the background, and therefore we leave them unindexed.} The vectors $U, f$ and $g$ are related to the hydrodynamical quantities in the following way:
\begin{equation}
\begin{split}
U &= 
\begin{pmatrix}
U_1 \\[3pt]
U_2
\end{pmatrix}
=
\begin{pmatrix}
w \gamma^2 - p \\[3pt]
w \gamma^2 v
\end{pmatrix} \, , \\
f &=
\begin{pmatrix}
w \gamma^2 v \\[3pt]
w \gamma^2 v^2 + p
\end{pmatrix}, \quad
g = 
\frac{d-1}{r}
\begin{pmatrix}
w \gamma^2 v \\[3pt]
w \gamma^2 v^2
\end{pmatrix} \, ,
\end{split}
\label{ufgevolve}
\end{equation}
where $w$ is the enthalpy density, $p$ the fluid pressure, 
$v$ the radial velocity, $\gamma = (1 - v^2)^{-1/2}$ the Lorentz factor, and $d$ is the number of spatial dimensions. The last term accounts for the divergence of the spherical flow.

These coupled nonlinear partial differential equations govern the time evolution of the local temperature and velocity fields of the expanding fireball. They form the basis for the numerical scheme, which is described explicitly in Appendix \ref{app:scheme}.

Throughout this paper, we perform numerical simulations in dimensionless units. The purpose of the simulations is to offer a verification of the validity of various analytical expressions. The simulations can be matched to physical cases by the following reparametrization: 
\begin{equation}
\begin{split}
    &p_{\rm phys} = \frac{g_\star \pi^2}{90} T_{\rm phys}^4 \, p_{\rm num} ,\\
 &E_{\text{inj, phys}} = \frac{g_\star \pi^2}{90} T_{\rm phys}^4 \left(\frac{L_{\rm fb, phys}}{L_{\rm fb, num}}\right)^3 \times E_{\text{inj, num}} 
    \end{split}
    \label{eq:SimToPhysUnits}
\end{equation}
where the physical quantities come with the label ``${\rm phys}$'' while the numerical ones are labeled by ``${\rm num}$.'' 

\begin{figure}[t]  
\centering
\includegraphics[width=0.45\textwidth]{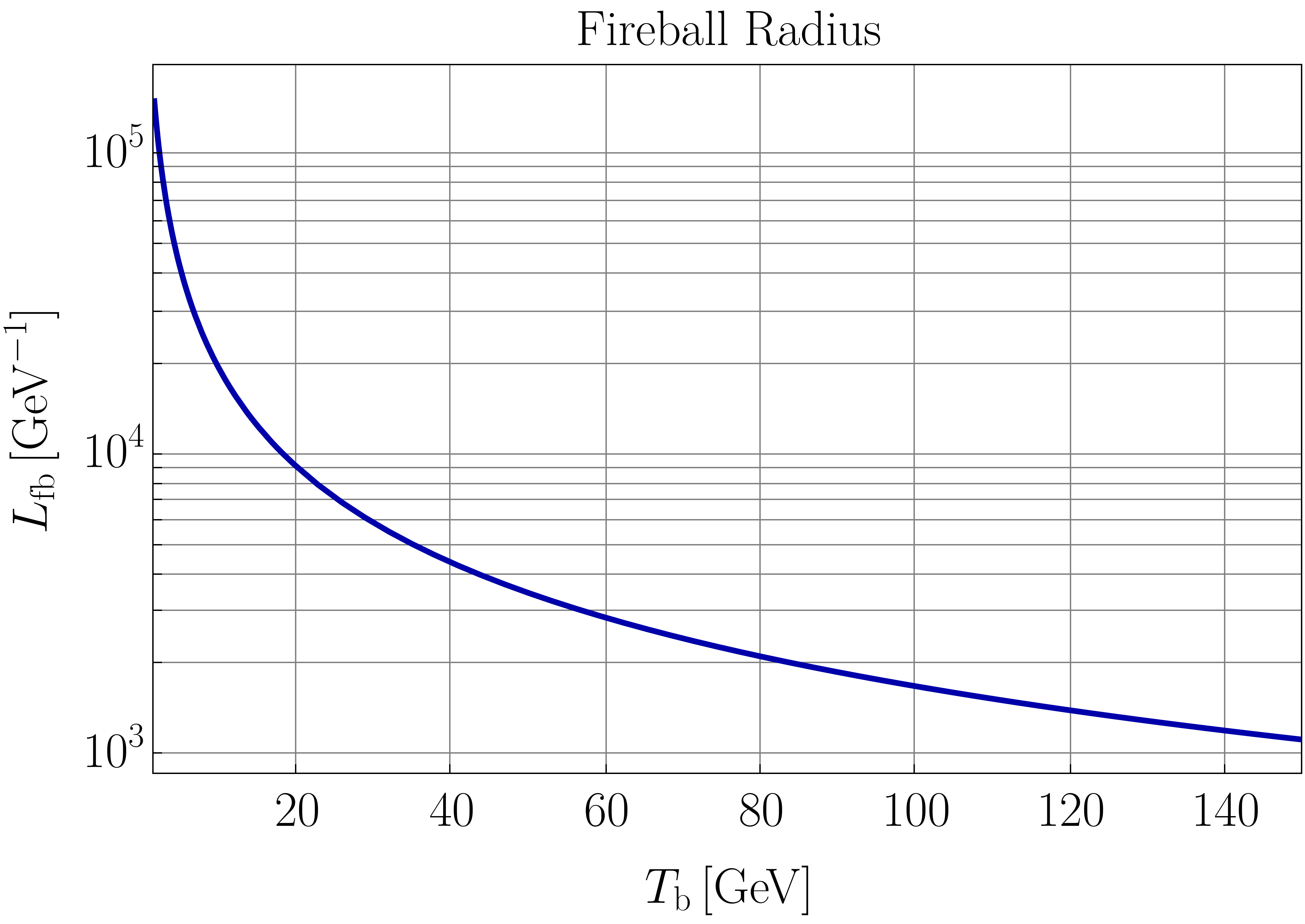}
    \caption{\justifying 
     Dependence of the initial fireball radius $L_{\rm fb}$ on the background temperature, from Eq.~\eqref{eq:Lfb}. This sets the initial size of the over-pressured sphere of plasma surrounding the PBH after it explodes and instantaneously injects energy $E_{\rm inj}=M_{\rm thres}(T_{\rm b})$.}
    \label{fig:LPM}
\end{figure}

\subsection{Timescales involved in the energy injection}

Hydrodynamics involves an inherent response timescale, which can be understood as the typical timescale over which waves are formed when a spatial gradient of pressure is turned on. Schematically, the relevant hydrodynamical equation takes the form $\partial_t v \sim (\partial_r p)/p$. Assuming some significant increase in the temperature between $T(r= L_{\rm LPM})$ and $T_{\rm b}$, concentrated within a region of length of order $L_{\rm LPM}$, this equation allows one to extract the timescale of wave formation: $ L_{\rm LPM}/c_s$.

Consequently, from the point of view of the fluid response, any amount of energy which is injected on a timescale $ \ll L_{\rm LPM}/c_s $ is injected \emph{instantaneously}. As we have seen in Sec.~\ref{sec:Hawking_rad}, the evaporation of a PBH is a runaway process in which the rate of energy injection increases with time. There is thus unavoidably some fraction of the mass of the PBH which will be injected in an \emph{instantaneous} fashion. We call this fraction the \emph{threshold mass} $M_{\rm thres}(T)$ and we obtain its value by equating the remaining PBH lifetime with the typical LPM length-scale: $\tau (M_{\rm thres})  = L_{\rm LPM}$. For PBHs with mass $M \lesssim 10^9 \, {\rm g}$, we may use the analytic expression in Eq.~(\ref{TauSmallM}) for $\tau (M)$; combining with Eq.~\eqref{eq:LPM_length} for $L_{\rm LPM}$, we find
\begin{equation}
\label{eq:Mthres}
M_{\rm thres} (T_{\rm b}) = \big[ 3 A f_{\rm max} \, L_{\rm LPM} (M_{\rm thres}, T_{\rm b}) \big]^{1/3} ,
\end{equation}
where $A = 6.036 \times 10^{72} \, {\rm GeV}^4$ and $f_{\rm max} = 15.522$ were introduced in Sec.~\ref{sec:Hawking_rad} and hold for the case of Hawking emission of strictly SM degrees of freedom. Given the dependence of $L_{\rm LPM}$ on $M_{\rm thres}$, Eq.~(\ref{eq:Mthres}) must be solved numerically. 

In this scenario, the evaporating PBH will inject energy $E_{\rm inj}=M_{\rm thres}(T_{\rm b})$ on a time scale shorter than the plasma response time scale, thus instantaneously heating a sphere of radius $L_{\rm LPM}$ around the PBH. Therefore, we can exactly apply Eq.~\eqref{eq:TLPMdef} to compute the temperature of the fireball: 
\begin{equation}
\begin{split}
    \label{eq:T_hs_init}
 T_{\rm fb} & (K, T_{\rm b}) \\ & = T_{\rm LPM}(M=M_{\rm thres}, E_{\rm inj}=K M_{\rm thres}, T_{\rm b})\\
 &= \left[ \frac{ 45}{2 \pi^3} \frac{ K \, M_{\rm thres} (T_{\rm b})}{ g_* (T_{\rm fb}) \, L_{\rm LPM}^3 (M_{\rm thres}, T_{\rm b})} \right]^{1/4} . 
     \end{split}
\end{equation}
Note that we have generalized to $E_{\rm inj}=KM_{\rm thres}(T_{\rm b})$, such that, for the instantaneous case we have 
\begin{equation}
    T_{\rm fb}^{\rm inst}(T_{\rm b})\equiv T_{\rm fb}(K=1, T_{\rm b}).
\end{equation}

\begin{figure}[t!]
\centering
    \includegraphics[width=0.45\textwidth]{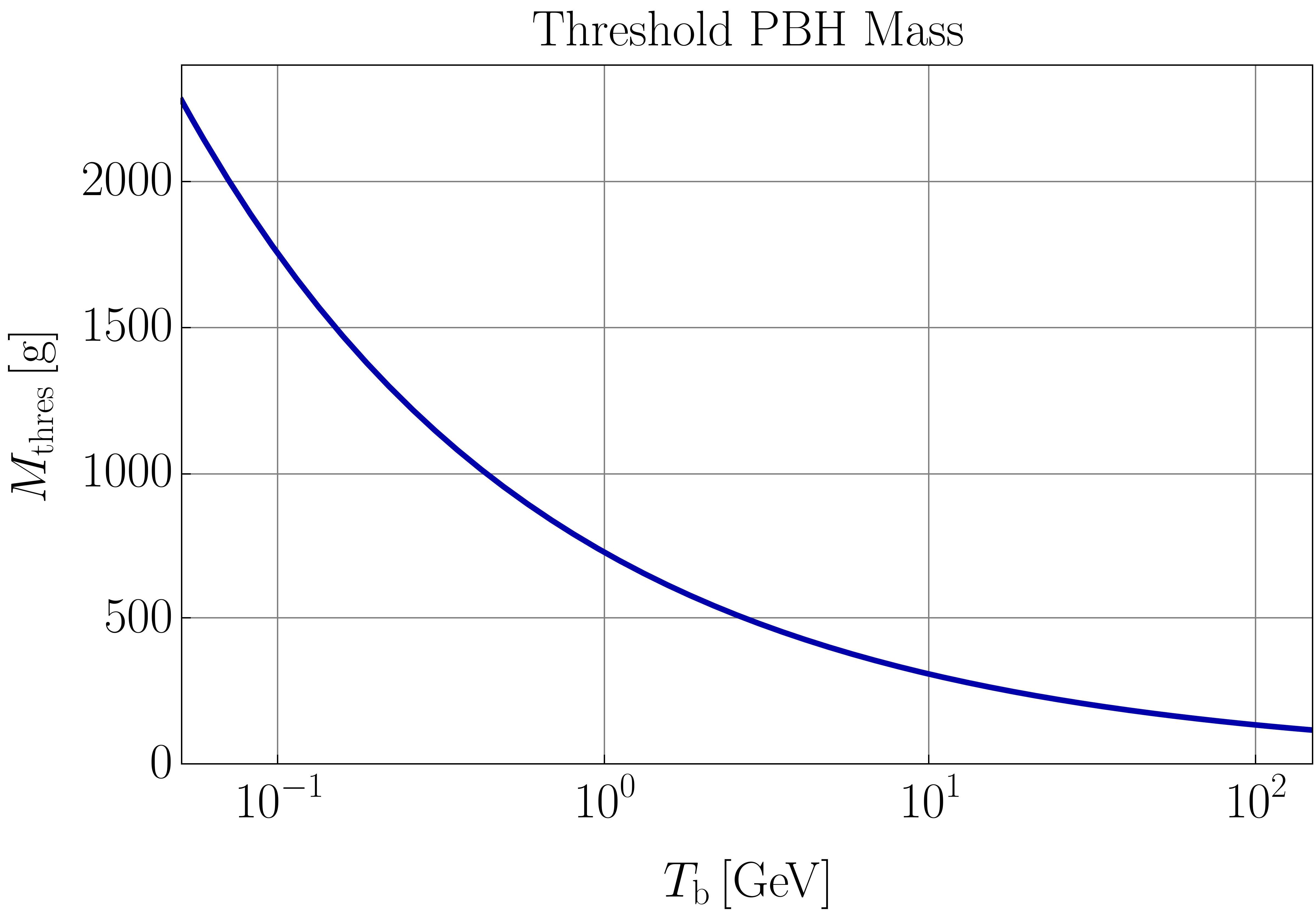} \includegraphics[width=0.45\textwidth]{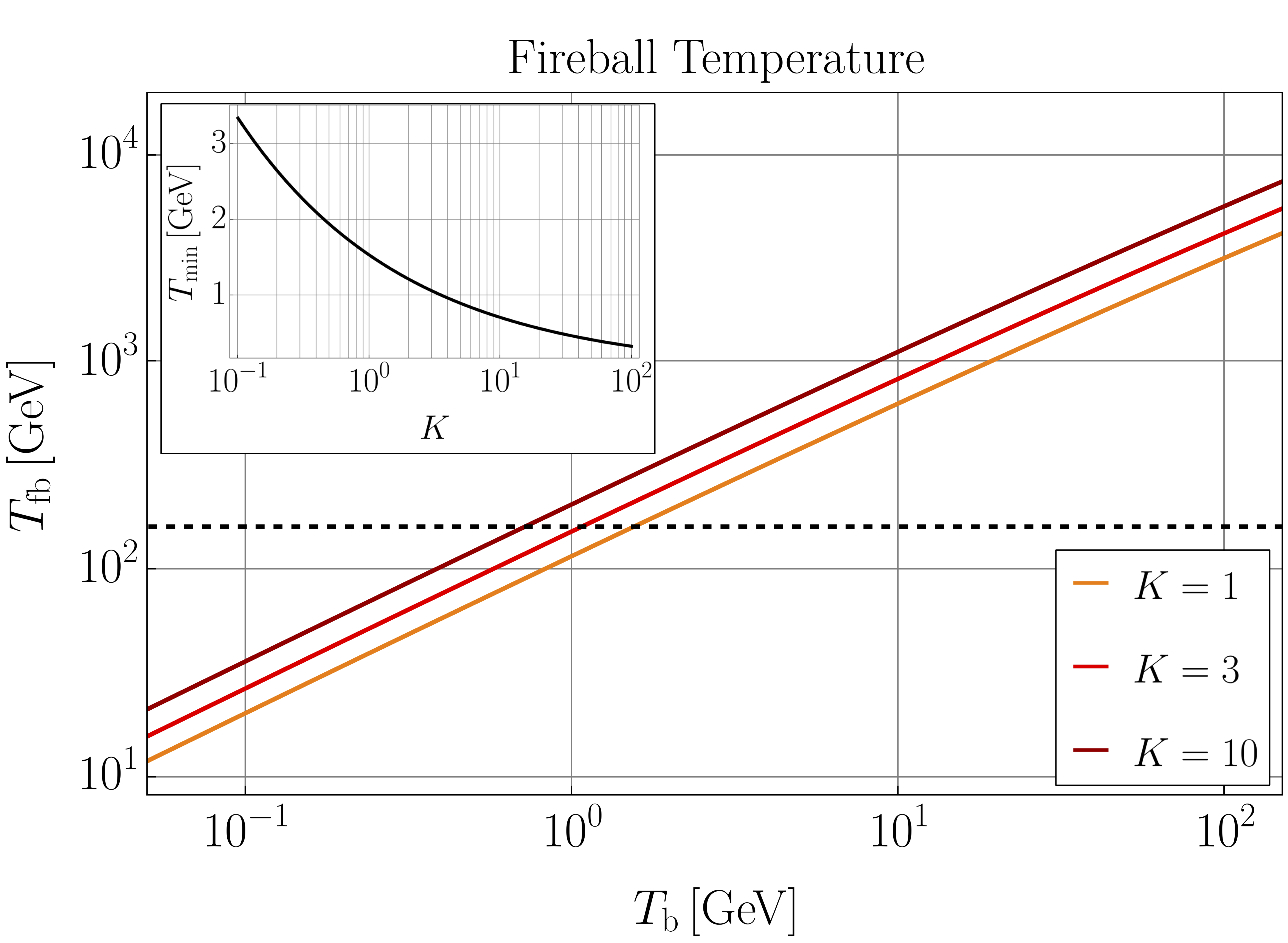}
   \caption{
   \justifying
    ({\it Top}) Threshold PBH mass for instantaneous energy injection from Eq.~(\ref{eq:Mthres}) as a function of cosmological background temperature $T_{\rm b}$. ({\it Bottom}) Initial temperature of the fireball $T_{\rm fb}$ as a function of background temperature $T_{\rm b}$ at the time of the PBH explosion. The parameter $K$ sets the amount of injected energy, with $K=1$ corresponding to the purely instantaneous injection case. The inset plot shows the minimum background temperature $T_{\rm min}$ such that a PBH explosion will create a fireball with $T_{\rm fb}>T_{\rm EW}\simeq162 \, {\rm GeV}$ (See Sec.~\ref{sec:Pheno_impl} for discussion of EW symmetry breaking). 
    }
    \label{fig:Mthres}
\end{figure}

As we will show with simulations and discuss in detail in Sec.~\ref{sec:InjectionRateComparison}, taking $K>1$ allows us to apply the instantaneous formalism (i.e. Eq.~\eqref{eq:T_hs_init}) to approximate the behavior of the fireball in more realistic \emph{extended} energy injection scenarios. Setting $K>1$ takes into account local heating from the ``slow'' emission phase, which occurs \emph{prior} to the final instantaneous explosion and formation of the shock and will effect the temperature profile around the PBH at the time of the explosion. See Fig.~\ref{fig:Mthres} for plots of $M_{\rm thres}$ and $T_{\rm fb}$ as functions of $T_{\rm b}$.

We can now define the initial radius of the fireball for the instantaneous injection scenario:
\begin{equation}
    \label{eq:Lfb}
    L_{\rm fb}(T_{\rm b})=L_{\rm LPM}(M=M_{\rm thres}, T_{\rm b})\,,
\end{equation}
where we have evaluated $L_{\rm LPM}$ from Eq.~\eqref{eq:LPM_length} at $M_{\rm thres}(T_{\rm b})$, which is found by numerically solving Eq.~\eqref{eq:Mthres}. See Figure~\ref{fig:LPM}.

As we will see, analyzing the impact of the energy emitted before $M = M_{\rm thres}(T_{\rm b})$ requires longer and more involved simulations than are necessary for the instantaneous injection case. We therefore split the analysis into three regimes: (i)~the slow injection of energy with constant power $P$, discussed in Sec.~\ref{sec:long_before}, (ii)~strictly instantaneous energy injection in Sec.~\ref{sec:inst}, in which a fireball rapidly converts into a strong shock and the surrounding plasma is unaffected by earlier energy injection, and (iii)~finally, a treatment that incorporates both the extended PBH-like injection phase and the final explosion in Sec.~\ref{sec:PBH-like_inj}. In this final case, a fireball persists for a finite time before relaxing into a shock once the Hawking emission of energetic particles ends. As discussed in Sec.~\ref{sec:InjectionRateComparison}, in the light of these simulations, we find $K \sim {\cal O} (5)$ for scenarios of interest.

\section{Hydrodynamic transport of energy long before the explosion}
\label{sec:long_before}

In this section, we estimate the velocity of the fluid around the PBH long before the final PBH explosion, in the regime in which the evaporation is slow and the power emitted by the PBH is approximately constant. To streamline the presentation, we first present some simplified analytical considerations, before refining them with numerical simulations. 

\begin{figure*}[t!]
        \centering
        \includegraphics[width=0.5\linewidth]{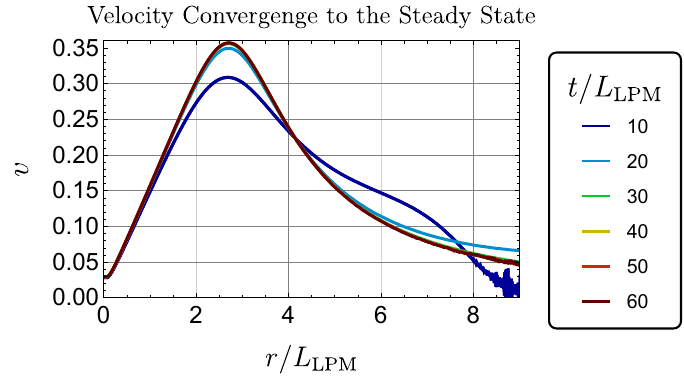}
        \includegraphics[width=0.4\linewidth]{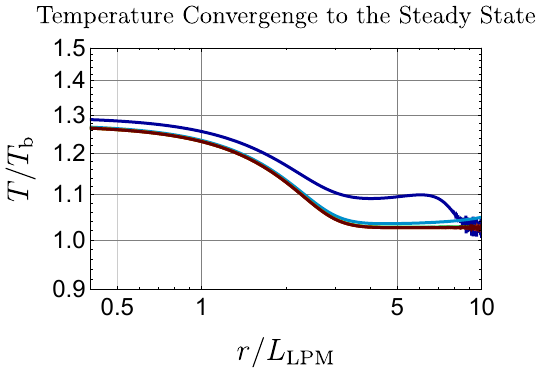}
    \caption{\justifying\label{fig:far_from_expl_1} {\textbf{Convergence to the Steady State profile}}  ({\it Left})  Behavior of the fluid velocity for the case of slow energy injection, earlier than the PBH explosion. The injected power in this simulation is fixed to $P_{\rm num} = 450$ and it is injected on a radius being $2L_{\rm LPM}$. ({\it Right}) Same for the temperature. One observes that the velocity and the temperature profiles each reach a time-independent \emph{steady state} solution.
    }
\end{figure*}

Neglecting the temperature gradient, the hydrodynamical equations in the presence of a localized constant source simplify to 
\begin{equation}
\label{eq:toyeq}
    %p_{\rm PBH}
    {\cal P} (r)\sim \rho_{\rm b} \bigg(\partial_r v + \frac{(d-1) v}{r}\bigg) \, ,
\end{equation}
where $v$ is the velocity of the plasma and 
%%%%
\begin{equation}
{\cal P}(r) \equiv  \frac{3 P}{4\pi L_{\rm LPM}^3} \Theta(L_{\rm LPM} -r) 
\end{equation}
is the power density emitted by the PBH, which we assume to be slowly varying in time for $t \ll \tau (M)$, uniform within a sphere of radius $r= L_{\rm LPM}$, and zero beyond it. From now on, we fix $d=3$ and, as a first approximation, we consider non-relativistic fluid flow, thereby setting $\gamma = 1$ in the fluid equations. One can solve for the fluid velocity $v$ by integrating Eq.~\eqref{eq:toyeq} over a ball of radius $r > L_{\rm LPM}$,
\begin{equation}
   \frac{4\pi}{3} {\cal P} L_{\rm LPM}^3 \approx P \sim 4\pi\rho_{\rm b} \int^r d \tilde{r} \, \tilde{r}^{2}\bigg(\partial_{\tilde{r}} v + \frac{2v}{ \tilde{r}}\bigg)  \, ,
\end{equation}
where we assumed a constant temperature profile for simplicity. For $ r > L_{\rm LPM}$, the solution for a constant source is given by 
\begin{equation}
\label{eq:far_from_exp_1}
  v(r) = \frac{C}{r^{2}}   \qquad \left( r > L_{\rm LPM} \right) ,  
\end{equation} 
where the constant $C$ has to be determined by matching at $r \sim  L_{\rm LPM}$. At $r \sim L_{\rm LPM}$, the velocity as a function of $r$ is maximal, and therefore we have $\partial_r v \sim 0$ and the hydrodynamical equation simplifies to ${\cal P}(r)\sim \rho_{\rm b}  (d-1) v/r$. We then find that the velocity is given by 
\begin{equation}
\label{eq:max_velo}
    v(r= L_{\rm LPM}) \approx P \frac{L_{\rm LPM}^{-2}}{8\pi\rho_{\rm b}}  \, . 
\end{equation}
 This expression is of course only valid as long as $v \ll 1$ and the other terms in the hydrodynamical equation can be neglected. Notice that this last equation allows us to restore the units: 
 \begin{equation}
 \label{eq:Power_trans}
     P_{\rm phys} = \frac{\rho_{\rm b, phys} L_{\rm LPM, \rm phys}^2}{\rho_{\rm b, num}L_{\rm LPM, \rm num}^2} \times P_{\rm num} \, .
 \end{equation}

\begin{figure}[t!]
        \centering
        \includegraphics[width=0.9\linewidth]{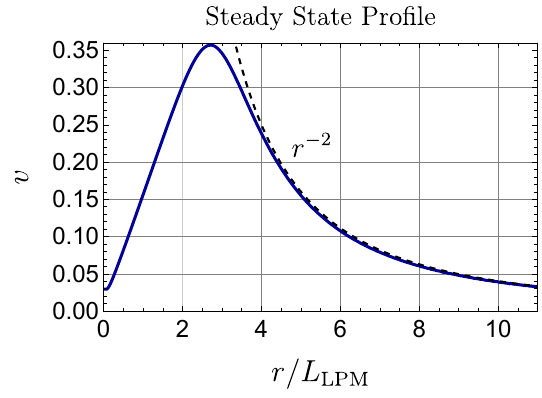}
    \includegraphics[width=0.9\linewidth]{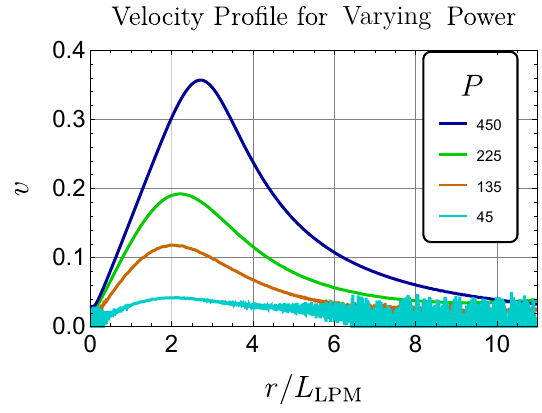}\caption{\justifying\label{fig:far_from_expl}  ({\it Top})  Behavior of the fluid velocity for the case of slow energy injection, earlier than the PBH explosion. The energy is injected homogeneously over a radius of $2 L_{\rm LPM}$. One observes that the falloff of the velocity is very well fitted by a $1/r^2$ curve, as obtained analytically in Eq.~\eqref{eq:far_from_exp}. ({\it Bottom}) Velocity curves for different values of the power injected by the PBH.  }
\end{figure}

\begin{figure}[t!]
        \centering
        \includegraphics[width=0.9\linewidth]{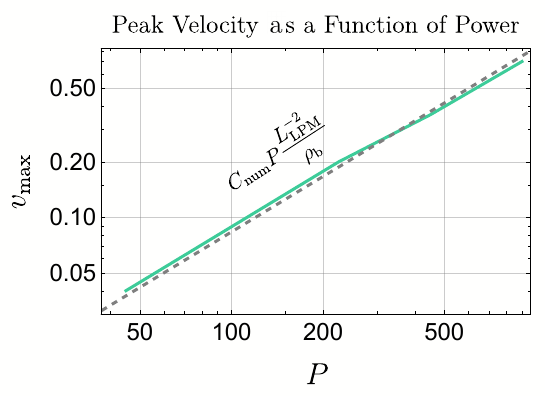}
    \caption{\justifying\label{fig:far_from_expl_trend}  Dependence of maximal fluid velocity $v_{\rm max} \sim v(r=L_{\rm LPM})$ on the power emitted by the PBH during the slow-injection phase for $L_{\rm LPM }=2$ and $\rho_B = 1.5$. We observe that the dependence is very close to linear and we can extract $C_{\rm num} \approx 0.005$, see Eq.~(\ref{eq:vslowinj}), which is presented by a dashed line on the plot.
    }
\end{figure}

To confirm these expectations, we now turn to numerical simulations of the system, using the hydrodynamical code presented in Appendix \ref{app:scheme}. Numerically, we inject the energy homogeneously within a radius of order $2L_{\rm LPM}$ and we set $\rho_{\rm b, \rm num} = 1.5$. Using the relation in Eq.~\eqref{eq:Power_trans}, the power corresponds to $P_{\rm phys}=[45,450] \times (\rho_{b,\rm phys} L_{\rm LPM, \rm phys}^2/6) \sim 10^{39} \, {\rm GeV \, s^{-1}}$. This corresponds, via Eq.~\eqref{eq:PowerLawFit}, to emission from a PBH with mass $M\simeq\mathcal{O}(10^6\,{\rm g})$. We obtain the following criterion,
 \begin{equation}
 \frac{P/\rho_{\rm b}}{8\pi L_{\rm LPM}^{2}} \lesssim 1/\sqrt{3}\, , 
 \end{equation}
which, as we will see, is the regime in which neither a shock wave nor a fireball form. We present the results in Figs.~\ref{fig:far_from_expl_1} and \ref{fig:far_from_expl}. First, we observe in Fig.~\ref{fig:far_from_expl_1} that the fluid velocity profile tends to a steady state, where the profile is smooth everywhere and decays as expected as $r^{-2}$ for $r> 2L_{\rm LPM}$. Second, as can be observed in Fig.~\ref{fig:far_from_expl}, the fluid velocity profile presents a peak at $r \sim {\cal O} (L_{\rm LPM})$. Additionally, we observe that Eq.~\eqref{eq:max_velo} is well verified if we replace the coefficient $1/8\pi$ with some constant $C_{\rm num}$:
 \begin{equation}
 \label{eq:vslowinj}
    v(r= L_{\rm LPM}) \approx C_{\rm num} P \frac{L_{\rm LPM}^{-2}}{\rho_{\rm b}}  \, , \qquad  C_{\rm num}  \approx 0.005 \, . 
\end{equation}
We infer the value $C_{\rm num} \approx 0.005$ from the scaling shown in Fig.~\ref{fig:far_from_expl_trend}. Notice that the numerically obtained value is close to the analytical estimate $1/8\pi \sim 0.03$. As discussed in Sec.~\ref{sec:Peclet}, the scaling of $v (r)$ as in Eq.~(\ref{eq:vslowinj}) yields a P\'{e}clet number for this scenario ${\rm Pe}_{\rm slow} \gg 1$, indicating that advection strongly dominates diffusion in this regime.

\begin{figure*}
    \centering
    \includegraphics[width=0.42\linewidth]{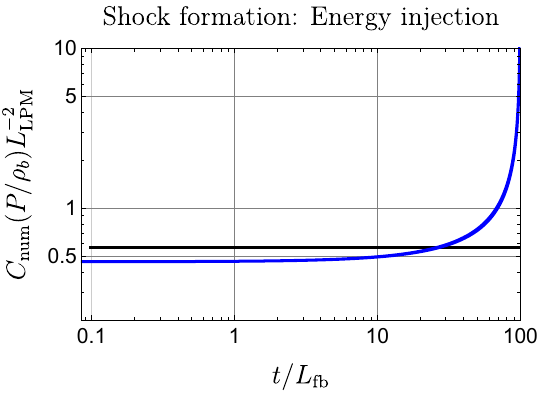}
    \includegraphics[width=0.5\linewidth]{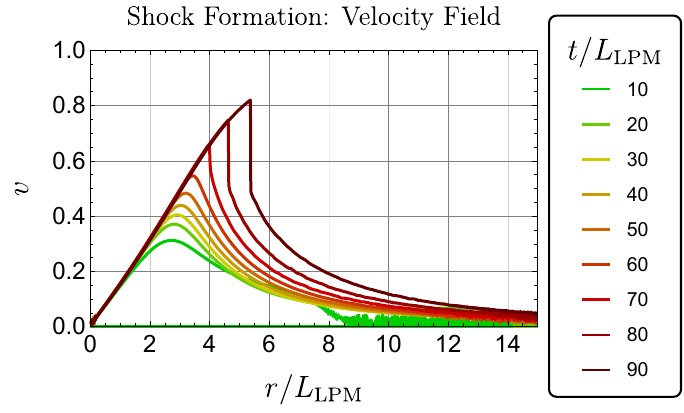}

    \caption{\justifying {\textbf{Shock formation for increasing power}} ({\it Left}) Normalized injection function as a function of time for an exploding PBH. The energy is injected homogeneously on a radius of $2L_{\rm LPM}$. The black line shows $c_s = 1 / \sqrt{3} \approx 0.57$, and we expect a transition between the steady state described in Sec. \ref{sec:hydro_motion} and the shock formation regime described in Sec. \ref{sec:inst}. ({\it Right}) Corresponding velocity curves for different times between $t/L_{\rm fb} \in [0, 100]$. Each curve is separated by $t/L_{\rm fb} = 10$. 
    }
    \label{fig:ShockForm}
\end{figure*}

Before closing this section, we present a numerical example. For an initial PBH mass $M_i\sim 10^6 $ g, with Hawking temperature $T_H \sim 10^7 \, {\rm GeV}$, Eq.~\eqref{eq:vslowinj} indicates that $v(r \sim L_{\rm LPM}) \sim 0.1$. This velocity is below the speed of sound, and hence the energy injection will not produce a shock. The P\'{e}clet number for this scenario is $\mathrm{Pe}_{\rm slow} \sim {\cal O} (10^2)$, which is deep in the advection regime. However, the same PBH energy injection rate at earlier times, with background temperature $T_{\rm b} \gg 10^3 \, {\rm GeV}$ and hence larger values of $\rho_{\rm b}$, would correspond to a P\'{e}clet number ${\rm Pe} < 1$ and would be dominated by diffusive transport. Furthermore, as indicated in Fig.~\ref{fig:PowerTimeTemp}, close to the time of the PBH explosion, the power injected by the PBH grows very rapidly by many orders of magnitude, which quickly drives the fluid to supersonic velocities $v(r\sim L_{\rm LPM}) \to 1$. In this regime, which will be discussed in detail in the following sections, an energy $E_{\rm inj} \sim M_{\rm thres} \sim \mathcal O(10^2 \, {\rm g})$ is injected instantaneously, and the P\'eclet number at $r\sim L_{\rm LPM}$, computed using Eq.~\eqref{eq:Pc_fb}, is $\mathcal O(10^4)$. Thus, the behavior of the fluid surrounding the PBH changes with time and passes through distinct phases as the universe expands and cools and the PBH Hawking emission ramps up: diffusive transport dominates at very early times, then advective transport with no shock formation, and then the rapid formation and propagation of a shock front. 

Lastly, we numerically study the formation of the shock. As noted in Eqs.~(\ref{eq:no_shock_formation})--(\ref{eq:shock_formation}), the ratio $(P / \rho_{\rm b}) L_{\rm LPM}^{-2}$ determines whether or not a shock forms in the plasma. We verify this claim in Fig.~\ref{fig:ShockForm} by simulating a case of PBH emission in which the increasing power crosses the two regimes. In the top panel, we show the power, where $C_{\rm num}(P/\rho_{\rm b}) L_{\rm LPM}^{-2}$ begins below the speed of sound and then evolves to the opposite regime. We observe indeed that in the first moments of the simulation, the system tends to the steady state with a $1/r^2$ velocity trend. However, at later times, when the power grows larger, the velocity field quickly evolves into a shock.

\begin{figure*}[t!]
    \centering
    \includegraphics[width=0.32\linewidth]{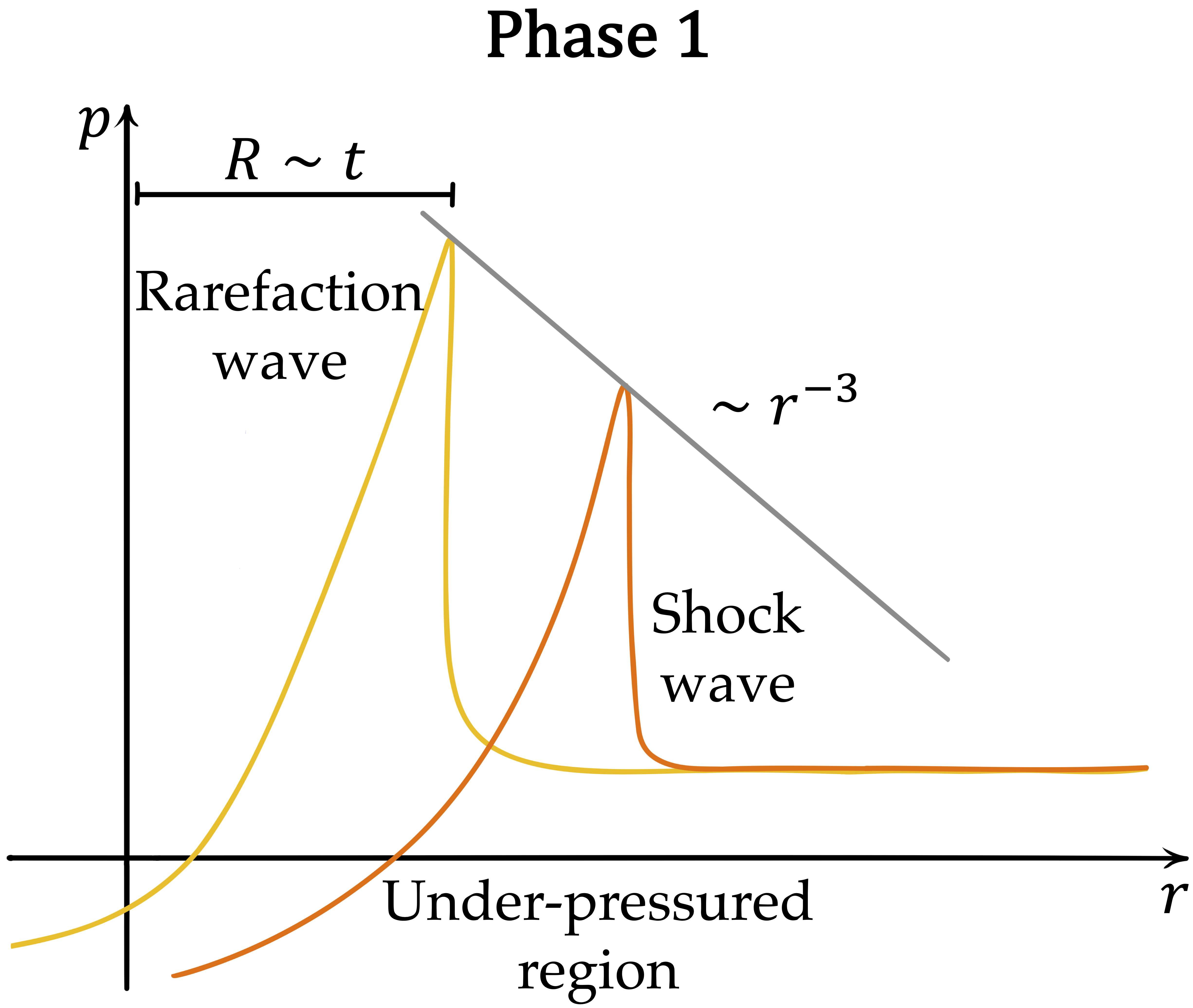}
    \hfill
    \includegraphics[width=0.32\linewidth]{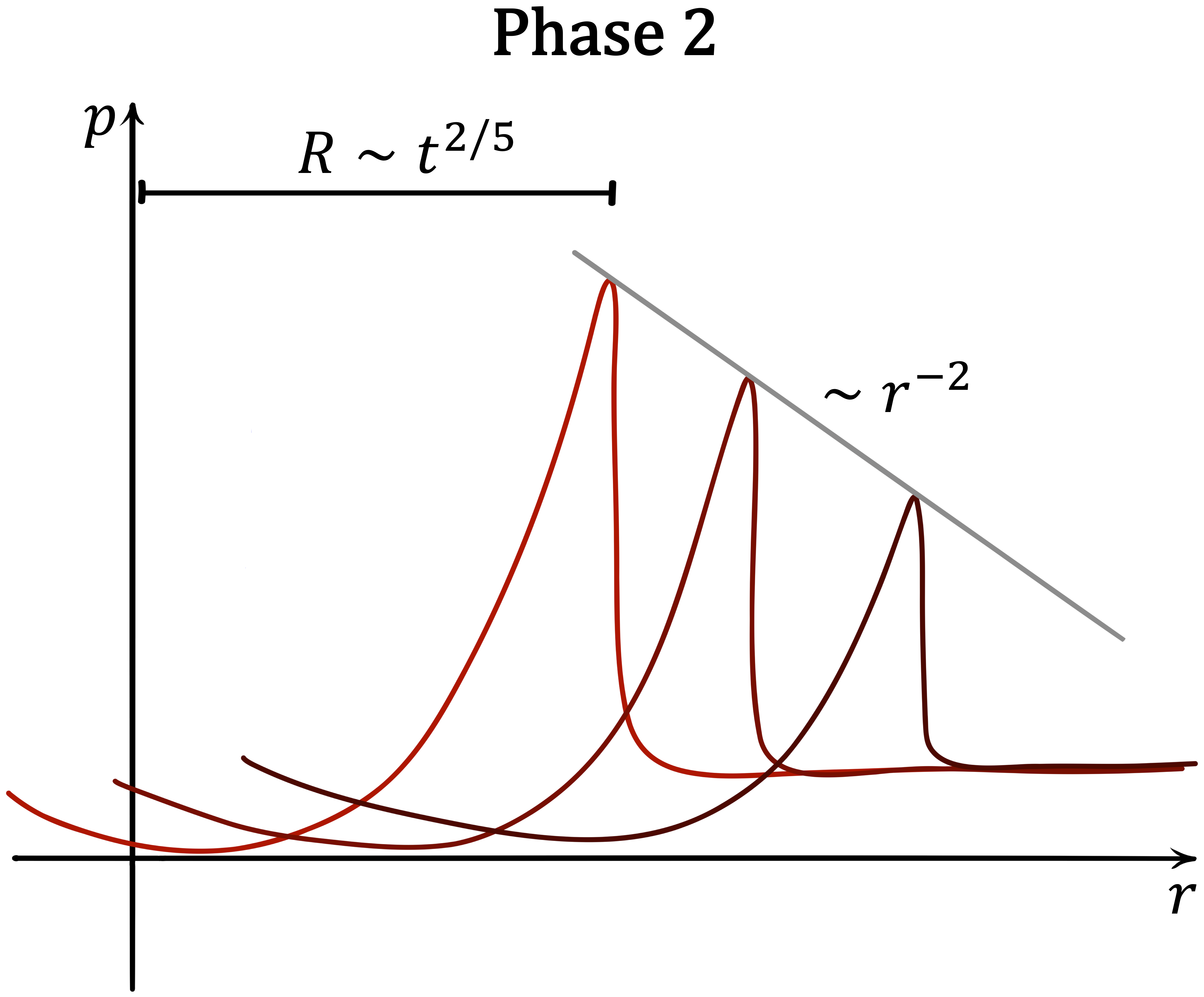}
    \hfill
    \includegraphics[width=0.32\linewidth]{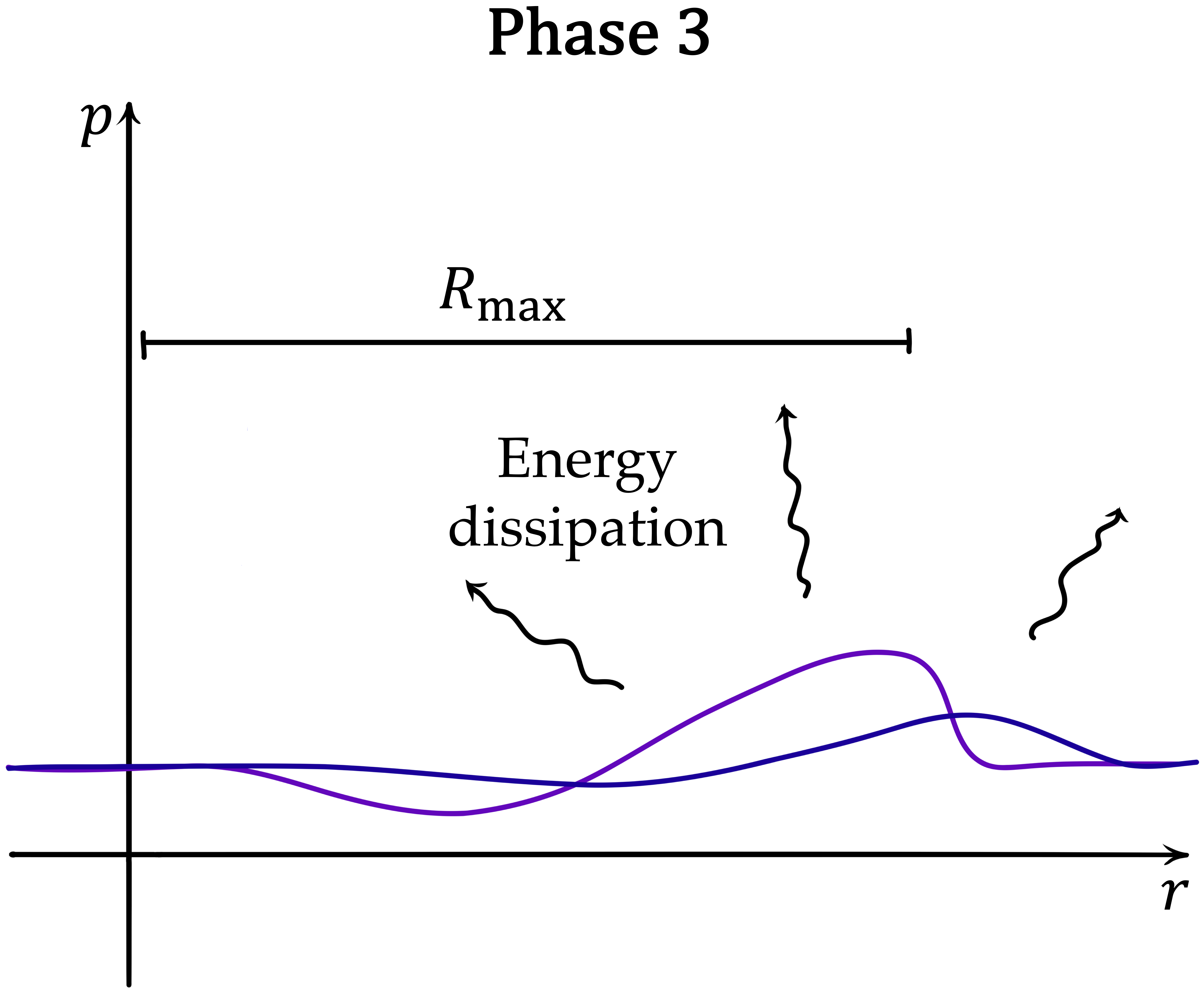}
    \caption{\justifying A sketch of the three phases of the evolution of the shock wave following instantaneous energy injection.  (\emph{Left}) Phase 1: McKee--Blandford regime, Friedlander waveform with $\Gamma\gg 1$. (\emph{Center}) Phase 2: Sedov-Taylor blast with $\Gamma\simeq 1$. (\emph{Right}) Phase 3: Wave is non-relativistic and dissipates energy beyond some maximum radius $R_{\rm max}$.
    }
    \label{fig:sketch_inst}
\end{figure*}

\section{Hydrodynamical evolution of the shock wave: instantaneous energy injection}
\label{sec:inst}

The case of instantaneous energy injection is conceptually simple and analytically tractable.  At very far distances from the PBH ($r\gg L_{\rm LPM}$), the background plasma is undisturbed at temperature $T_{\rm b}$. Traveling radially inward toward the PBH, the temperature profile\footnote{Notice that since the pressure is directly determined by the temperature, the latter can be obtained by the former.} is expected to take the form of a \emph{shock}, in which the pressure jumps discontinuously over a typical length scale set by the mean free path $\sim 1/T$ (as we will discuss later on) to a larger pressure value. Continuing radially inward toward the PBH, this shock is followed by a rarefaction wave, along which the temperature drops continuously and can even become smaller than the background.

\subsection{Overview of the instantaneous picture}

Before moving to the detailed description of our analysis, we provide a broad perspective and summarize our results. In the case of instantaneous energy injection, the fluid profiles around the PBH follow three distinct phases, as sketched in Fig.~\ref{fig:sketch_inst}. In Phase 1, the \emph{Blandford-McKee} (BM) regime~\cite{10.1063/1.861619}, an ultrarelativistic shock waveform expands following a \emph{Friedlander}~\cite{10.1098/rspa.1946.0046,2022arXiv220604106R} profile with $v\simeq 1$. In Phase 2, the front of the wave starts to slow down, and the velocity of the shock tends toward the sound speed in the plasma, entering the non-relativistic \emph{Sedov-Taylor} (ST) regime of blast waves~\cite{RezBook}. Finally, in the Phase 3, the dissipative effects inherent to the viscous plasma become dominant and the shock is smoothed and eventually washed out.

\subsection{Phase 1: Formation and propagation of the relativistic shock (Blandford-McKee regime)}
\label{sec:phaseA}

In this section we describe Phase 1 of the fluid profile (see Fig.~\ref{fig:sketch_inst}), which includes discussion of shock formation and the shape of the shock. Shock evolution is governed by the hydrodynamical equations detailed in the previous section.

\paragraph{Dependence of shock on initial fireball.}

Let us consider the limit $r\gg L_{\rm LPM}$, or, equivalently, the late-time evolution of the wave. It is known that the shock wave far away from the explosion only depends on the ratio $E_{\rm inj}/\rho_{\rm b}$, where $\rho_{\rm b}$ is the background energy density. In particular, the profile does not depend on the exact pressure profile of the fireball nor on the exact initial size $R_0$~\cite{10.1063/1.861619}. This picture is confirmed in Appendix~\ref{app:checks}, in which we study the shock evolution for different initial fireball sizes and shapes, and observe that the evolution of the shock wave becomes gradually independent of the initial size and exact shape of the small region in which energy has been injected. 

The fact that the profile of the pressure at late times depends only on one parameter $E_{\rm inj}/\rho_{\rm b}$ suggests a simple and analytical description of the temperature, fluid velocity, and pressure profiles. The pressure profile contains two qualitatively different regions: 1) the shock, where the hydrodynamical quantities jump in a discontinuous manner and which can be described by matching conditions, and 2) a hydrodynamical rarefaction wave which extends behind the shock and along which the pressure gradually relaxes.

\begin{figure*}[t!]
        \centering
        \includegraphics[width=0.43\linewidth]{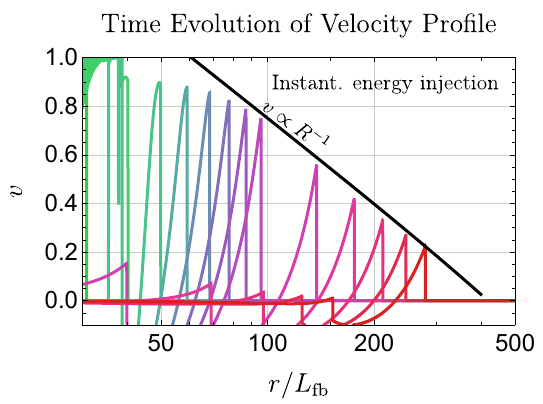}
        \includegraphics[width=0.56\linewidth]{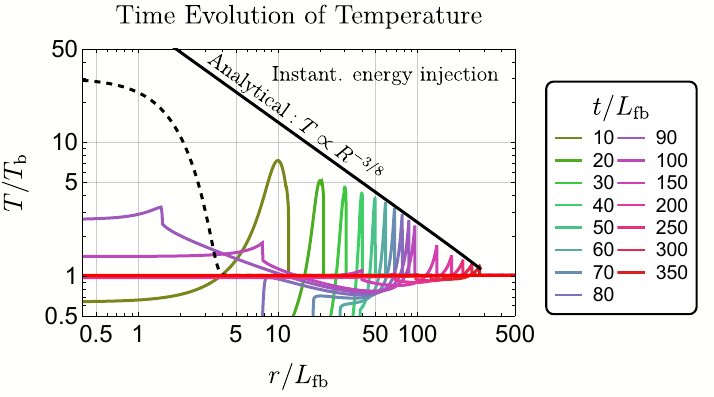}
    \caption{\justifying\label{fig:realistic_simu}  {\textbf{Instantaneous energy injection}}. Simulation of the outward spatial propagation of the wave, normalized by the LPM length $L_{\rm LPM}$ given by Eq.~(\ref{eq:LPM_length}).  Colored curves show temperature profiles at different times for the case of $T_{\rm fb} \sim 30 \, T_{\rm b}$, corresponding to $E_{\rm inj} \sim 7 \times 10^7$ (which corresponds physically to $E_{\rm inj} \sim 10^3 \text{ g}$), which is a representative case for our study. On this plot, the gray dashed line is the initial temperature profile initialized in the code. The scaling $v\propto R^{-1}$ is a numerical fit to the peaks of the velocity profiles.
    }
\end{figure*}

\paragraph{The shock evolution: analytical results.}
Let us begin by explaining the physics of the shock. The shock is a discontinuity in the hydrodynamical quantities which can be described by its position $R$, its velocity $\dot R$, or equivalently its boost factor $\Gamma \equiv \sqrt{1/(1-(\dot R)^2)}$, and by the value that the hydrodynamical variables take on each side of the discontinuous jump. Enforcing energy and momentum conservation  across the shock yields the \emph{matching conditions} that we present in full detail in Appendix~\ref{sec:relativistic_jump}. A key relation is the one between the pressure values on each side of the shock
\begin{equation}
p_{\rm sh}
=
\frac{8}{3}\,\tilde\gamma_{\rm b}^2 \, p_{\rm b}= \frac{8}{3}\,\Gamma^2 \, p_{\rm b}\, ,
\label{eq:p_sh_cont}
\end{equation}
where the tilde refers to quantities in the shock frame and quantities without tildes are in the plasma frame (rest frame of PBH). As indicated in Sec.~\ref{sec:notation} the subscript ``b'' for \emph{background} refers to the quantity in front of the shock (before it gets hit by the shock), while ``sh'' refers to the quantity behind the shock (after it gets hit by it the shock), and $\gamma = \sqrt{1/(1-v^2)}$ is the boost factor of the fluid, where $v$ is the fluid velocity. 

The matching conditions used above to obtain $p_{\rm sh}$ as a function of $v_{\rm sh}$ (the velocity of the fluid just behind the shock in the PBH frame) do not by themselves predict the behavior of the shock. To determine the evolution of the velocity of the shock,  one needs to rely on energy conservation arguments. The total energy density (in the PBH frame) is given by $T^{00}\approx 4{\gamma_{\rm}}^2 v_{\rm} p$. Thus, the energy of the wave instantaneously contained in the sphere of radius $R_{\rm }$ is given by\footnote{Notice that this integral includes contributions from regions away from the shock. They are however subdominant in the energy budget and most of the energy is concentrated in the shock wave.}
\begin{equation}
   E_{\rm inj} \approx E_{\rm}( R_{\rm }, t) = \int_{0}^{R_{\rm }} dr \, 16\pi\, r^2 \, \gamma^2_{\rm }(r,t) \, p_{\rm } (r,t) ,
\label{eq:EinjIntegral1}
\end{equation}
which is a quantity that must be approximately conserved in time. Using now the fact that the thickness of the shock region scales like $R/\Gamma^2$~\cite{10.1063/1.861619} and the fact that the integral in Eq.~(\ref{eq:EinjIntegral1}) is dominated by the region of the peak, we have 
\begin{equation}
\begin{split}
\label{eq:E_inj_ins}
    E_{\rm inj} &= \int^{R} dr \, 16\pi r^2\, \gamma^2_{\rm}(r,t) \, p_{\rm} (r,t) \\
    &\sim 16\pi\gamma_{\rm sh}^2 p_{\rm sh} (t) R_{\rm }^3 \, / \Gamma^2 \\
    &= 16\pi\gamma_{\rm sh}^2  R_{\rm }^3 \times \frac{8}{9}\rho_{\rm b} \, ,
    \end{split}
\end{equation}
where we have used $\gamma_{\rm sh}^2 =2 \Gamma^2$ and $p_{\rm sh} = 8\Gamma^2p_{\rm b}/3 $ (see Appendix~\ref{sec:relativistic_jump}). In this approximate integral, we have used the fact that the integral is dominated by the region around the shock, concentrated in a distance $R^2/\Gamma$. Moreover,  Eq.~\eqref{eq:E_inj_ins} and the fact that $E_{\rm inj}$ is a constant at late times implies the important relation between the boost factor and the radius of the shock:
\begin{equation}
    \Gamma^2 \propto p_{\rm sh} \propto R_{\rm }^{-3} \, ,
\end{equation}
while dimensional analysis dictates that 
\begin{equation}
\label{eq:gamm_est}
    \Gamma^2 \propto \frac{E_{\rm inj}}{p_{\rm b}} R_{\rm }^{-3} \, .
\end{equation}
To compute the exact prefactor in Eq.~\eqref{eq:gamm_est}, one needs to solve for the shape of the rarefaction wave. Ref.~\cite{10.1063/1.861619} demonstrates that the hydrodynamical equations in the presence of a shock and behind a shock wave take an approximately self-similar form, described by
\begin{align}
\label{eq:prof}
p_{\rm sh}(T, r, t) = \frac{8}{9}\rho_{\rm b} \, \Gamma(R)^2\,  \chi(r,t)^{-17/12}\,,
\end{align}  
where the approximate self-similar variable $\chi (r,t)$ takes the form
\begin{align}
\chi (r,t)= (1+8 \Gamma^2)(1-r/t) \, .
\end{align}

In Appendix~\ref{app:checks}, we compare these analytical profiles with the numerical simulations, finding that they differ by at most a factor of 20\% to 30\% close to the rarefaction wave. Plugging the profile of Eq.~\eqref{eq:prof} into Eq.~\eqref{eq:E_inj_ins}, we can perform the integration exactly and obtain
\begin{equation}
\label{eq:Gamma2}
\Gamma^2 \equiv \frac{1}{1-(\dot R_{\rm})^2} \approx  \frac{17E_{\rm inj}}{24\pi p_{\rm b} R_{\rm }^3} \, ,\qquad  p_{\rm sh} =  \frac{17E_{\rm inj}}{9\pi  R_{\rm }^3}\,.
\end{equation}    
These relations will play an important role in the analysis below. (See also Ref.~\cite{10.1063/1.861619} for further developments.) Finally, the position of the shock is  found to be 
\begin{equation}
    R_{\rm } = \int dt \, v(t) \approx \int dt \bigg(1- \frac{1}{2\Gamma^2(t)} \bigg) \approx t \bigg(1- \frac{1}{8\Gamma^2}\bigg) \, .
\end{equation}

\paragraph{The shock evolution: numerical results.}

In the simulations, we numerically solve the hydrodynamical equations presented in Eq.~\eqref{eq:hydro_eq} using a Kurganov-Tadmor scheme, which is meant to resolve and dynamically maintain shocks, \emph{i.e.}, hydrodynamical discontinuities. We provide the details of our numerical scheme in Appendix~\ref{app:scheme}. In simultaneous simulations, we initialize the fireball as a $\text{tanh}(z/w)$ profile, where $w$ is the initial size of the fireball of the order $2L_{\rm LPM}$ (as for example the dashed black curve in the right panel of Fig.~\ref{fig:realistic_simu}). The exact initial size however has no impact on the large-$r$ behavior. In Fig.~\ref{fig:realistic_simu}, we show on the left panel the evolution of the fluid velocity as a function of $r$, and on the right panel the evolution of the temperature $T$ as a function of $r$. In this specific case, we study the explosion of a PBH at $T = T_{\rm b}$, injecting enough energy to heat the fireball to $T_{\rm fb} \sim 30 \, T_{\rm b}$, as can be verified in the plot of $T_{\rm fb} (T_{\rm b})$ in Fig.~\ref{fig:Mthres}, with the curve $K = 1$ and $T_{\rm b} \sim 100$ GeV, for an explosion close to the EW transition. The profile of initial temperature that we inject corresponds to an energy of order $E_{\rm inj} \approx  10^8 \times T_{\rm b} (T_{\rm b} L_{\rm fb})^3$, which leads to the  initial boost factor of the shock $\Gamma_{0}\sim 10^3$.

As the swept-up mass increases, the shock progressively decelerates and becomes acoustic when $v_{\rm} \sim c_s = 1 / \sqrt{3}$; this marks the transition to Phase 2. We present a simple and illustrative numerical example. The temperature can be obtained from our simulations by rescaling the pressure via
\begin{equation}
T(R) \equiv \left( \frac{ 90 \, p (R) }{ g_\star\pi^2} \right)^{1/4}  .
\label{Tshdef}
\end{equation}

\begin{figure}[t!]
        \centering
\includegraphics[width=0.95\linewidth]{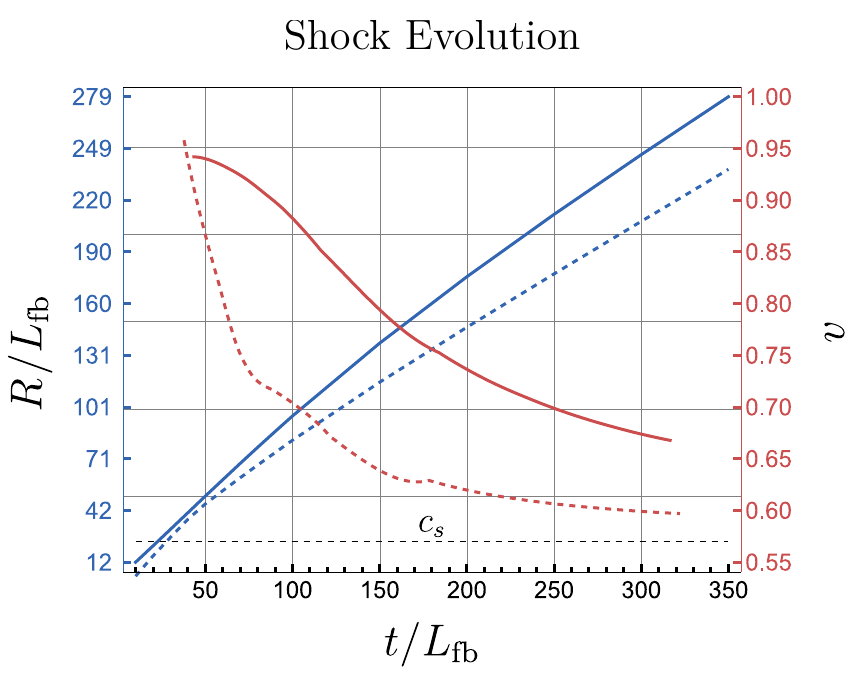}
        \caption{\justifying 
        \label{fig:velo_shock} {\textbf{Evolution of the shock front and rarefaction wave:}} Shock front (solid lines) and rarefaction wave (dashed lines) position ($R$, blue) and velocity ($v$, red) as a function of time. We observe that, whereas the shock is initially highly supersonic with $v_{\rm} \gg c_s$ and $R_{\rm }\sim t$, it relaxes after some time to sonic velocities $v_{\rm } \sim c_s$. The velocity of the rarefaction wave settles much more quickly to $v_{\rm rear}\sim c_s = 1 / \sqrt{3}$. }
     \end{figure}

Looking at the behavior of the velocity of the shock wave (solid red line) in Fig.~\ref{fig:velo_shock}, and using  
\begin{equation}
  R \, \Gamma^{2/3} \simeq R_0\,\Gamma_{0}^{2/3} \sim \mathcal{O}(200)\, L_{\text{fb}} \, ,
\end{equation}
we observe that the shock wave becomes sonic at a distance
\begin{equation}
    R_{\rm sonic} \simeq  \mathcal{O}(200)\, L_{\text{fb}}  \, . 
\end{equation}
We therefore conclude that in this simulation, the shock wave traveled a distance of order 200 times the initial size of the fireball before becoming sonic. This is consistent with the shock front (solid blue line in Fig.~\ref{fig:velo_shock}) following an ultrarelativistic trajectory with $R_{\rm } \sim t$ up to $t/L_{\text{fb}}\sim 100$ and then beginning to slow down. The rear of the rarefaction wave reaches near-acoustic propagation more quickly: $v_{\rm rear}\to c_s$ by $t/L_{\rm fb}\simeq200$ (dashed red line in Fig.~\ref{fig:velo_shock}). 

On the right panel of Fig.~\ref{fig:realistic_simu}, the solid black line shows the temperature immediately behind the shock, which is the maximal temperature, as a function of $r$. In this sense, it constitutes the envelope of the hydrodynamical profiles. This envelope is computed using Eq.~\eqref{eq:Gamma2}. We also verify in Fig.~\ref{fig:realistic_simu} (right panel, solid black line) that this analytical approximation is a very good fit for $T_{\rm sh}$ at rather large radius $R \gtrsim 50 \, L_{\rm fb}$. 
     
\paragraph{Pressure profile.} In the simulations, the pressure, temperature and velocity profiles display a distinctive behavior. The relativistic shock front rises sharply to a peak overpressure, followed by an exponential relaxation as the perturbation propagates through the expanding plasma. As the perturbation expands into the surrounding medium, the profile transitions into an underpressure, where the pressure drops below the local background value. This underpressure persists longer and remains weaker than the initial positive peak, reflecting the rapid rarefaction of the plasma, as is typical of a Friedlander-type waveform.

\paragraph{The determination of $R_{\rm max}$.}

From the above analysis we can determine the approximate distance $R_{\rm max}$ over which the shock can propagate before becoming weak and non-relativistic. Two distinct criteria can be used to define this approximate distance. We may define $R_{\rm max}$ as the radius at which the pressures become comparable, $p_{\rm b} \sim p_{\rm sh}$, or when $\Gamma \to 1$. These two criteria give essentially the same result. Using the former, we obtain
\begin{equation}
    R_{\rm max}\big|_{\rm inst} \sim \bigg(\frac{8}{3} \times \frac{ 17E_{\rm inj}}{ 9\pi  p_{\rm b}}\bigg)^{1/3} \, . 
    \label{eq:Rmax}
\end{equation}
On the other hand, if we use the latter criterion we have
\begin{equation}
    R_{\rm max}\big|_{\rm inst} \sim \bigg(\frac{ 17E_{\rm inj}}{24\pi  p_{\rm b}}\bigg)^{1/3} \, ,
    \label{eq:Rmax_2}
\end{equation}
which differs by a factor $1.6$. This distance determines the typical size of the region of plasma that will be impacted by the PBH explosion.  

Before closing this section, we address a potential caveat regarding shock-wave propagation. In several situations, shock waves can be unstable and may decay. We discuss shock stability in Appendix \ref{app:instabilities}, where we argue that the shock is indeed stable on the scale of $R_{\rm max}$.

\subsection{Phase 2: Non-relativistic shock wave (Sedov-Taylor regime)}

When the Lorentz factor of the shock becomes close to $\Gamma \sim \sqrt{3/2}$, corresponding to $\dot R \sim \sqrt{1/3}$, the shock becomes sonic and the scaling of the relevant quantities changes~\cite{RezBook}. In the simulation shown in Fig.~\ref{fig:realistic_simu}, this transition between an ultrarelativistic and a sonic shock occurs for 
\begin{equation}
    R \simeq  \mathcal{O}(200) \, L_{\text{fb}}  \, . 
\end{equation}
For non-relativistic velocities, but for $v > c_s$, the shock radius must satisfy~\cite{RezBook}
\begin{equation}
R_{\rm }(t) = C \left( \frac{E_{\rm inj} t^2}{\rho_{\rm b}} \right)^{1/5},
\end{equation}
where $ C$ is a dimensionless constant (typically $C \approx 1.15$~\cite{RezBook}). The shock velocity decreases in time as
\begin{equation}
\begin{split}
\dot R =  \frac{2}{5} \frac{R_{\rm }}{t} &= \frac{2}{5} C \left( \frac{E_{\rm inj}}{\rho_{\rm b} t^3} \right)^{1/5} \\
&= \frac{2}{5}C^{-1/2} \bigg(\frac{E_{\rm inj}}{\rho_{\rm b} R^3}\bigg)^{1/2}.
\end{split}
\end{equation}

\begin{figure}[t!]
        \centering
\includegraphics[width=1\linewidth]{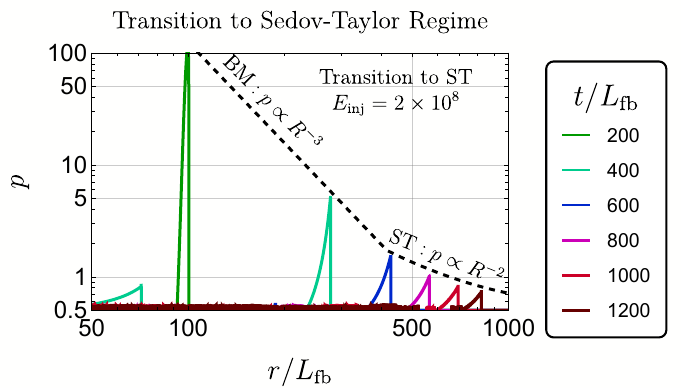}
        \caption{\justifying 
        \label{fig:transition} {\textbf{Transition from the MB to the ST regime}}: Simulation of the transition from the Blandford-McKee regime (BM) with $p \propto R^{-3}$ to the Sedov-Taylor (ST) regime with $p \propto R^{-3}$. We used $p_b = 0.5$ and $E_{\rm inj} = 2 \times 10^8$. 
        }
     \end{figure}

Returning to Fig.~\ref{fig:velo_shock}, this behavior of the velocity is approached briefly before the velocity tends toward $v \to c_s$, where it has to become constant (for the shock to remain stable). Since the tail of the rarefaction wave propagates also at the speed of sound, the thickness of the wave tends to a constant. This implies that the main source for the weakening of the wave is the volume dilution.
Using energy conservation arguments similar to those above, one concludes that in this regime, the pressure downstream will evolve as
\begin{equation}
\label{eq:STrscaling}
    p_{\rm sh} \propto R^{-2} \, .
\end{equation}
During and after this Sedov-Taylor phase, the shock becomes weak and is more similar to an ordinary wave, which is a small perturbation propagating in a fluid. The transition from the MB regime to the ST regime is displayed in \ref{fig:transition}, confirming the scaling in Eq.\eqref{eq:STrscaling}. 

\subsection{Phase 3: Dissipation of the wave}

The simulations that we have performed are pure hydrodynamical simulations and do not include viscosity. For the early universe QGP, we expect a shear viscosity $\eta \sim 0.1 s$, where $s$ is the comoving entropy density. The shock wave will eventually dissipate its energy and weaken due to viscosity and entropy production: the sound is absorbed by the plasma, where it is dissipated as heat. In this section, we estimate the dissipation timescale of the shock wave once it becomes sonic and enters the weak-shock regime.

Following Ref.~\cite{LandauFluid} (see also Refs.~\cite{Fogaca:2013cma, Zhu_Wu_Zhou_Yang_Xu_2024}), the rate of heat production, which corresponds to the energy loss of the shock, is given by
\begin{equation} \label{eq:energy_diss}
\begin{split} 
     |\dot E| &\approx \bigg(\frac{4\eta}{3} +\zeta\bigg) \int dr \, 4\pi \,r^2  \bigg(\frac{dv}{dr}\bigg)^2 \\
     &\sim 4\pi \bigg(\frac{4\eta}{3} +\zeta\bigg) R_{\rm }^2 \delta \frac{v_{\rm }^2}{\delta^2} \sim  \bigg(\frac{4\eta}{3} +\zeta\bigg)\frac{4\pi R_{\rm }^2 }{\delta} \, ,
     \end{split}
\end{equation}
where $\delta$ is the microscopic thickness of the shock (which is studied in Appendix~\ref{sec:thickness_shock}), $\eta$ is the shear viscosity, and $\zeta\sim \eta$ is the bulk viscosity. The microscopic thickness of the shock is the distance across which the pressure (or the temperature) jumps from $p_{\rm b}$ to $p_{\rm sh}$.

The damping of the wave is given by an exponential $e^{- \gamma_{\rm damp} r}$, where 
\begin{equation}
    \gamma_{\rm damp} \sim \frac{ |\dot E|}{2E_{\rm inj} } \, .
\end{equation}
Meanwhile, the mean free path of particles in the plasma may be approximated by
%%%%
\begin{equation}
    l_{\rm mfp} \sim \frac{1}{\rho_{\rm b}} \left( \frac{ 4 \eta}{3} + \zeta \right) .
    \label{eq:lmfp}
\end{equation}
Considering that the thickness of the shock $\delta$ is approximately equal to $l_{\rm mfp}$, as discussed in Appendix~\ref{sec:thickness_shock}, we find 
\begin{equation}
    \gamma_{\rm damp}  \sim  \frac{4\pi R^2\rho_{\rm b} }{ E_{\rm inj}}  \sim  \frac{4\pi R^2 T_b^4 }{ E_{\rm inj}} \, .
\end{equation}
This implies that viscous dissipation becomes important and the shock will dissipate on a time scale of order $1/\gamma_{\rm damp}$ as wave energy is converted to heat, accompanied by entropy production, or given by 
\begin{equation} \label{eq:Ldiss}
    L_{\rm diss} \sim \bigg(\frac{E_{\rm inj}}{T_b^4}\bigg)^{1/3} \, . 
\end{equation}

For the case of a PBH of mass $M_{\rm PBH} \sim 10^6$ g exploding around the EW transition, $L_{\rm diss} \sim 10^7 \text{GeV}^{-1} \sim 10^3 L_{\rm LPM} \gg R_{\rm max}$.

\section{PBH-like injection}
\label{sec:PBH-like_inj}

In the previous section, we introduced a simplified picture of the evolution the shock wave generated by a PBH explosion, which we modeled as instantaneous injection of energy $E_{\rm inj}=M_{\rm thres}(T_{\rm b}) \gg \rho_{\rm b} \, L_{\rm LPM}^3$ into a plasma with  temperature $T_{\rm b}$. However, the energy released by the PBH \emph{prior} to its final instantaneous explosion, which we referred to in Sec.~\ref{sec:long_before} as the `slow injection' regime, would also contribute to establishing a temperature profile around the PBH. In this section, we present a more realistic scenario which we call the \emph{PBH-like} injection picture. We perform numerical simulations and present analytical descriptions of an \emph{extended} period of energy injection by the PBH, which includes both the final explosion and some portion of the `slow' injection phase leading up to it. 

The basic picture is as follows. The PBH injects a total amount of energy $E_{\rm tot, inj}$ over a timescale $t_{\rm inj}\gg L_{\rm LPM}$, which is no longer instantaneous on hydrodynamic timescales. This creates a dynamically evolving and expanding \emph{fireball}, which we may compare with the simplified fireball picture from the instantaneous case in Sec.~\ref{sec:inst}, which was initialized with some size and immediately evolved into a shock with no period of growth. In the PBH-like injection case, after time $t_{\rm inj}$ has elapsed and the PBH has fully evaporated, the fireball then detaches from the origin and evolves into an outwardly propagating shock front with a trailing rarefaction wave---in a similar progression to the instantaneous case. We find that shocks can form at early times, well before the final instantaneous explosion. Another main result in this section is to derive the maximum radius $R_{\rm max}$, which we will show is larger than the maximum radius for the instantaneous case. 

We note that throughout this section, the fireball is now dynamically evolving in time unlike in the instantaneous case, where it was effectively initialized instantaneously with radius $L_{\rm fb}=L_{\rm LPM}$ from Eq.~\eqref{eq:Lfb} and temperature $T_{\rm fb}^{\rm inst}=T_{\rm fb}(K=1)$ from Eq.~\eqref{eq:T_hs_init}. The fireballs discussed in the context of PBH-like injection expand to larger values than $L_{\rm LPM}$. Note that throughout this section, $L_{\rm fb}$ refers to the definition in Eq.~\eqref{eq:Lfb} which equates $L_{\rm fb}=L_{\rm LPM}$, \emph{not} to the size of the dynamically evolving fireball for the PBH-like case. 

\subsection{Description of the injection function and simulation details}

The power emitted by a PBH leading up to its explosion is discussed in Sec.~\ref{sec:Hawking_rad} and explicitly given by Eq.~\eqref{eq:TotalInjectionTime}. To simulate the PBH-like injection scenario, which involves injecting energy over a time scale $t_{\rm inj}\gg L_{\rm LPM}$, we use the following parametrization of the injected power:
\begin{equation}
\label{eq:PBH_inject}
P (t) = \frac{D}{[1 - (t / \tilde{\tau} ) ]^{2/3}} \\
 \Theta \left(t-\tilde{\tau}(1-nC)\right) \Theta \left(-t+\tilde{\tau}\right),
\end{equation}
where we fix $C=10^{-5}$ and $n\leq10^4$ is a tunable parameter. 
 
The Heaviside theta functions in Eq.~\eqref{eq:PBH_inject} fix the energy-injection time interval
\begin{equation}
\label{eq:deltat}
t_{\rm inj} = \tilde{\tau} - \tilde{\tau}(1-n C) = nC\tilde{\tau}.
\end{equation}
In our simulations, because of limited numerical resolution, we set the maximum injection time to $t_{\rm inj}= 100 \, L_{\rm LPM}$. Thus, taking the largest value of $n=10^4$, we fix the parameter 
\begin{equation}
    \label{eq:tautilde}
    \tilde{\tau} = 10^3 L_{\rm LPM}=10^3 \tau(M_{\rm thres})
\end{equation}
via Eq.~\eqref{eq:deltat}, where $\tau (M_{\rm thres})$ is the remaining lifetime of a PBH with mass $M_{\rm thres}$. Note that henceforth $\tilde{\tau}$ and $C$ are fixed, and tuning $n<10^4$ allows us to simulate shorter injection windows.\footnote{Note that the emitted power has a singularity for $t=\tilde{\tau}$, while the total deposited energy found by integrating the power is always finite. Typically, the singularity is avoided by truncating the PBH's evolution when it reaches the Planck scale at $M\simeq M_{\rm pl} = 4.33 \times 10^{-6}\, {\rm g}$. (See Fig.~\ref{fig:PowerTimeTemp}.) In the case of our simulations, the singularity is avoided by discretization.}

The total amount of energy injected over an interval $t_{\rm inj}$ is:
\begin{equation}
   E_{\rm tot, inj} 
   = \int_0^{\infty} dt P(t) = 3 D (nC)^{1/3}\, \tilde{\tau}.
   \label{Einjtotalmax1}
\end{equation}
The constant $D$ is therefore fixed by
\begin{equation}
    D=\frac{E_{\rm tot, inj}}{3\tilde{\tau}(nC)^{1/3}}.
\end{equation}
The approach to simulating extended PBH-like energy injection involves first numerically determining the value of $D$ that corresponds to a desired $E_{\rm tot, inj}$ given a fixed value of $t_{\rm inj}$, and then simulating the injection according to Eq.~\eqref{eq:PBH_inject}.

\begin{figure}[t!]
        \centering
        \includegraphics[width=0.9\linewidth]{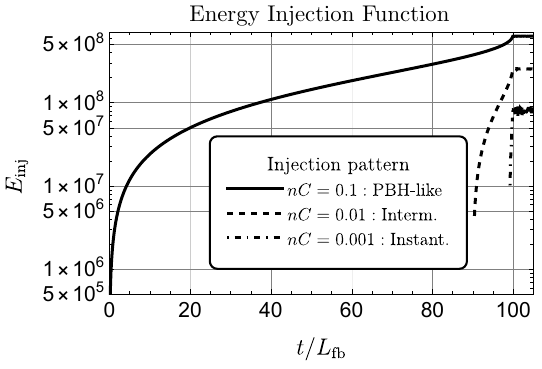}
\includegraphics[width=0.9\linewidth]{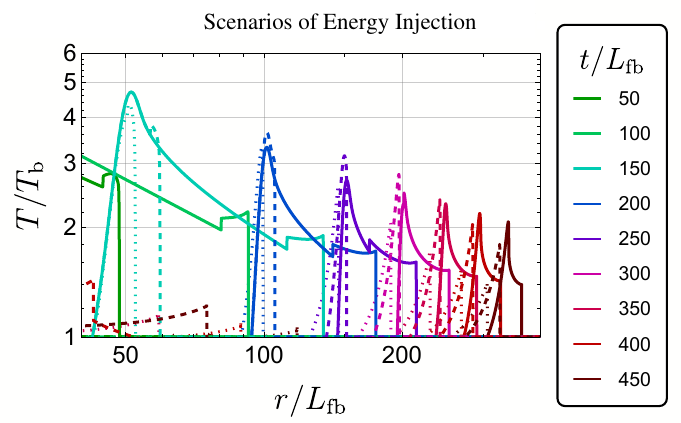}
        \caption{\justifying 
        \label{fig:PBH_inject_2} 
        {\textbf{Comparison of different injection time intervals for the \emph{same} PBH explosion.}} ({\it Top}) Cumulative energy injected as a function of time in the case of PBH-like injection for three values of $t_{\rm inj}$: 1) $t_{\rm inj}=100L_{\rm LPM}$, $nC = 0.1$ (solid), 2) $t_{\rm inj}=10L_{\rm LPM}$, $nC = 0.01$ (dashed), and 3) $t_{\rm inj}=1L_{\rm LPM}$, $nC = 0.001$ (dot-dashed). Integrating the injected power for each case, one obtains:  $E_{\rm inj}\sim 6 \times 10^8$ (solid), $2.6 \times 10^8$ (dashed), and $7 \times 10^7$ (dot-dashed) over $T_{\rm b} = 0.46$.  While $nC = 0.001$ retains only the energy injected at the very final stage and mimics the instantaneous case, $nC = 0.1$ includes a larger, extended window of PBH emission. The case $nC = 0.01$ is intermediate. ({\it Bottom}) Temperature profiles for the same three cases described in the upper panel. Note the Friedlander waveform shape of the instantaneous $t_{\rm inj}=1\,L_{\rm LPM}$ case and the development of a shell in the $t_{\rm inj}=100\,L_{\rm LPM}$ extended case. 
        }
\end{figure}

In our simulations, one of the goals is to compare the \emph{instantaneous} injection with a extended \emph{PBH-like} injection. Varying $n$ allows us to capture two regimes: 1) instantaneous injection if $t_{\rm inj}=nC \tilde{\tau} = L_{\rm LPM}$, which corresponds to $n=10^2$ and 2) extended injection with $n > 10^2$. Recall that energy injection by a PBH with mass $M_{\rm thres}$ and remaining lifetime $\tau\simeq L_{\rm LPM}$ is instantaneous on hydrodynamic timescales, and the total amount of energy deposited is given by $M_{\rm thres}$ in Eq.~\eqref{eq:Mthres}. The upper panel of Fig.~\ref{fig:PBH_inject_2} shows energy injection profiles for three values of $t_{\rm inj}$, corresponding to PBH-like ($nC = 0.1$), intermediate ($nC = 0.01$), and instantaneous ($nC = 0.001$). The ratio of injected energy in the instantaneous case to the PBH-like case shown in Fig.~\ref{fig:PBH_inject_2} is $(7\times10^7)/(6\times10^8)=0.12$. We would expect analytically to find a ratio of 
\begin{equation}
    \frac{M_{\rm thres}}{M(\tau=100L_{\rm LPM})}  = \left(\frac{1}{100}\right)^{1/3}=0.22.
\end{equation}
We therefore find that the energy injected for the PBH-like case in Fig.~\ref{fig:PBH_inject_2} is consistent with our analytical expectation to within a factor of $\lesssim2$.  

In the lower panel of Fig.~\ref{fig:PBH_inject_2} we compare the instantaneous and extended PBH-like injection intervals in the context of the same PBH explosion. The dash-dotted lines in these simulations correspond to the case of instantaneous energy injection, since the energy is injected over the timescale of $L_{\rm LPM}$. In contrast, dashed and solid lines show examples of more extended PBH-like injection with $nC=0.01$ and $nC=0.1$, which correspond to $t_{\rm inj}=10L_{\rm LPM}$ and $t_{\rm inj}=100L_{\rm LPM}$, respectively. We observe two main differences: 1) In the case of extended injection, the shock develops a finite shell thickness, of the order of the length $10^{-1} \, t_{\rm inj}$, even though (as we will see in Fig.~\ref{fig:shell}) this thickness is itself dynamical. 2) The shock in the extended case propagates further before undergoing viscous damping because simulations of the extended injection scenario account for more injected energy than the instantaneous-injection approximation. 

In the simulations, we have used the following normalization: the region in which the energy is injected is fixed to $R/L_{\rm fb} = 2$ while the background temperature is given by $T_{\rm b} = 0.46$. We report the dimensionless value of the energy that we injected in the caption of the figures.

As noted above, our simulations of PBH-like injection are restricted to $t_{\rm inj}\leq 100 \, L_{\rm LPM}$ by our numerical resolution, which fixes $\tilde{\tau}=10^3L_{\rm LPM}$ in Eq.~\eqref{eq:PBH_inject}. This is well within the regime to form a shock immediately. For larger initial values of $t_{\rm inj}=\tau$, however, we can generically study the full duration of the fireball growth. For sufficiently large $t_{\rm inj}$, the PBH emission will start off in the regime of Sec.~\ref{sec:energy_inj}, in which slow emission prohibits shock formation and temperature is essentially steady, as expected from Eq. \eqref{eq:no_shock_formation}.

In that case, as soon as the emitted power is large enough to enter the regime of Eq. \eqref{eq:shock_formation}, the fireball will begin to expand and a shock discontinuity will form. In Appendix~\ref{eq:fireball_growth}, we study the transition between these two regimes, the ``no shock'' and the ``shock'' formation. We find that the criterion for the shock formation in Eq.~\eqref{eq:shock_formation} is very well satisfied.

\begin{figure*}[t!]
    \centering
    \includegraphics[width=0.24\linewidth]{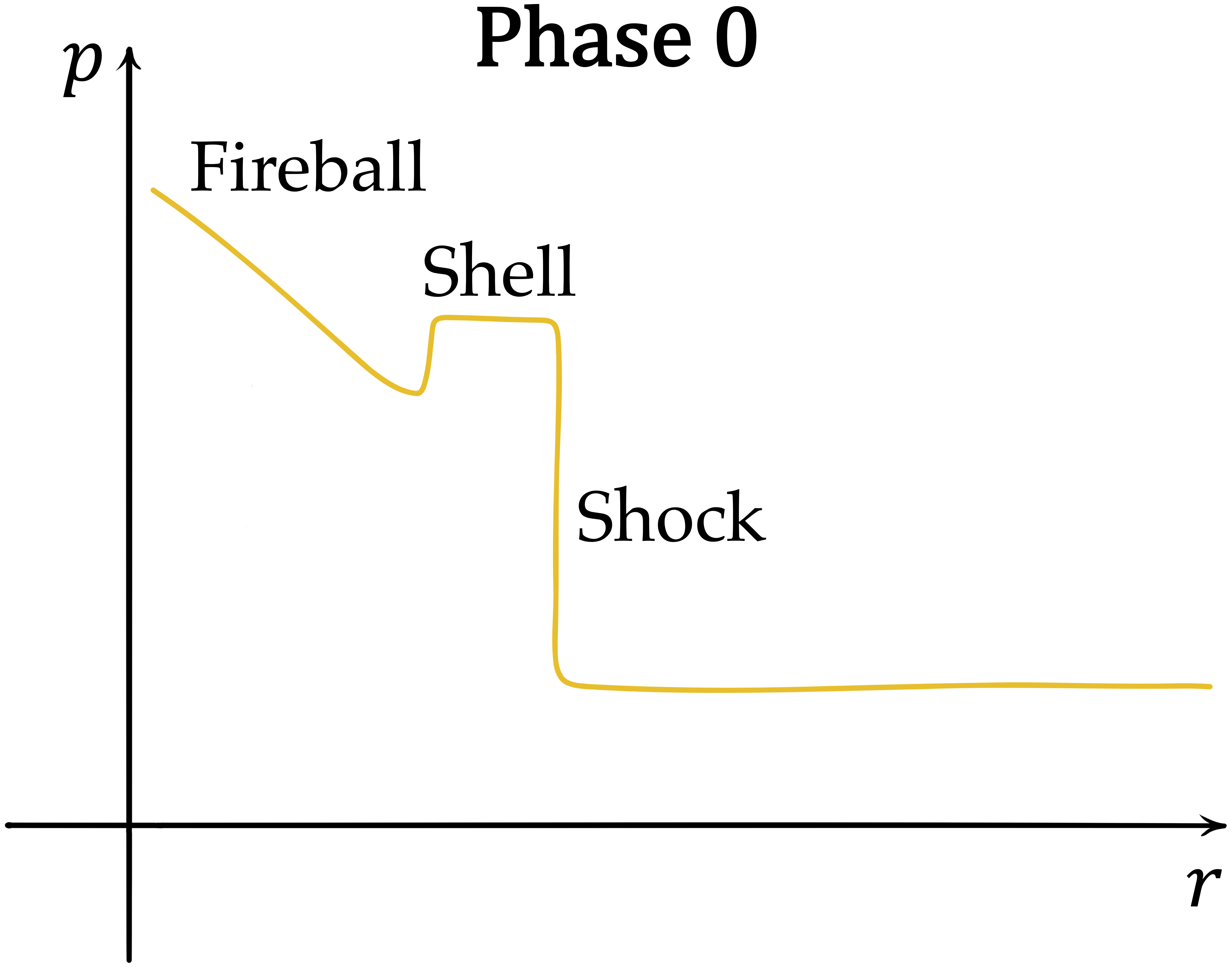}
    \includegraphics[width=0.24\linewidth]{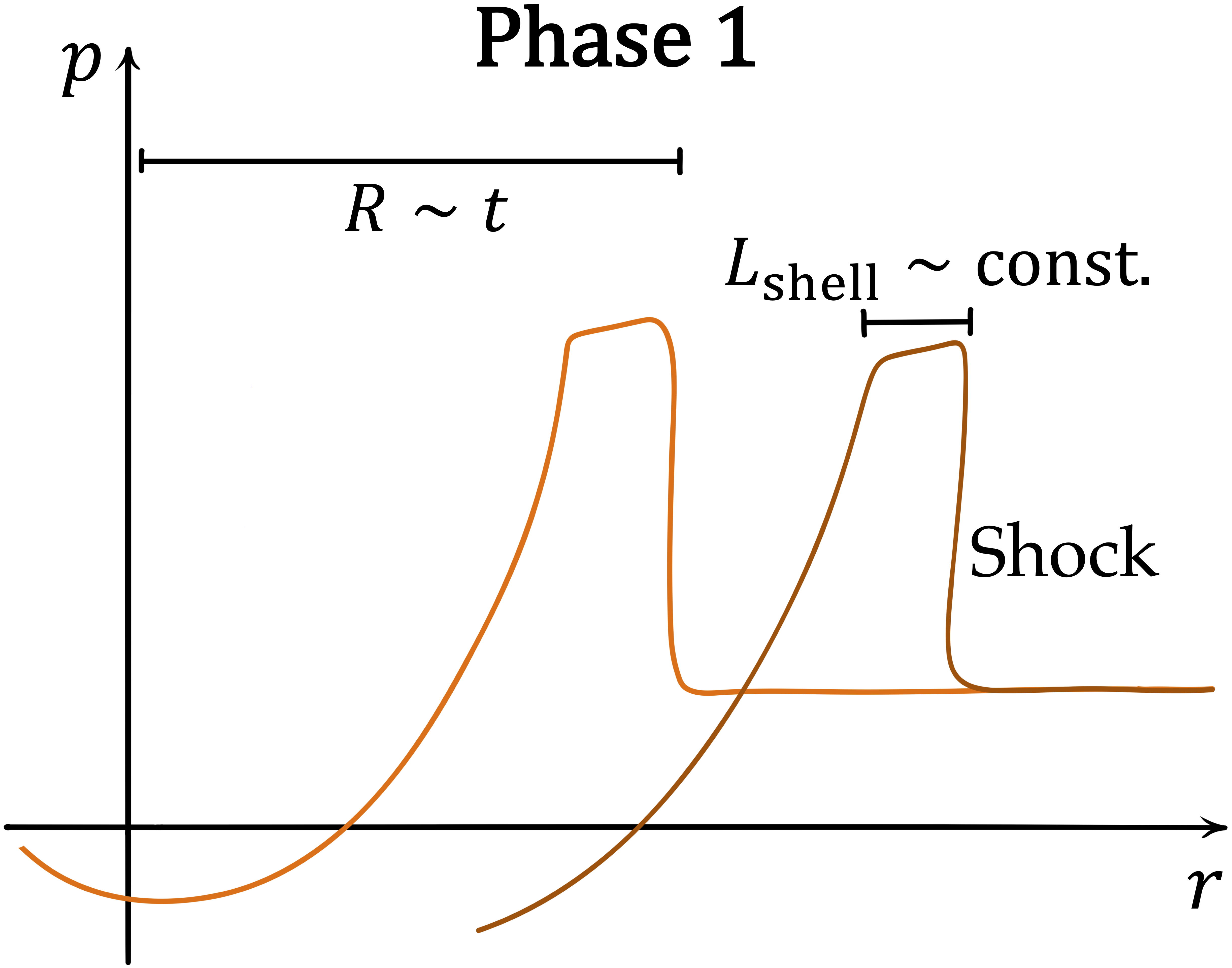}
    \includegraphics[width=0.24\linewidth]{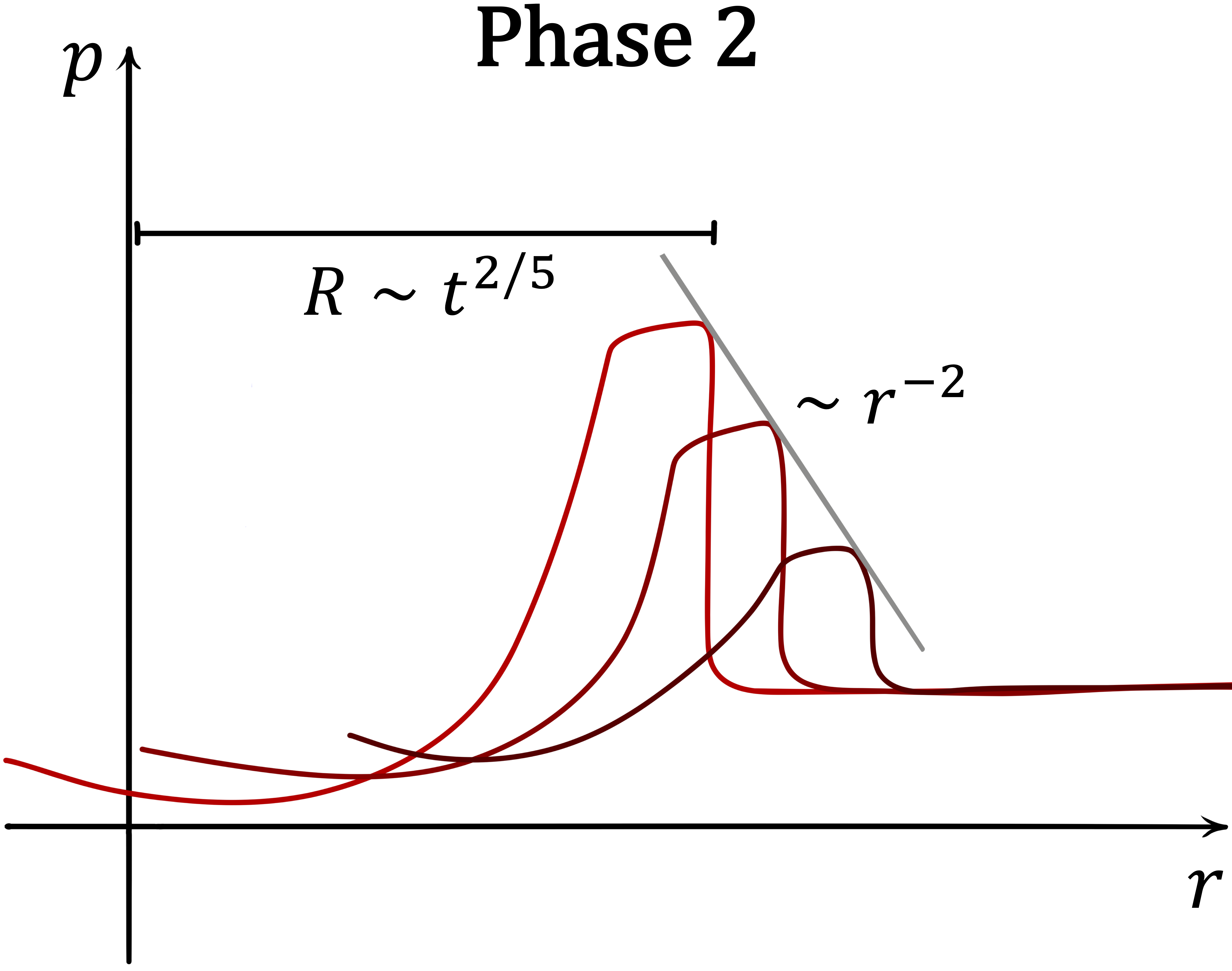}
    \includegraphics[width=0.24\linewidth]{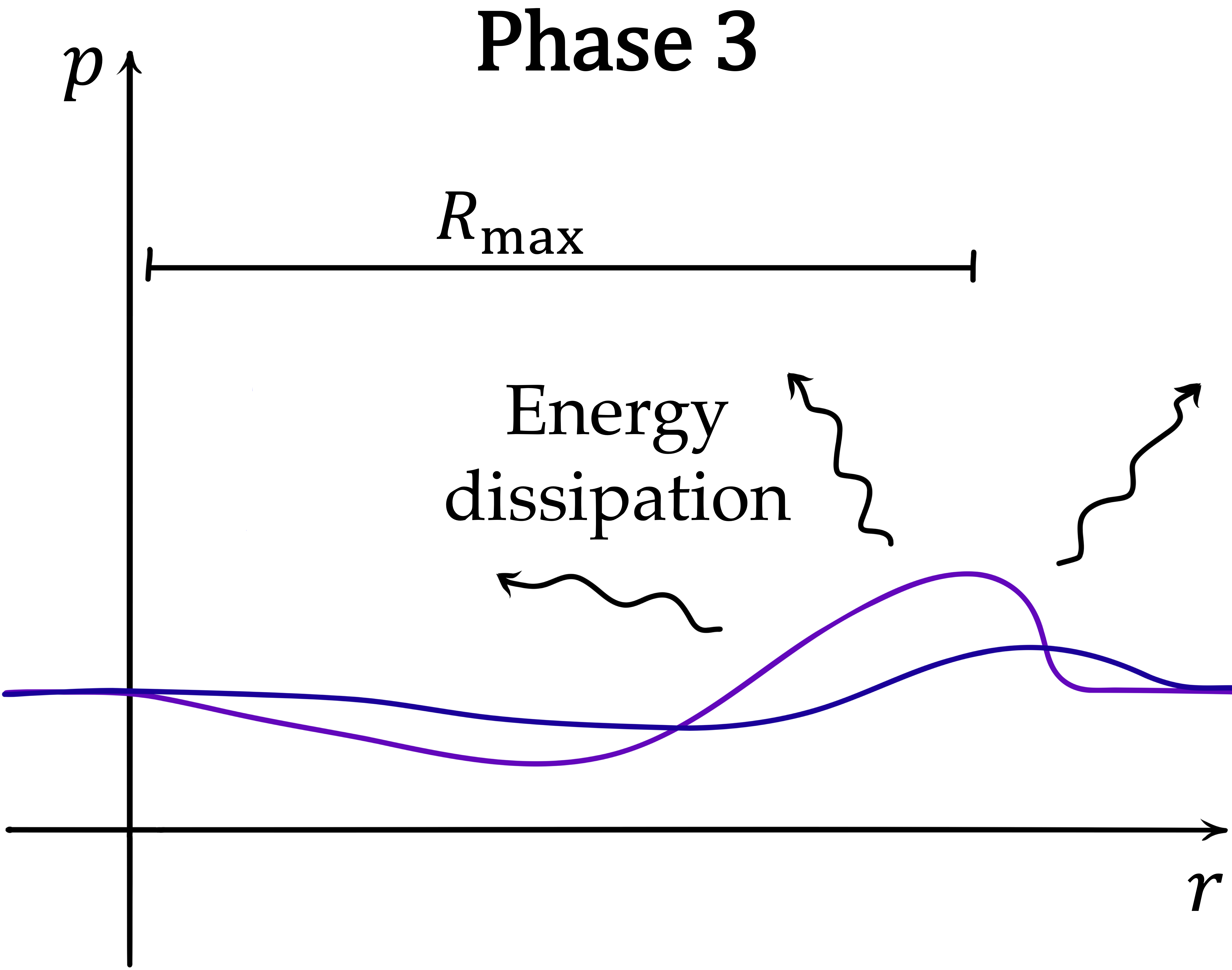}
    \caption{\justifying  A sketch of the four phases of shock wave evolution following PBH-like energy injection. (\emph{Top left}) Phase 0: PBH Hawking emission and expansion of the fireball. (\emph{Top right}) Phase 1: Ultra-relativistic shock wave propagation with $\Gamma\gg1$. (\emph{Bottom left}) Phase 2: Sedov-Taylor blast with $\Gamma\simeq 1$. (\emph{Bottom right}) Phase 3: Damping of the wave by viscous dissipation for $R>R_{\rm max}$. Note the formation of a \emph{shell} of finite width behind the shock, which differs from the instantaneous case.}
    \label{fig:sketch_PBH}
\end{figure*}

\subsection{Overview}

The main difference between the PBH-like injection scenario and the instantaneous one lies in the resulting pressure and temperature profiles. In the instantaneous case, the shock is directly followed by a rarefaction wave, creating a Friedlander waveform. However, in the case of PBH-like injection, the shock is instead followed by a \emph{shell} of thickness $L_{\rm shell} < R_{\rm max}$, where the pressure is roughly constant. The shell is then %finally 
terminated by a rarefaction wave. This shell can be directly observed in the simulations by looking, for example, at Fig.~\ref{fig:realistic_simu_PBH}.

The PBH-like injection scenario can be divided into four distinct phases, as illustrated in Fig.~\ref{fig:sketch_PBH} and described below: \\
\begin{itemize}
\item {\bf \emph{Phase 0}} corresponds to the period of active energy injection, during which the fireball expands and drives a relativistic shock whose structure is determined by the Rankine--Hugoniot jump conditions and by the way injected power is partitioned between internal and kinetic energy. 
\item {\bf \emph{Phase 1}} describes the subsequent evolution once the PBH has evaporated; the shock propagates as a thin shell where energy is (approximately) conserved.
\item {\bf \emph{Phase 2}} describes the transition as the shock wave decelerates to mildly relativistic speeds, during which it enters the non-relativistic Sedov-Taylor regime.
 
\item {\bf \emph{Phase 3}} occurs at sufficiently late times, as diffusive processes broaden and dissipate the wave.

\end{itemize}

\subsection{Phase 0: A growing fireball}
\label{sec:PBH_phase0}

Phase 0 is specific to the PBH-like injection and did not appear in the instantaneous discussion. Indeed, the heating of the plasma starts long before the PBH fully disappears. The energy is pumped into the plasma at an increasing rate and creates a hot bubble around the PBH that expands at nearly the speed of light.

\begin{figure}[t!]
        \centering
       \includegraphics[width=1.\linewidth]{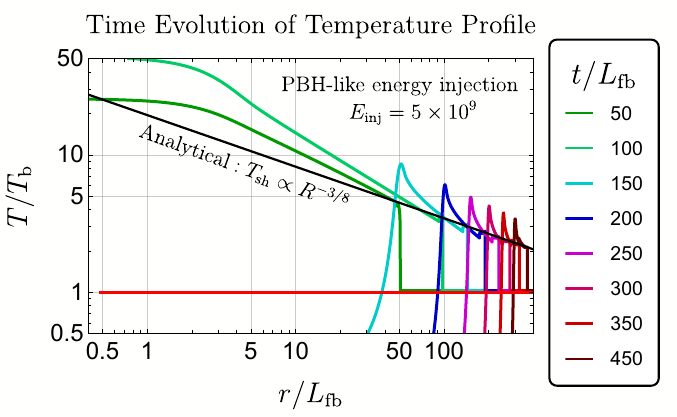}
       \includegraphics[width=1.\linewidth]{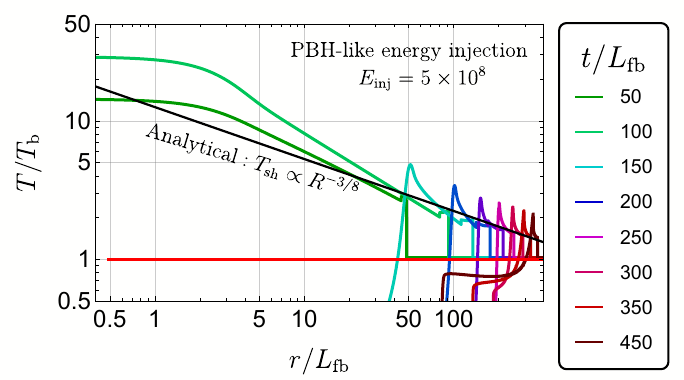}
    \caption{\justifying\label{fig:realistic_simu_PBH}  {\textbf{PBH-like energy injection}}. ({\it Top}) Simulation of the outward spatial propagation of the wave, normalized by the LPM length $L_{\rm fb}$. The final explosion of the PBH is set at $t = 100 L_{\rm fb}$ The total injected energy is given by $E_{\rm inj} = 5 \times 10^{9}$ with $T_{\rm b} = 0.46$ (or $p_{\rm b} =0.5$), leading to $E_{\rm inj}/(L_{\rm fb}^3p_{\rm b}) = 10^{10}$. The black line is a fit of the envelope of the temperature at the shock, which follows a law $T_{\rm sh}(R) \propto R^{-3/8}$ (corresponding to $p_{\rm sh}(R) \propto R^{-3/2}$) and which is justified in Eq.~\eqref{eq:PBH_inj_law} if $L_{\rm shell} \propto R$.
   ({\it Bottom}) Same simulation for $E_{\rm inj} = 5 \times 10^8$ over a $T_{\rm b}= 0.46$ (or $p_{\rm b} =0.5$), leading to $E_{\rm inj}/(L_{\rm fb}^3p_{\rm b}) = 10^{9}$. 
    }
\end{figure}

One example of such a situation can be observed in Fig.~\ref{fig:realistic_simu_PBH}, in which the dark green curve at $t/L_{\rm fb}= 50$ shows such a \emph{fireball} profile, which is also depicted schematically in the upper left panel of Fig.~\ref{fig:sketch_PBH}. By $t/L_{\rm fb}= 100$ in Fig.~\ref{fig:realistic_simu_PBH}, however, the PBH has already disappeared, and one sees afterward that the temperature behind the shell has become colder (see $t/L_{\rm fb}=150$ curve), suggesting a Friedlander-like underpressure.

\begin{figure}[t!]
        \centering
       \includegraphics[width=0.9\linewidth]{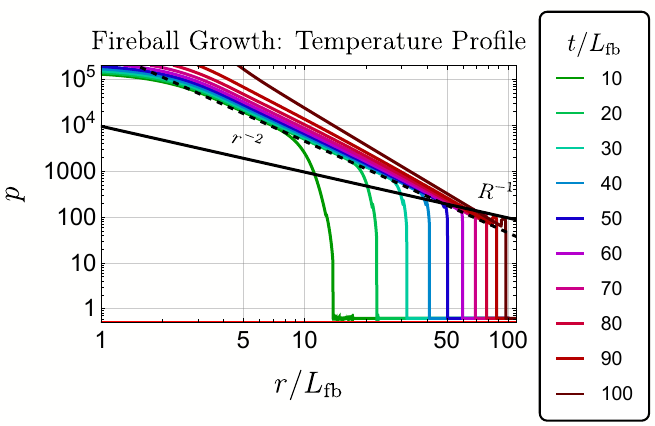}
    \caption{\justifying\label{fig:realistic_simu_fb}  {\textbf{Fireball growth}}.  Simulation of the Phase 0 with $E_{\rm inj}=5 \times 10^9$, $T_{\rm b} = 0.46$ within the time period $t/L_{\rm fb} \in [0,100]$ where each curve (from blue to red) corresponds to a time interval of $t/L_{\rm fb} = 10$. The PBH explosion occurs at $t \sim 100 L_{\rm fb}$, corresponding to the last curve, the dark red. The temperature profile along the fireball (at a given time), decrease like $r^{-2}$ (corresponding to  $T_{\rm sh}(R) \propto R^{-1/2}$) before to be terminated by the shock. This is represented by the dashed line on the plot. As for the shock wave expansion, we found that in the context of the fireball growth and when the shell has been formed, the pressure after the shock follows the behavior  $p_{\rm sh}(R) \propto R^{-1}$ (corresponding to  $T_{\rm sh}(R) \propto R^{-1/4}$) which is presented by the black line. 
    }
\end{figure}

Going back to the dark green curve of Fig.~\ref{fig:realistic_simu_PBH}, we observe that the fireball profile includes a core of roughly constant temperature $T_{\rm fb}$ with a width of roughly $R \sim L_{\rm fb}$. This $T_{\rm fb}$ grows over time. In Fig.~\ref{fig:realistic_simu_fb}, we show a simulation focused on the fireball growth with the pressure profile. The profile of the pressure  scales like $r^{-2}$ (or $r^{-1/2}$ for the temperature) and is then expanding at a velocity close to the speed of light. The value of the shocked pressure $p_{\rm sh}(R,t)$ scales $R^{-1}$. 

\begin{figure}[t!]
        \centering
    \includegraphics[width=0.9\linewidth]{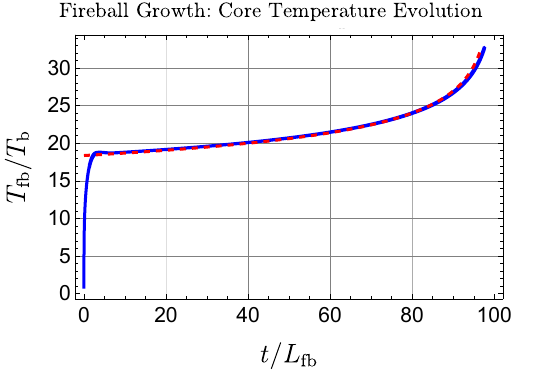}  \\[0.2cm]
    \includegraphics[width=0.95\linewidth]{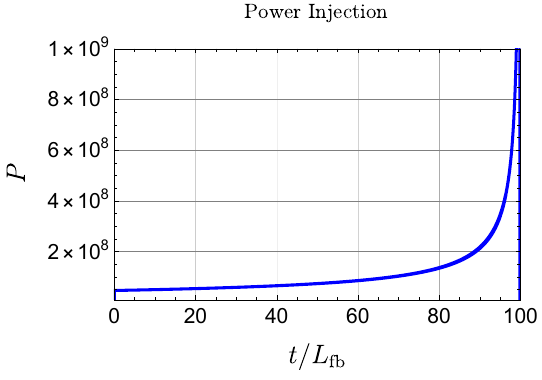} \\[0.2cm]
    \includegraphics[width=0.95\linewidth]{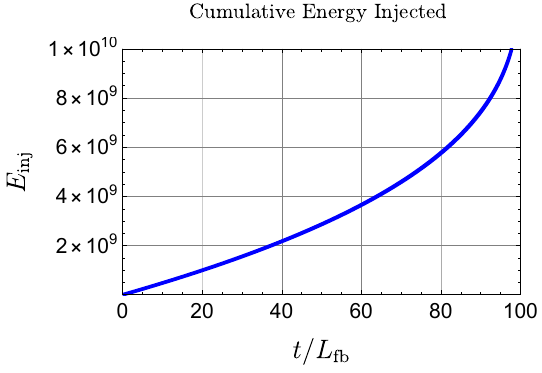}\caption{\justifying\label{fig:realistic_simu_fb_2} {\textbf{PBH-like energy injection}}. Simulation of the Phase 0 with a total energy injection $E_{\rm inj} =5 \times 10^9$.  ({\it Top}) Temperature of the fireball at the origin. We find that after a short period of very fast temperature increase (which is spurious and specific to the simulation) the temperature smoothly and slowly. The dashed red line is a fit to $T \propto (1-(t/t_{\rm inj}))^{-1/6}$. (Middle) Power emitted by the PBH as a function of time. ({\it Bottom}) Cumulative energy injected by the PBH. }
\end{figure}

We track the value of the temperature at the core, $r \sim L_{\rm LPM}$, and present the results on Fig.~\ref{fig:realistic_simu_fb_2}. We observe that the temperature at the core $T_{\rm fb}$ increases like 
\begin{equation}
\label{eq:Tfb}
 T_{\rm fb} / T_{\rm b} \propto \left( 1 - (t / t_{\rm inj}) \right)^{-1/6} \, 
\end{equation} 
for times $t\lesssim t_{\rm inj}$. We emphasize that the quick increase in $T_{\rm fb}/T_{\rm b}$ at early times, which can be observed in Fig.~\ref{fig:realistic_simu_fb_2}, is a spurious feature of the simulation characteristics. This growth lasts until the final explosion of the PBH, after which it saturates at roughly a value computed in Eq.~\eqref{eq:ThsEqn}, where we can consider $E_{\rm inj} \approx M_{\rm thres}(T_{\rm b})$. This computation allows one to fix the prefactor in Eq.~\eqref{eq:Tfb}. Beyond the core at constant temperature $T_{\rm fb}$, the temperature starts to fall as $T\propto R^{-1/2}$ up to the forward shock.

Interestingly, we observe that in the context of the fireball growth and after the shell has formed, the temperature behind the shock scales as $T_{\rm sh} (R,t) \propto R^{-3/8}$, corresponding to $p_{\rm sh}(R,t) \propto R^{-3/2}$.\footnote{Let us emphasize that this is not a profile, but a specific value of a pressure at a given time.} See the fit (black line) in Fig.~\ref{fig:realistic_simu_PBH}. This trend can be understood from an analytical point of view. To determine the evolution of the pressure immediately behind the shock, we apply the same energy conservation considerations as in Sec.~\ref{sec:inst}, in the instantaneous case. During the fireball growth phase, the shock wave propagates as a thin shell within which energy is (approximately) conserved, with 
\begin{equation}
E_{\rm inj}(t) = \int dr \, 16\pi r^2 \, \gamma_{\rm }(r,t)^2 \, p (r,t)  \, . 
\end{equation}

To evaluate this integral, we recall that the shock front is located at $R$ and we adopt an ansatz for the pressure dependence on the radius: $p(r) \propto p_{\rm sh} (r/R)^{-n}$. Similarly, the boost factor $\gamma(r) \propto \gamma_{\rm sh} (r/R)^{m}$. We obtain thus
\begin{equation}
\begin{split}
\int dr \, &16\pi r^2 \, \gamma^2_{\rm }(r,t)\, p (r,t)  \\
&=16\pi R^2\int^R dr \,  p_{\rm sh} \gamma^2_{\rm sh}(r/R)^{2+2m-n} \,  \\
&=  \frac{16\pi}{3+2m -n} p_{\rm sh} \gamma^2_{\rm sh} R^3 \, , 
\end{split}
\end{equation}
if $3+2m -n > 0$ (notice that $n,m$ are themselves functions of time). This condition is seen to be numerically satisfied. Now we can use that $p_{\rm sh} = 8\Gamma^2p_{\rm b}/3 $ and $\gamma^2_{\rm sh} = 2 \Gamma^2$ to conclude that 
\begin{equation} \label{eq:p_early_fb}
p_{\rm sh}(R,t)
=
\sqrt{\frac{3+2m-n}{12\pi}}\;
\sqrt{\frac{p_{\rm b}\,E_{\rm inj}(t)}{R^{3}}}.
\end{equation}
The cumulative energy injected is presented in the bottom panel of Fig.~\ref{fig:realistic_simu_fb_2}. For a long time, $E_{\rm inj}(t)$ is a quite weak function of time (except very close to the PBH final explosion). We thus conclude that $p_{\rm sh}(R,t)$ should go like a power of $R^{-h}$, with $h$ slightly smaller than $1.5$. This analytical conclusion is consistent with the numerical fit of $p_{\rm sh}(R,t)\propto R^{-3/2}$. 

In Appendix~\ref{eq:fireball_growth} we derive a relativistic energy-balance description of the fireball growth and show that, in the appropriate limit, it reproduces the scaling relation obtained in the main text.

\subsection{Phases 1-3: Evolution after the explosion}

We now enter the phase in which the PBH has evaporated and energy injection has ceased. As the source turns off, the temperature at the origin quickly drops and the shock detaches from the source and begins to expand freely. 

\paragraph{Analytic considerations.}

To determine the evolution of the pressure immediately behind the shock, we follow the same energy-conservation argument used in Sec.~\ref{sec:inst}. During the propagation phase, the shock wave propagates as a thin shell within which energy is (approximately) conserved, with 
\begin{equation}
\begin{split}
E_{\rm inj} &= \int dr \, 6\pi r^2 \, \gamma^2_{\rm }(r,t) \, p (r,t)  \\
&\sim 16\pi \, \gamma^2_{\rm sh}(R,t) \, p_{\rm sh} (t) \, R_{\rm }^2 \, L_{\rm shell} \, . 
\end{split}
\end{equation}
Whereas in the instantaneous case, the length $L_{\rm shell}$ could be determined by analytical arguments, this is no longer the case for PBH-like injection. Instead, the value of $L_{\rm shell}$ can be determined directly from the simulations.

    \begin{figure}[t!]
        \centering
\includegraphics[width=0.9\linewidth]{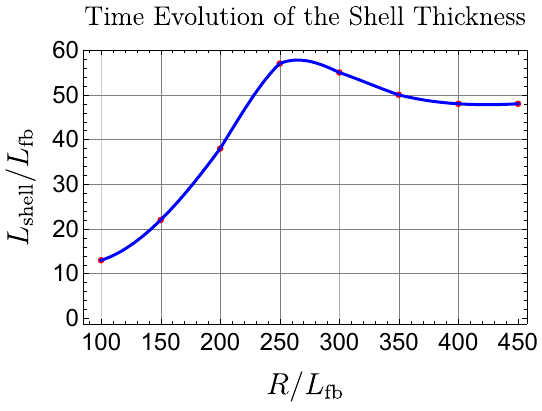}
        \caption{\justifying 
        \label{fig:shell}  Evolution of the shell thickness in Phases 1-2, \emph{after} the PBH evaporation with a total energy injection $E_{\rm inj} =5 \times 10^9$. We observe that $L_{\rm shell}$ first scales linearly with $R\sim t$ and then reaches a saturation value, after which it remains nearly constant. The red dots designate the data points that could be extracted from the simulations, while the blue line is an interpolation of the data points.  
        }
     \end{figure}

In Fig.~\ref{fig:shell}, we display the evolution of the shell thickness. We find that the thickness scales like $L_{\rm shell} \propto t \sim R$ initially, and subsequently reaches an approximately constant value, which is of the order of $\mathcal {O}(0.2) \, R_{\rm max}$. Using that $p_{\rm sh} = 8\Gamma^2p_{\rm b}/3 $ and $\gamma^2_{\rm sh} = 2 \Gamma^2$, we then obtain
\begin{equation} 
\begin{split}
E_{\rm inj} &\sim 16\pi\frac{16\Gamma^4p_{\rm b}}{3} R_{}^2 L_{\rm shell} \\ &\quad \Rightarrow \> \Gamma^2 \sim \sqrt{\frac{3E_{\rm inj}}{256\pi p_{\rm b} R_{\rm }^2 L_{\rm shell}}}  \, ,
\end{split}
\end{equation} 
which thus gives
\begin{equation} 
p_{\rm sh} (R)\big|_{\text{PBH-like}} \sim \sqrt{\frac{E_{\rm inj} p_{\rm b} }{12\pi R_{\rm}^2 L_{\rm shell} (R) }} \, .
\label{eq:PBH_inj_law}
\end{equation}

This result, valid for PBH-like injection, should be compared with the result for instantaneous injection in Eq.~\eqref{eq:Gamma2}, which we repeat here for convenience
\begin{equation}
 p_{\rm sh}(R)\big|_{\text{inst}} =  \frac{17E_{\rm inj}}{9\pi  R_{\rm }^3}\, .
\end{equation} 
Interestingly, the PBH-like injection result for $p_{\rm sh} (R)$ depends explicitly on $p_{\rm b}$. One expects this dependence to disappear once the constant value of $L_{\rm shell}$ is taken into account. One would thus expect $L_{\rm shell} \propto p_{\rm b}$. For illustration, we temporarily set $p_{\rm b} = 1$ to remove this dependence, and one thus observes that, for a fixed energy injection $E_{\rm inj}$:  
\begin{equation}
  p_{\rm sh}(R)\big|_{\text{inst}   } \gg p_{\rm sh}(R)\big|_{\text{PBH-like}   }\,.
\end{equation} 
This is an expected result, as injecting the same total energy over a longer timescale produces a weaker shock than an instantaneous release. Physically, however, the energy injected in the PBH-like case should be much larger than the instantaneous case, since the PBH-like case includes both part of the slow injection and the instantaneous final explosion. As a result, for a given PBH evaporation event, one expects 
\begin{equation}
  p_{\rm sh}(R)\big|_{\text{inst}   } < p_{\rm sh}(R)\big|_{\text{PBH-like}  } 
\end{equation} 
for any $R$. We have verified this fact numerically by considering a PBH-like injection with different window functions. In Fig.~\ref{fig:PBH_inject_2}, we show three different simulations, with the same PBH mass history but with three different injection time windows, which we parametrize with different values of of $nC$. 

Let us now discuss the numerical results. The behavior of the $L_{\rm shell} (R)$ dictates the evolution of $ p_{\rm sh}$ for the PBH-like case. When $L_{\rm shell} (R) \propto R$, \emph{i.e.}, while the shell is thickening, the pressure behind the shock scales like $p_{\rm sh} \sim R_{\rm }^{-3/2}$. Comparing this prediction with the temperature profiles on Fig.~\ref{fig:realistic_simu_PBH}, we observe that the profiles are well fitted by $p_{\rm sh} \sim R_{\rm }^{-3/2}$. In a second phase, the thickness of the shell reaches a constant value, and the pressure behind the shock scales like $p_{\rm sh} \sim R^{-1}_{\rm}$, which is shown in Fig.~\ref{fig:realistic_simu_fb}.

\paragraph{Estimation of $R_{\rm max}$.}

Numerically, from the simulation presented in Fig.~\eqref{fig:realistic_simu_PBH}, we obtain
\begin{equation}
R_{\rm max}^{\rm num}\big\vert _{\rm PBH-like}\sim \mathcal{O}(400)\, L_{\rm fb} \, .
\label{RmaxPBHlikeestimate}
\end{equation}
where we used as a criterion for $R^{\rm num}_{\rm max}$ the radius for which the shock becomes sonic.

To obtain an analytical estimate of the maximal radius, we require that the evolution terminates when $p_{\rm b} \sim p_{\rm sh}$, and obtain 
\begin{equation}
\label{eq:Rmax_for_PBH}
    R_{\rm max} \big\vert_{\rm PBH-like} \approx \bigg(\frac{E_{\rm tot, inj} }{12\pi p_{\rm b} L_{\rm shell}}\bigg)^{1/2} \, ,
\end{equation}
where $E_{\rm tot, inj}$ 
is the energy injected on a timescale close to $R_{\rm max}$ itself given in Eq.~\eqref{Einjtotalmax1}. 

This scaling of $R_{\rm max}$ has been verified using the simulations. These simulations confirm the predicted scaling of $R_{\rm max}$ with the injected energy if one approximates $L_{\rm shell} \sim 0.5 \, t_{\rm inj} \sim 50 \, L_{\rm fb}$, where $L_{\rm fb}$ is the size of the fireball before it detaches from the origin. That is, after the source turns off, the temperature at the origin quickly drops and the wave starts to expand. 

Following Phases 0 and 1 in the PBH-like injection case, the evolution during Phases 2 and 3 proceeds in the same way as in the instantaneous-injection scenario.

\subsection{Comparison between the instantaneous and the PBH-like injection}
\label{sec:InjectionRateComparison}

For the instantaneous case (see in particular Eq.~\eqref{eq:Rmax} in Sec.~\ref{sec:phaseA}), we computed the distance $R_{\rm max}$ over which the shock propagates in the ultra-relativistic regime. For convenience, we repeat the result here:
\begin{equation}
\label{eq:Rmax_instant}
\begin{split}
    &R_{\rm max}\big|_{\rm inst} \sim C_{\rm inst}\bigg(\frac{
    %M_{\rm thres}(T_{\rm b})
    E_{\rm inj}
    }{p_{\rm b}}\bigg)^{1/3} \, , \\ &C_{\rm inst} = \bigg(\frac{8}{3} \times\frac{17}{24\pi} \bigg)^{1/3} \approx 0.84\, ,
    \end{split}
\end{equation}
where, for instantaneous injection, $E_{\rm inj}=M_{\rm thres}(T_{\rm b})$. For the same PBH explosion, but taking into account the PBH-like energy injection, one expects a larger value for $R_{\rm max}$. One may parameterize $R_{\rm max}$ for the PBH-like injection scenario as
\begin{equation}
    R_{\rm max}\big|_{\rm PBH-like} \sim C_{\rm inst}\bigg(\frac{K M_{\rm thres}(T_{\rm b})}{p_{\rm b}}\bigg)^{1/3}  \, ,
    \label{eq:RmaxPBHlikeK}
\end{equation}
where we have taken $E_{\rm inj}=K M_{\rm thres}$ as in Eq.~\eqref{eq:T_hs_init}, and $K>1$ is a numerical coefficient. Thus, in this section we finally estimate the parameter $K$ introduced in Eq.~\eqref{eq:T_hs_init} which allows us to generalize $T_{\rm fb}^{\rm inst}$ and $R_{\rm max}^{\rm inst}$ to the extended PBH-like injection scenario.

Let us now return to the estimate of $R_{\rm max}$ in the PBH-like case. We can re-write Eq.~(\ref{eq:Rmax_for_PBH}) as
\begin{equation}
    R_{\rm max}\big|_{\rm PBH-like} = \bigg(\frac{E_{\rm tot, inj} }{12\pi p_{\rm b} a R_{\rm max}}\bigg)^{1/2} \, ,
\end{equation}
where we define $a$ through $L_{\rm shell}=a\,R_{\rm max}$ and recall that $E_{\rm tot, inj}$ is not the total energy injected, but the energy injected over a timescale of the order of the shock expansion. This simplifies to
\begin{equation}
    \begin{split}
    R_{\rm max}\big|_{\rm PBH-like} & \approx \bigg(\frac{E_{\rm tot, inj} }{12\pi p_{\rm b} a}\bigg)^{1/3}\\ 
    &=  C_{\rm inst}\bigg(\frac{K M_{\rm thres}(T) }{ p_{\rm b} }\bigg)^{1/3}\, ,
    \end{split}
    \label{eq:RmaxPBHlikeK2}
\end{equation}
where in the second line we substituted the quantity 
\begin{equation}
    \label{eq:K_def}
    K = \frac{E_{\rm tot, inj}}{12\pi  a C^3  M_{\rm thres}}\,.
\end{equation}

We have seen in Fig.~\ref{fig:shell} that the constant $a$ is of order $a \sim \mathcal{O}(0.1)$, which we obtain by considering that $R_{\rm max} \sim (300-400) L_{\rm fb}$ for the simulation appearing in Fig.~\ref{fig:shell}. For a PBH exploding around $T\sim 100$ GeV, one obtains that $E_{\rm tot, inj}\big|_{\rm max}/M_{\rm thres} \sim \mathcal{O}(10)$. This corresponds to  
\begin{equation}
\label{eq:value_of_K}
    K \sim \mathcal{O}(5) \, , 
\end{equation}
which ultimately increases $R_{\rm max}$ by a factor of $2$ when compared with the instantaneous case. This level of enhancement is confirmed by numerical simulations. We note that this estimate for $K$ is likely an underestimate, since we cannot simulate the entire energy injection process, from the initial formation of a shock until the final explosion. The estimate $K \sim {\cal O} (5)$ corresponds to a simulation duration of $t_{\rm inj}=100 \, L_{\rm LPM}$, which captures energy injection over an interval 100 times longer than the final instantaneous explosion. The scaling of $P(t)$ from Eq.~\eqref{eq:PBH_inject} implies that extending simulations to earlier times (and hence larger values of of $t_{\rm inj}$) will likely lead to moderate increases in $E_{\rm tot, inj}$ and thus to $R_{\rm max}$. 

\section{Implications for electroweak symmetry restoration}
\label{sec:Pheno_impl}

If the temperature of the ambient plasma is below $T_{\rm EW}\approx 162\, {\rm GeV}$, the EW symmetry is broken. Furthermore, if the maximal temperature reached in the shock wave exceeds $T_{\rm EW}$, then the outwardly propagating shock front and the trailing rarefaction wave delimit a moving region of hot plasma within which EW symmetry is \emph{restored}. Thus, the expanding shock-wave generated by the PBH explosion is contained within a shell with two ``bubble-like walls'': the wall restoring the symmetry at the shock front and the rear wall of the rarefaction wave breaking it again. This is illustrated in Fig.~\ref{fig:realistic_simu}. As we have seen in Fig.~\ref{fig:velo_shock}, these two walls are not symmetric and do not expand at the same velocity.

The EW symmetry is restored when $T_{\rm sh} > T_{\rm EW}$. 
Achieving $T_{\rm fb}(K, T_{\rm b})>T_{\rm EW}$ is a fundamental requirement for EW restoration from a PBH explosion. The lowest background temperature $T_{\rm min}(K)$ such that a PBH explosion which instantaneously injects $E_{\rm dep}=K M_{\rm thresh}$ will be hot enough to restore the EW symmetry is found by numerically solving
\begin{equation}
    T_{\rm fb}(K, T_{\rm min})=T_{\rm EW},
\end{equation} 
where $T_{\rm fb}$ is given by Eq.~\eqref{eq:T_hs_init} and we recall that $K=1$ corresponds to the instantaneous case while $K\simeq4$ gives a good approximation of the PBH-like case. See the inset of Figure~\ref{fig:Mthres}b for a plot of $T_{\rm min}(K)$. We find $T_{\rm min}=1.54 \, {\rm GeV}$ for the instantaneous case $K=1$.

Using Eq.~\eqref{eq:Gamma2}, imposing $p_{\rm sh} = p_{\rm EW}$, and solving for $R$, one can also compute the maximal distance over which an EW wall is maintained:
\begin{equation}
\label{eq:Rmax_instanT_bub}
    R^{\rm EW}_{\rm max}\big|_{\rm inst} \sim 0.84\bigg(\frac{M_{\rm thres}(T_{\rm b})}{p_{\rm EW}}\bigg)^{1/3} \, . 
\end{equation}
A detailed study of this EW restoration effect as a novel baryogenesis mechanism is presented in a companion paper~\cite{BAUinprep}, where we compute both the baryon number produced by the expanding shock from a single PBH explosion and the total baryon asymmetry of the universe due to a population of PBHs exploding after the electroweak phase transition.

\paragraph{Backreaction of the wall.}
Let us now discuss the possible backreaction that the EW wall could have on the shock wave propagation. So far we have considered the shock wave from the purely hydrodynamical point of view. However, the wall at the position of the shock is very similar to an EW \emph{inverse} phase transition~\cite{Barni:2024lkj}. Across this bubble wall, the SM particles change mass (since the value of the Higgs field differs on each side of the wall), and this change of mass is known to backreact on the wall expansion itself. The pressure from the SM particles \emph{losing} their mass will be of order~\cite{Azatov:2022tii}
\begin{equation}
    \begin{split}
    p_{\rm SM} & \sim - T_{\text{b}}^2 v_{\text{EW}}^2 \left( \frac{y_t^2}{8} + \frac{g^2 + g'^2}{32} + \frac{g^2}{16} \right)\\
    & \approx -0.17 \, T_{\text{b}}^2 v_{\text{EW}}^2.
    \end{split}
\end{equation}

This pressure is negative and the shock is pulled forward by the particles losing their mass. However, this pressure remains typically smaller than the pressure inside the shock as long as the shock remains relativistic. Consequently, these corrections can likely be neglected for the velocity of the wall and its position, and thus also for the determination of $R_{\rm max}$.

\paragraph{Entropy production by the EW bubble wall.}

One might worry that the entropy production due to the EW profile inside the shock could broaden the shock. Here we argue that this effect is also negligible. Indeed, the local entropy production due to the wall will scale like~\cite{Espinosa:2010hh, Ai:2025bjw}
\begin{equation}
\begin{split}
&\partial_\mu (s u^\mu) \sim \eta_{\rm EW} (\partial_z \phi)^2,\\
&\eta_{\rm EW} \sim \frac{0.01}{ T_{\rm b} \delta} (v_{\rm EW}/T_{\rm b})^4 \, , 
\end{split}
\end{equation}
where $\delta$ is the thickness of the shock. In these expressions, the parameter $\eta_{\rm EW}$ is an effective parameter capturing the production of entropy inside the EW bubble wall and extracted from simulations~\cite{Moore:1995ua}. This implies that the energy loss associated with this entropy production is given by
\begin{align}
    |\dot E| &\sim -T_{\rm b} \dot S = -T_{\rm b} \dot R \int dr \, 4\pi \,r^2  \frac{ds}{dr}
    \\ \nonumber
   & \sim 
    -T_{\rm b} \eta_{\rm EW} (\partial_z \phi)^2 4\pi R^2 \delta \dot R
    \\ \nonumber
   & 
    \sim - 0.01 \frac{v_{\rm EW}^6}{T_{\rm b}^4\delta^2}  4\pi R^2 \, ,
    \label{eq:energylossentropy}
\end{align}
where we used in the last line $\dot R \sim 1$ and $ds/dr \sim \eta_{\rm EW} (\partial_z \phi)^2$ only inside the shock itself. Comparing this energy loss with that arising from ordinary viscosity, we observe that the entropy from the EW bubble is subdominant if 
\begin{equation}
    \bigg(\frac{4\eta}{3} +\zeta\bigg) \gg 0.01 \frac{v_{\rm EW}^6 }{T_{\rm b}^4 \delta}  \, . 
\end{equation}
 This inequality is expected to be fulfilled in our case. We can thus largely neglect the entropy produced due to the bubble wall with respect to the entropy due to viscosity.

\section{Conclusion}
\label{sec:conclusion}

In this paper, we have performed an in-depth study of primordial black hole explosions in the early-universe quark-gluon plasma. The main result of our work is that PBH explosions generate relativistic shock waves whose propagation can be reliably described by hydrodynamics and whose maximal extent is only moderately affected by the time profile of the energy injection. The generation of such shocks, derived here for the first time, may have important implications for several aspects of early-universe cosmology---including baryogenesis and big bang nucleosynthesis---and can inform future detailed studies to further clarify the role PBHs may have played within them.

The physical mechanism can be summarized as follows. When the PBH evaporates, it emits a large flux of high-energy particles which deposit their energy over a characteristic distance $L_{\rm LPM}$, determined using the Landau--Pomeranchuk--Migdal formalism. This localized energy deposition reheats the plasma around the PBH and generates pressure gradients that drive fluid motion. Long before the final explosion, the energy is injected at an approximately constant rate and the fluid profile approaches a quasi-steady-state configuration. At the moment of the final PBH explosion, however, the injection rate rapidly increases and the resulting pressure gradients generate a strong blast wave which propagates through the plasma as a relativistic shock followed by a rarefaction wave. 

To study this phenomenon we combined hydrodynamical simulations with analytical modeling. The numerical simulations allow us to follow the full nonlinear evolution of the plasma and determine the structure of the shock, while the analytical treatment provides physical insight into the scaling relations governing the expansion. We also develop a novel relativistic generalization of the bubble-driven shock formalism of Ref.~\cite{2013ApJ...768..113M}, which describes the growth of the fireball while the PBH is still injecting energy into the plasma.

Using these tools, we showed that the shock propagation proceeds through three regimes: an initial ultra-relativistic phase, a transition to a non-relativistic Sedov--Taylor blast wave, and finally the dissipation of the wave through viscous effects in the plasma. We determined the maximal distance over which the shock propagates before entering the non-relativistic regime, and found that this distance is only weakly sensitive to the detailed time profile of the energy injection. In particular, incorporating the realistic PBH-like injection history increases the propagation length by only a factor of order $2$ relative to the instantaneous approximation.

Our study relied on hydrodynamical simulations, which by construction neglect, first, microscopic effects such as viscosity and diffusion and, second, the impact of the Higgs wall profile sustained by the hydrodynamical background. We have argued that these effects remain subdominant in the regime of interest considered in this paper. A dedicated numerical study of these effects could nevertheless provide a more precise assessment and is left for future work.

We conclude by summarizing the possible phenomenological consequences of expanding shock waves. In addition to the novel baryogenesis mechanism explored in our companion paper~\cite{BAUinprep}, other possible effects include rapid energy injection from PBH emission during big bang nucleosynthesis (which could modify existing PBH bounds) and the generation of cosmological magnetic fields during shock propagation. Furthermore, in close analogy to the bubbles produced during first-order phase transitions, expanding shocks naturally generate quadrupolar stress-energy perturbations that can source gravitational waves, potentially detectable by current and forthcoming experiments.

%%%%%%%%%%%%%%%%%%%%
\section*{Acknowledgments}
We gratefully acknowledge Yago Bea, Jorge Casaderrey, Peter Fisher, Benjamin Lehmann, David Mateos, and Yuber Ferney Perez Gonzalez for enlightening discussions. MV sincerely thanks Xander Nagels for assistance in the development of the hydrodynamical code, and acknowledges CERN TH Department for hospitality while this research was being carried out. ST was supported by the Office of High Energy Physics of the US Department of Energy (DOE) under Grant No.~DE-SC0012567, and by the DOE QuantISED program through the theory consortium ``Intersections of QIS and Theoretical Particle Physics'' at Fermilab (FNAL 20-17). ST is additionally supported by the Swiss National Science Foundation project number PZ00P2\_223581, and acknowledges CERN TH Department for hospitality while this research was being carried out. This project has also received funding from the European Union's Horizon Europe research and innovation programme under the Marie Sk{\l}odowska-Curie Staff Exchange grant agreement No.~101086085 -- ASYMMETRY. Portions of this research were conducted in MIT's Center for Theoretical Physics --- A Leinweber Institute and supported by the Office of High Energy Physics within the Office of Science of the U.S.~Department of Energy under grant Contract Number DE-SC0012567. We also gratefully acknowledge support from the Amar G.~Bose Research Grant Program at MIT.

\newpage 
\appendix

 \section{Matching conditions for a shock}
\label{sec:relativistic_jump}

\begin{figure*}
    \centering
    \includegraphics[width=0.6\textwidth]{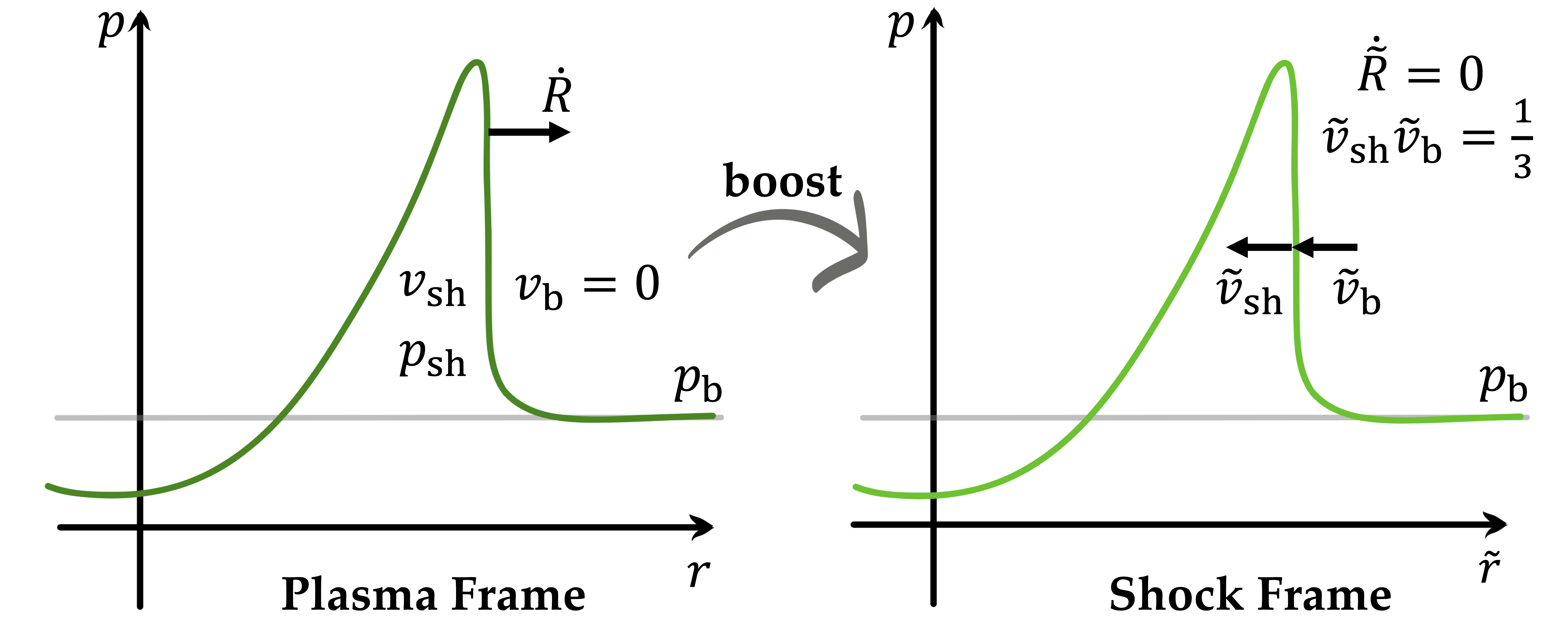}
    \caption{Sketch of the matching conditions across the shock.
   }
    \label{fig:sketch_inst_2}
\end{figure*}

In the main body of this paper, we have made extensive use of the jump conditions across a shock and the results coming from energy and momentum conservation. In this Appendix, we spell out the relevant the details.

To determine the hydrodynamic discontinuity across the expanding shock, we relate the quantities defined in two different frames: i) The \emph{PBH frame} is the rest frame of the PBH and of the unshocked ambient plasma far ahead of the shock; in this frame the upstream fluid has 4--velocity $u^{\mu}=(1,0,0,0)$, and ii) the \emph{shock frame} is the frame in which the shock front is instantaneously at rest (quantities evaluated in this frame carry a tilde). In this frame the ambient fluid flows toward the shock with velocity $\tilde v_{\rm b}$ and Lorentz factor $\tilde\gamma_{\rm b}=(1-\tilde v_{\rm b}^2)^{-1/2}$, while the shocked shell flows away from the shock with velocity $\tilde v_{\rm sh}$ and Lorentz factor $\tilde\gamma_{\rm sh}=(1-\tilde v_{\rm sh}^2)^{-1/2}$. A sketch of the different quantities introduced in this section is presented on Fig.~\ref{fig:sketch_inst_2}.

Conservation of the energy--momentum tensor across an infinitesimally thin discontinuity yields, in the shock frame, the relativistic Rankine--Hugoniot conditions~\cite{RezBook}
\begin{align}
w_{\rm b}\,\tilde\gamma_{\rm b}^2\,\tilde v_{\rm b} 
&= 
w_{\rm sh}\,\tilde \gamma_{\rm sh}^2\,\tilde v_{\rm sh},
\label{eq:sh_cont_1}
\\
w_{\rm b}\,\tilde\gamma_{\rm b}^2\tilde v_{\rm b}^2 
+ p_{\rm b}
&=
w_{\rm sh}\,\tilde \gamma_{\rm sh}^2\tilde v_{\rm sh}^2 
+ p_{\rm sh}\,.
\label{eq:sh_cont_2}
\end{align}
For the relativistic plasma of interest, we have $w_{\rm b}=4p_{\rm b}$ and $w_{\rm sh}=4p_{\rm sh}$. Eliminating $p_{\rm sh}$ between Eqs.~\eqref{eq:sh_cont_1}--\eqref{eq:sh_cont_2} gives a quadratic relation between $\tilde v_{\rm b}$ and $\tilde v_{\rm sh}$,
\begin{equation}
\tilde v_{\rm sh}
=
\frac{1}{6\tilde v_{\rm b}}
\left[
  1+3\tilde v_{\rm b}^2
  \pm\sqrt{(1+3\tilde v_{\rm b}^2)^{2}-12\tilde v_{\rm b}^2}
\right]\,.
\label{eq:v_quadratic}
\end{equation}
This equation admits two solutions for $\tilde v_{\rm sh}$ as a function of $\tilde v_{\rm b}$; however, only one of these two mathematical branches corresponds to a physical strong shock. This physically relevant solution is characterized by
\begin{equation}
\tilde v_{\rm b}\,\tilde v_{\rm sh}=\frac{1}{3}\,.
\label{eq:v_sh_cont}
\end{equation}

Using $w_{\rm b}=4p_{\rm b}$, $w_{\rm sh}=4p_{\rm sh}$ and Eq.~\eqref{eq:v_sh_cont}, the jump conditions imply
\begin{equation}
p_{\rm sh}
=
3p_{\rm b}\,\tilde v_{\rm b}^2\tilde\gamma_{\rm b}^2
\left(
1-\frac{1}{9\tilde v_{\rm b}^2}
\right)\,,
\label{eq:p_sh_cont_full}
\end{equation}
where $\tilde v_{\rm b}$ is the ambient velocity in the shock frame. In the ultrarelativistic limit $\tilde v_{\rm b}\rightarrow 1$, this reduces to
\begin{equation}
p_{\rm sh}
=
\frac{8}{3}\,\tilde\gamma_{\rm b}^2p_{\rm b}= \frac{8}{3}\,\Gamma^2p_{\rm b}\,,
\end{equation}
where we used that, since the plasma in the PBH frame is at rest in front of the shock, the modulus of $\tilde v_{\rm b}$ is equal to $\dot R$. The corresponding relation for the shocked shell in the shock frame Lorentz factor is
\begin{equation}
\tilde \gamma_{\rm sh}^{2}
=
\frac{9(\tilde\gamma_{\rm b}^{2}-1)}{8\tilde\gamma_{\rm b}^{2}-9}\,.
\label{eq:gamma_sh_cont}
\end{equation}
The relations between the different Lorentz factors in the different frames may be summarized as follows:

\begin{itemize}

\item The upstream fluid is at rest in the PBH frame, so
\begin{equation}
\gamma_{\rm b} = 1\,.
\label{eq:gamma_up}
\end{equation}

\item The shock moves through this upstream fluid with velocity $\dot R$ in the PBH frame, which is equal in magnitude with the upstream velocity in the shock frame:
\begin{equation}
\dot R = \tilde v_{\rm b},
\qquad
\Gamma = \tilde\gamma_{\rm b}\,.
\label{eq:Gamma_equiv}
\end{equation}

\item The downstream Lorentz factor in the shock frame is related to the upstream one by Eq.~\eqref{eq:gamma_sh_cont}.

\item The downstream Lorentz factor in the PBH frame is obtained by boosting:
\begin{equation}
\gamma_{\rm sh}
=
\tilde\gamma_{\rm b}\,\tilde\gamma_{\rm sh}\,
(1+\tilde v_{\rm b}\tilde v_{\rm sh})\,.
\label{eq:gamma_boost1}
\end{equation}

Using \eqref{eq:v_sh_cont}, this becomes
\begin{equation}
\gamma_{\rm sh}
=
\frac{4}{3}\,\tilde\gamma_{\rm b}\,\tilde\gamma_{\rm sh}\,.
\label{eq:gamma_boost2}
\end{equation}

Combining Eqs.~\eqref{eq:gamma_sh_cont} and \eqref{eq:gamma_boost2} yields
\begin{equation}
\gamma_{\rm sh}^{2}
=
\frac{16\,\tilde\gamma_{\rm b}^{2}\bigl(\tilde\gamma_{\rm b}^{2}-1\bigr)}
     {8\tilde\gamma_{\rm b}^{2}-9}\,.
\label{eq:gamma_sh_down}
\end{equation}

\end{itemize}

 \section{A shock capturing numerical scheme}
\label{app:scheme}

The numerical solution of nonlinear conservation laws is notoriously challenging due to the presence of advective behavior, rarefaction and compression waves, and discontinuities such as shock fronts. In our physical setup, the dynamical evolution of the system is expected to generate \textit{strong shocks} as a natural consequence of steep pressure gradients and nonlinear wave interactions. Therefore, a numerical method capable of capturing discontinuities accurately without introducing unphysical oscillations or excessive numerical diffusion is essential. More information about the comparison of the different schemes is presented in Ref.~\cite{RezBook}. 

Traditional first-order schemes, such as the Lax--Friedrichs method, are in general robust but suffer from large artificial viscosity, which tends to smear sharp discontinuities and wash out small-scale features of the flow. A common way to overcome this limitation is to incorporate the local nonlinear structure of the solution into the numerical flux through the use of \textit{limiters}. These hybrid approaches combine the stability of first-order upwind methods with the accuracy of higher-order central 
schemes.

The Kurganov--Tadmor (KT) scheme provides an elegant alternative to traditional Riemann solvers. It belongs to the class of central-upwind methods and achieves second-order accuracy in both space and time without requiring the explicit solution of Riemann problems. Instead, the KT scheme approximates the local propagation speeds and introduces a constructed dissipation term that depends on the maximal eigenvalue of the local flux Jacobian. This allows the scheme to capture sharp discontinuities with minimal smearing, while maintaining numerical stability and 
non-oscillatory behavior.

The numerical implementation is based on the second-order Kurganov--Tadmor (KT) central-upwind scheme~\cite{KURGANOV2000241} for systems of conservation laws of the form
\begin{equation}
\partial_t U + \partial_x f(U) = S(U,x)\,,
\end{equation}
where $U$ denotes the conserved variables, $f(U)$ the flux function, and $S$ a geometric source term arising from spherical symmetry. The KT scheme provides a high-resolution, Riemann-solver-free method that maintains second-order accuracy while avoiding spurious oscillations near discontinuities. It achieves this by coupling a non-oscillatory reconstruction of the fluid variables with a local estimate of the maximal propagation speed.

In its semi-discrete form, the update for the cell-averaged quantity $U_j(t)$ is written as
\begin{equation}
\frac{d U_j}{dt} = -\frac{H_{j+1/2} - H_{j-1/2}}{\Delta x} + S_j\,,
\end{equation}
where $H_{j+1/2}$ is the numerical flux through the interface between cells $j$ and $j+1$. Following Kurganov and Tadmor, the flux is expressed as
\begin{equation}
\begin{split}
H_{j+1/2} &= 
\frac{f(U^+_{j+1/2}) + f(U^-_{j+1/2})}{2}\\
&\quad\quad  - \frac{a_{j+1/2}}{2}\left(U^+_{j+1/2} - U^-_{j+1/2}\right),
\end{split}
\end{equation}
with $a_{j+1/2}$ denoting the local maximal signal velocity:
\begin{equation}
a_{j+1/2} = \max\!\left\{
\left|\frac{\partial f}{\partial U}(U^+_{j+1/2})\right|,
\left|\frac{\partial f}{\partial U}(U^-_{j+1/2})\right|
\right\}\,.
\end{equation}
The staggered states $U^-_{j+1/2}$ and $U^+_{j+1/2}$ are reconstructed from neighboring cell averages using a piecewise-linear profile:
\begin{equation}
U^-_{j+1/2} = U_j + \frac{\Delta x}{2}(U_x)_j, \quad
U^+_{j+1/2} = U_{j+1} - \frac{\Delta x}{2}(U_x)_{j+1}\,,
\end{equation}
where the slopes $(U_x)_j$ are limited using the minmod function to suppress 
numerical oscillations:
\begin{equation}
\begin{split}
&(U_x)_j \\
&\>= \text{minmod}\!\left(
\theta\,\frac{U_j - U_{j-1}}{\Delta x},
\frac{U_{j+1} - U_{j-1}}{2\Delta x},
\theta\,\frac{U_{j+1} - U_j}{\Delta x}
\right)\,,
\end{split}
\end{equation}
with $\theta \in [0.1, 1]$ controlling the level of numerical dissipation.

In the present implementation, this reconstruction is applied independently to the conserved quantities $U_1$ and $U_2$. The corresponding fluxes $f(U_1,U_2)$ are defined analytically, and the maximal characteristic speed $a_{j+1/2}$ is computed from the local velocity and sound speed,
\begin{equation}
a_{j+1/2} = \max\!\left(
\left|\frac{v \pm c_s}{1 \pm v c_s}\right|
\right)\,,
\end{equation}
ensuring numerical stability in relativistic flow regimes. The evolution proceeds via explicit time-stepping with uniform grid spacing $\Delta x$ and timestep $\Delta t$, constrained by the Courant--Friedrichs--Lewy (CFL) condition.

This formulation combines the simplicity and robustness of central schemes with the accuracy of high-resolution methods, providing a stable and efficient framework for simulating nonlinear wave propagation and shock dynamics in spherically symmetric systems.

\section{Stability, isotropy, and numerical validation}

In this appendix we collect several supplementary analyses that support the results presented in the main text. We first discuss the stability of the shock front against corrugation and Rayleigh--Taylor instabilities (Sec.~\ref{app:instabilities}). We then demonstrate that the reheated region around the PBH is isotropic, justifying the use of spherically symmetric hydrodynamics (Sec.~\ref{app:homogen}). Finally, we present a series of numerical checks: independence of the late-time profiles from the initial fireball shape and size, comparison of the analytical self-similar profiles with the simulations, and verification of the $R_{\rm max}^{\rm EW}$ scaling (see Appendix~\ref{app:checks}).

\subsection{The possibility of instabilities}
\label{app:instabilities}

In this appendix, we discuss possible instabilities that could jeopardize both the instantaneous and the PBH-like energy injection scenaria. It is known that shock fronts can be unstable to perturbations of their surface (\emph{corrugation} modes)~\cite{LandauFluid}, which can subsequently develop into turbulence. Shocks that become subsonic relative to the upstream flow are \emph{non-evolutionary} and therefore unstable~\cite{LandauFluid}, so that the shock rapidly decays. By contrast, a supersonic shock is evolutionary. This implies that as long as the shock remains highly relativistic ($\Gamma\gg 1$), it is stable to corrugation.

Other instabilities may also affect the shock front. One example is the \emph{Rayleigh-Taylor} instability, which can arise when a decelerating shock propagates through a stratified medium. This effect has been studied for the case of ejecta propagating in the galactic medium~\cite{Duffell:2014vsa, Duffell:2015eua}. Such astrophysical situations share many similarities with the one we have presented in this paper. One can compare the temperature profiles shown in Fig.~\ref{fig:PBH_inject_2} (for instance the red curves) with Fig.~3 of~\cite{Duffell:2015eua}. In the red curves in Fig.~\ref{fig:PBH_inject_2}, we observe a shock followed by a shell and a rarefaction wave terminating in an underpressure, together with the appearance of a secondary inner shock at smaller radius $R\sim10\,L_{\rm fb}$. 

Using 2+1 dimensional simulations (instead of the 1+1 dimensional simulations used in our study), the authors of Refs.~\cite{Duffell:2014vsa, Duffell:2015eua} found that the inner shock is typically unstable. The instability is driven by the effective gravitational force felt in the decelerating reference frame of the inner shock. In general, the configuration becomes unstable when the pressure gradient has the opposite sign to the density gradient. 

It is probable that the inner shock found in our simulations suffers from the same type of instabilities. However, the behavior of this inner shock is not the main object of our study, and we leave its further investigation to future research. Importantly, in the 2+1 simulations no instabilities were observed in the forward front (at large $R$). Consequently, we expect that the front shock in our scenario remains stable.

\subsection{An isotropic heated region}
\label{app:homogen}

Here we further comment on the homogeneity of the heated sphere of volume $4\pi L_{\rm LPM}^3/3$. Most particles emitted by a PBH in our mass range $M\simeq 10^6\,{\rm g}$ are color-charged, and emission is specifically dominated by energetic quarks. The particles emitted by the PBH reheat a cone with opening angle~\cite{Dokshitzer:2001zm,Asaka:2003vt}, 
\begin{equation}
\theta_{\rm LPM}^2 \sim \frac{L_{\rm LPM}T_{\rm b}}{l^2_{g, \rm{mfp}} E_g^2} \, , 
\end{equation}
where $l_{g, \rm{mfp}}$ is the mean free path of the emitted soft gluon in the transverse direction and $E_g \sim \sqrt{E T_{\rm b}}$ is the energy of the emitted gluon~\cite{Dokshitzer:2001zm}. Taking, as a conservative estimate, the mean free path of the gluon to be $l_{g, \rm{mfp}}\sim 1/T_{\rm b}$, one obtains 
\begin{equation}
\theta_{\rm LPM}^2 \sim \frac{\sqrt{T_H/T_{\rm b}} \, T_{\rm b}}{E} \sim \sqrt{T_{\rm b}/E} \,. 
\end{equation}
This implies that the surface of the section of the heated cone scales like
\begin{equation}
   \mathcal{S}_{\rm cone} \sim  4\pi L_{\rm LPM}^2\theta_{\rm LPM}^2 \sim 4\pi L_{\rm LPM}^2 \sqrt{T_{\rm b}/E} \,  ,
\end{equation} 
which is to be compared with the surface of the entire sphere $\mathcal{S}_{\rm full} \sim  4\pi L_{\rm LPM}^2$,
\begin{equation}
    \frac{\mathcal{S}_{\rm cone}}{\mathcal{S}_{\rm full} } \sim \sqrt{T_{\rm b}/E}  \, . 
\end{equation}
One can conclude that if the inverse fraction of the surface of the reheated sphere covered by one emitted particle $\mathcal{S}_{\rm full} /\mathcal{S}_{\rm cone}$ is much smaller than the total number of particles emitted by the PBH, then the temperature on the surface at $r \sim L_{\rm LPM}$ is \emph{isotropic}. 

To get a sense of the scales, let us consider some numbers. We use the example of a PBH of an initial mass $M_i \sim 10^6$ g exploding close to the EW temperature and injecting an energy of order $E_{\rm inj }\sim 10^{3}\, {\rm g}$ distributed among roughly $10^{15}$ emitted hard particles. More precisely, for a $10^3\,{\rm g}$ PBH explosion, $4.37\times10^{15}$ quarks are emitted and $2.96\times10^{14}$ gluons are emitted. In this case, the inverse fraction of the surface of the reheated sphere covered by one emitted particle $\mathcal{S}_{\rm full} /\mathcal{S}_{\rm cone}  \sim 10^4$, which is much smaller than the number of particles. The reheating is thus expected to be highly isotropic, which justifies the use of spherically symmetric hydrodynamics that we use throughout this paper.

\subsection{Numerical validation of analytical treatment}
\label{app:checks}

\paragraph{Independence of the initial profile and size for instantaneous injection.}

First, we verify that the hydrodynamical profiles far away from the heat source actually do not depend on the particular shape and size of this region if the energy is injected on a timescale much shorter than $R_{\rm max}$. We thus perform three simulations with different initial size of the heated region but with the same energy injected. We take the initial size of this region to be $R_0 = 2,4, 8$. We also try different initial profile shapes. Even if initially the pressure profiles differ strongly, at late times, they exactly superpose and become independent of $R_0$ and the initial shape. 

\paragraph{The analytical approximation of the profile.}

Secondly, we verify the analytical profiles obtained in Eq.~\eqref{eq:prof} by comparing them with proper simulations. We find that they differ by 20\% to 30\% in the vicinity of the rarefaction wave. Thus we find that the analytical estimates, close enough to the shock, are in general well reproduced by the numerical simulations. 

\paragraph{$R^{\rm EW}_{\rm max}$ scaling.}

Finally, we verify the scaling of the maximal distance over which the shock may restore EW symmetry $R^{\rm EW}_{\rm max}$ as a function of the energy injection $E_{\rm inj}$, (see Eq.~\eqref{eq:Rmax_instanT_bub} for the case of instantaneous injection). We find that the determination of $R^{\rm EW}_{\rm max}$ using the analytical estimate agrees within 10\% to 40\% with the one determined via numerical simulations. For example, using Eq.~\eqref{eq:Rmax_instanT_bub} with the values used in Fig.~\ref{fig:realistic_simu}, one obtains an analytical value $R_{\rm max} \approx 160 \,L_{\rm fb}$, while the numerical simulation suggests $R_{\rm max} \approx 250 \, L_{\rm fb}$.

\section{Thickness of the shock}
\label{sec:thickness_shock}
For particle physics purposes, including baryogenesis, the thickness of the shock is a crucial ingredient. Ideal hydrodynamics would predict an infinitely sharp discontinuity at the locus of the shock. This discontinuity is of course non-physical and it is smoothed out by the viscosity present naturally in the early universe plasma, giving a physical width to the shock. This width can be determined by including in the relativistic Rankine--Hugoniot conditions the effect of viscosity and dissipation: 
\begin{align}
w_{\rm b}\,\tilde\gamma_{\rm b}^2\,\tilde v_{\rm b} 
&= 
w_{\rm sh}\,\tilde \gamma_{\rm sh}^2\,\tilde v_{\rm sh}\,,
\\
w_{\rm b}\,\tilde\gamma_{\rm b}^2\tilde v_{\rm b}^2 
&+ p_{\rm b} - \left(\frac{4}{3}\eta_{\rm b} +\zeta_{\rm b}\right)\frac{dv_{\rm b}}{dx} \nonumber \\
&=
w_{\rm sh}\,\tilde \gamma_{\rm sh}^2\tilde v_{\rm sh}^2 
+ p_{\rm sh} - \left(\frac{4}{3}\eta_{\rm sh} +\zeta_{\rm sh}\right)\frac{dv_{\rm sh}}{dx} \, .
\end{align}
Following the discussion in \S 93 of Ref.~\cite{LandauFluid}, we obtain that the profile pressure across the shock is approximated by 
\begin{equation}
    p(z) \approx \frac{p_{\rm b}+ p_{\rm sh}}{2} + \frac{p_{\rm sh}- p_{\rm b}}{2} \,\text{tanh}(z/\delta) \, ,
\end{equation}
where $z$ is the perpendicular direction to the shock wave. The thickness $\delta$ of the front then evolves as 
\begin{equation}
    \delta \approx \frac{8 \eta / (2\rho) }{\Delta p(\partial^2V/\partial p^2)_s/V^2}\,.  
\end{equation}
where $\Delta p = p_{\rm sh}- p_{\rm b}$. 

There is an inherent ambiguity in this expression because it involves thermodynamic quantities such as energy density and pressure, which take different values ahead of and behind the shock. The analytical derivation does not specify on which side these quantities should be evaluated.\footnote{It is also well-known that across this pressure profile, the entropy density first increases very fast, then reaches a maximum and drops to a constant value $s_{\text{in}}$ which is higher than the value of the entropy density in the unshocked region, $s_{\text{in}} > s_{\rm sh} > s_{\rm b}$. This phenomenon is known as the \emph{overshot entropy}.}

Consequently, the thickness is inversely proportional to the jump of pressure across the wall and directly proportional to the viscosity coefficients. Combining these relations and taking $(\partial^2V/\partial p^2)_s \rho/V^2 \sim \rho^2/p^2  \sim 10$ in the ambient medium, one obtains that the thickness of the wall is given by 
\begin{equation}
    \delta \approx \frac{ \eta  }{\Delta p} \, . 
\end{equation}

We note that even though this formula has been deduced only in the limit of small pressure difference $\Delta p\lesssim p_{\rm b}$, which is largely violated for our strong shocks, it is expected to hold qualitatively for larger pressure differences~\cite{LandauFluid}. This expression however recovers the expectation that a weaker shock, related to smaller jump in pressure $\Delta p$, has a larger thickness $\delta$, and in the limit of $p_{\rm sh} \to p_{\rm b}$, the shock becomes totally smooth. 

Since the kinematic viscosity $\eta/\rho$ is given by roughly the mean free path of particles in the plasma $\eta/\rho \sim l_{\rm mfp} v_{\rm sh}$, one obtains 
\begin{equation}
    \delta \approx \frac{ l_{\rm mfp}  p_{\rm b} }{\Delta p}  \gtrsim l_{\rm mfp} 
\end{equation}
in the limit of weak shocks. Notice that this length scale is natural because it is the only length scale that can be formed out of the two competing quantities at play: the change of pressure $\Delta p$ and the dynamical viscosity $\eta$.

\section{Relativistic generalization of bubble-driven shocks and fireball growth}
\label{eq:fireball_growth}

In Sec.~\ref{sec:PBH_phase0}, we presented numerical results for the growth of the fireball (Phase~0) together with a simple analytical estimate for the evolution of the pressure immediately behind the shock, based on energy conservation. In this Appendix we derive a more general description of the fireball growth using a formalism inspired by Ref.~\cite{2013ApJ...768..113M}, which studied bubble--driven shocks in the non-relativistic interstellar medium. Here we extend this treatment to the relativistic regime relevant for the early-universe plasma. 

In the appropriate limit, this formulation reproduces the relation in Eq.~\eqref{eq:p_early_fb} used in the main text for the early fireball growth. In addition, the formalism allows us to investigate the regime close to the final stages of the PBH evaporation, where the acceleration of the shocked region, namely the growing fireball, becomes dominant in the energy budget. In this limit, the term associated with the acceleration of the fireball dominates, leading to a different scaling of the post-shock pressure. This regime is not explored extensively in our simulations because of limited computational resources.

\paragraph{Distribution of injected energy.}
We start from the same energy balance as in Ref.~\cite{2013ApJ...768..113M}, namely
\begin{equation}\label{eq:phaseA_balance}
P(t)\, dt = dU_{\rm sh} + dK_{\rm sh} + dU_{\rm b} + dK_{\rm b}\,.
\end{equation}
The power is distributed among
\begin{itemize}
    % \item[(i)] the change of internal energy of the relativistic plasma inside the fireball,
    % $dU_{\rm fb}$;
    \item[(i)] the change of internal energy of the shocked region within the fireball, $dU_{\rm sh}$,
    \item[(ii)] the change of bulk kinetic energy of the shocked region, $dK_{\rm sh}$, for fixed swept-up mass;
    \item[(iii)] the internal energy imparted to the newly swept-up ambient plasma as it crosses the relativistic shock, $dU_{\rm b}$, so as to satisfy the relativistic jump conditions; and
    \item[(iv)] the corresponding change of bulk kinetic energy of this newly shocked material, $dK_{\rm b}$.
\end{itemize}

Both the ambient environment and the fireball contain relativistic plasma with equation of state $\rho =3p$, and so the total internal energy is simply 
\begin{equation}
U_{\rm sh} \simeq 3 p_{\rm sh} V \,,
\end{equation}
where $V = (4\pi/3)R ^{3}$, and $R$ is the radius of the shock. Differentiating gives
\begin{equation}
dU_{\rm sh} 
=
3V \, dp_{\rm sh}
+ 3p_{\rm sh}\, dV 
=
3V \, dp_{\rm sh}
+ 12\pi p_{\rm sh} R ^{2}\, dR \, .
\label{eq:dU_int}
\end{equation}
Next we compute the bulk kinetic energy.
In the PBH frame,
 \begin{equation} \label{eq:eps_kin_shell}
\rho_{\rm k,sh}
= 4p_{\rm sh}\bigl(\gamma_{\rm sh}^{2} - 1\bigr)\, .
\end{equation} 
For a fireball of volume $V_{\rm fb}$ and uniform $p_{\rm sh}$ and $\gamma_{\rm sh}$ (which is a good a posteriori approximation according to the hydrodynamical simulation), the total kinetic energy in the limit $\gamma_{\rm sh}^{2} \gg 1$ is
\begin{equation}
K_{\rm sh}
= \rho_{\rm k,sh}\, V_{\rm fb}
\simeq 4p_{\rm sh} \gamma_{\rm sh}^{2} V_{\rm fb}\,.
\label{eq:Ksh_def}
\end{equation}

For the change of kinetic energy of the already swept-up material, the swept mass is held fixed. Since the density inside the fireball is approximately uniform, then  
$dV_{\rm fb}=0$.\footnote{Each fluid element acquires its bulk kinetic energy upon being shocked, which is already captured by $dK_{\rm b}$. Including a $dV_{\rm fb}\neq 0$ term in $dK_{\rm sh}$ would double-count the kinetic energy of the newly swept-up material. The internal energy sector does not suffer from the same issue: the $3\,p_{\rm sh}\,dV$ term in $dU_{\rm sh}$ represents the thermodynamic $p\,dV$ work done by the shell's internal pressure as it expands, which is physically distinct from $dU_{\rm b}$, the shock-heating of newly swept-up material.}
Thus, the kinetic--energy change at fixed swept mass is
\begin{equation}
dK_{\rm sh}
=
4 \gamma_{\rm sh}^{2} \,V_{\rm fb}\, dp_{\rm sh}
+
4p_{\rm sh} V_{\rm fb}\, d(\gamma_{\rm sh}^{2})\,.
\label{eq:dK_shell}
\end{equation}

Finally, the volume that the shock has swept in time $dt$ is
\begin{equation}
dV = 4\pi R^{2}   \, dR\,,
\end{equation}
and the energy per unit swept volume is
\begin{equation} \label{eq:dU_swept}
dU_{\rm b} = w_{\rm sh} \gamma_{\rm sh}^{2} dV 
= 16\pi p_{\rm sh} R^{2} \gamma_{\rm sh}^{2} dR\,,
\end{equation}
while similarly for the kinetic energy
\begin{equation} \label{eq:dK_swept}
dK_{\rm b}
= \frac{w_{\rm sh} \gamma_{\rm sh}^{2}}{3}\, dV \simeq \frac{16\pi}{3} p_{\rm sh} R^{2}
\gamma_{\rm sh}^{2} dR\,.
\end{equation}
Consequently, as the shock sweeps additional ambient plasma, this material must be both heated and accelerated to the shocked-region Lorentz factor, and this process absorbs energy from the driving luminosity.

Together Eqs.~\eqref{eq:dU_int}, \eqref{eq:dK_shell}
\eqref{eq:dU_swept}, and \eqref{eq:dK_swept} yield 
\begin{equation}
\begin{split}
P(t)\,dt
&=
\bigl(3V + 4\gamma_{\rm sh}^{2}V_{\rm fb}\bigr)\,dp_{\rm sh}
\\
&\quad +
4p_{\rm sh} V_{\rm fb}\, d(\gamma_{\rm sh}^{2})
+
4\pi p_{\rm sh}R^{2}\,\bigl(3 + 4\gamma_{\rm sh}^{2}\bigr)\,dR \,.
\end{split}
\label{eq:L_full}
\end{equation}
We use Eq.~\eqref{eq:p_sh_cont}, Eq.~\eqref{eq:Gamma_equiv} taking $\Gamma \gg 1$, and also denote the ratio between the fireball and the total shock volume as $\lambda$, \emph{i.e.}, $V_{\rm fb} = \lambda V$, to write
\begin{align}
P(t)\,dt
&\approx
\left[
16\,V\,p_{\rm b}\,\Gamma
+\frac{256}{3}\,\lambda\,V\,p_{\rm b}\,\Gamma^{3}
\right] d\Gamma
\\[6pt]
&\quad
+
\left[
\frac{8}{3}\,\Gamma^{2}\,V\,(3+8\lambda\Gamma^{2})
\right] dp_{\rm b}
\\[6pt]
&\quad
+
\left[
\frac{256}{3}\,\pi\,R^{2}\,\Gamma^{4}\,p_{\rm b}
\right] dR \,.
\end{align}
We set the upstream pressure to be the ambient QGP pressure $p_{\rm b}=p_{\rm b}$ and set its differential to $dp_{\rm b}=0$, and finally using $\Gamma=(1-\dot R^2)^{-1/2}$ we obtain the master equation:
\begin{equation}
P(t)
=
\frac{64\pi}{3}\,p_{\rm b}\,\frac{R^{2}\dot R}{(1-\dot R^{2})^{2}}
\left[
\left(
1+\frac{16}{3}\frac{\lambda}{1-\dot R^{2}}
\right)R\ddot R + 4
\right]\,. \label{eq:master}
\end{equation}

Regarding the fractional volume parameter $\lambda$, we notice that hydrodynamical simulations predict a fireball of constant thickness which corresponds to $1- \lambda \sim \mathcal{O}(0.1)$. Its precise value only affects the overall normalization of the solution and does not modify the scaling relations derived below.

\paragraph{Early fireball formation.}

Eq.~\eqref{eq:master} simplifies considerably during the early fireball-growth phase and before the runaway stage of the PBH evaporation. In this phase the shock front moves at nearly constant ultrarelativistic velocity ($\dot R \simeq 1$), so that its acceleration is negligible, $R\ddot R \ll 1$.  In this limit Eq.~\eqref{eq:master} reduces to
\begin{equation}
P(t)
=
\frac{256\pi}{3}p_{\rm b} R^{2}\dot R\,\Gamma^{4}\,.
\end{equation}
Using $\partial_t(R^3)=3R^2\dot R $ and integrating over time yields the injected energy
\begin{equation}
E_{\rm inj}(t)
=
\frac{256\pi}{9}p_{\rm b}\Gamma^{4}R^3 \,.
\end{equation}
Using the relativistic shock jump condition in Eq.~\eqref{eq:p_sh_cont} we have
\begin{equation}
E_{\rm inj}(t)
=
\frac{4\pi}{p_{\rm b}}\,p_{\rm sh}^2 R^3 \,.
\end{equation}
This relation implies
\begin{equation}
p_{\rm sh}(R,t)
=
\sqrt{\frac{E_{\rm inj}(t) \, p_{\rm b}}{4\pi R^3}}\,.
\label{eq:fireball_scaling}
\end{equation}
This result is consistent with the scaling argument used in the main text. Indeed, the integral expression used in Eq.~\eqref{eq:p_early_fb} contains an undetermined coefficient $(3+2m-n)$ that depends on the detailed radial profiles of $p(r)$ and $\gamma(r)$. For the self-similar solution relevant for the fireball growth, one finds $2m-n\simeq0$, which reduces Eq.~\eqref{eq:p_early_fb} to the same normalization obtained above.

Physically, the limit $R\ddot R\ll1$ corresponds to the regime in which the energy injected by the PBH is predominantly converted into internal energy of the shocked plasma and into heating of newly swept-up material. In the energy budget of Eq.~\eqref{eq:phaseA_balance}, this corresponds to neglecting the change of kinetic energy of the already-swept region, \emph{i.e.}, $dK_{\rm sh} \approx 0$, whose contribution is suppressed once the fireball moves with nearly constant relativistic velocity $\dot R \simeq 1$.

\paragraph{Final stages of PBH evaporation.}

We consider the complementary limit close to the PBH explosion. At this stage the injected power increases rapidly as $P(t)\propto (\tau-t)^{-2/3}$ (see Eq.~\eqref{eq:TotalInjectionTime}), and the pressure driving the shock therefore grows quickly. As a result, the velocity of the fireball increases with time, implying $\ddot R>0$. In this regime the acceleration term dominates the dynamics of the shock front, $R\ddot R \gg 1$. In this case the constant term inside the brackets of Eq.~\eqref{eq:master} can be neglected. Assuming again an ultra-relativistic shock with $\Gamma^2=(1-\dot R^2)^{-1}\gg1$, Eq.~\eqref{eq:master} becomes
\begin{equation}
P(t)
=
6\pi\lambda
\frac{p_{\rm sh}^{3}}{p_{\rm b}^{2}}
R^{3}\dot R\,\ddot R \,.
\end{equation}
Noting that $\dot R\,\ddot R
=
\frac{1}{2}\partial_t(\dot R^{2})$ and integrating over time yields
 \begin{equation}
E_{\rm inj}(t)
=
3\pi\lambda
\frac{p_{\rm sh}^{3}}{p_{\rm b}^{2}}
R^{3}\dot R^{2}\,.
\end{equation}
In the relativistic regime $\dot R\simeq1$ and using again Eq.~\eqref{eq:p_sh_cont}, we solve for the post-shock pressure
\begin{equation}
p_{\rm sh}(R,t)
=
\left(
\frac{E_{\rm inj}(t)p_{\rm b}^{2}}
{3\pi\lambda R^{3}}
\right)^{1/3}\, ,
\end{equation}
which implies the scaling $p_{\rm sh}(R)\propto R^{-1}$.

%%%%%%%%%%%%%%%%%%%%%%%%%%%%%%%%%%%%
\setcounter{equation}{0}
\setcounter{figure}{0}
\setcounter{table}{0}
\makeatletter
\renewcommand{\theequation}{S\arabic{equation}}
\renewcommand{\thefigure}{S\arabic{figure}}

\bibliography{biblio}
%\bibliographystyle{apsrev4-1}
%\bibliography{ref}

\end{document}